\documentclass[twocolumn,aps,pre,epsf, superscriptaddress]{revtex4-2}

\usepackage{amsmath}
\usepackage{amssymb}
\usepackage{epsfig}
\usepackage[T1]{fontenc}
\usepackage[utf8]{inputenc}
\usepackage{color}
\usepackage[normalem]{ulem}
\usepackage{tabularx}
\usepackage{enumitem}

\usepackage{graphicx}
\usepackage{subcaption}
\usepackage{ragged2e}
\DeclareCaptionJustification{justified}{\justifying}
\usepackage{diagbox} % diagonal separation of cell in a table

\usepackage{dcolumn}% Align table columns on decimal point
\usepackage{bm}% bold math
\usepackage{hyperref}% add hypertext capabilities
\usepackage{xcolor}
\newcommand\myworries[1]{\textcolor{blue}{#1}}
\renewcommand\myworries[1]{} 
\def\d{\mathrm{d}}
\begin{document}
	
\title{Geometric clusters in the overlap of the Ising model}

\author{Michail Akritidis}
\email{michail.akritidis@coventry.ac.uk}
\affiliation{Centre for Fluid and Complex Systems, Coventry University, Coventry, CV1 5FB, United Kingdom}
 
\author{Nikolaos G. Fytas}
\email{nikolaos.fytas@essex.ac.uk}
\affiliation{Institut für Physik, Technische Universität Chemnitz, 09107 Chemnitz, Germany}
\affiliation{Department of Mathematical Sciences, University of Essex, Colchester CO4 3SQ, United Kingdom}

\author{Martin Weigel}
\email{martin.weigel@physik.tu-chemnitz.de}
\affiliation{Institut für Physik, Technische Universität Chemnitz, 09107 Chemnitz, Germany}

\date{\today}
	
%==================================================================================
%                             ABSTRACT
%==================================================================================
	
\begin{abstract}
		
We study the percolation properties of geometrical clusters defined in the overlap space of two statistically independent replicas of a square-lattice Ising model that are simulated at the same temperature. In particular, we consider two distinct types of clusters in the overlap, which we dub soft- and hard-constraint clusters, and which are subsets of the regions of constant spin overlap. By means of Monte Carlo simulations and a finite-size scaling analysis we estimate the transition temperature as well as the set of critical exponents characterizing the percolation transitions undergone by these two cluster types. The results suggest that both soft- and hard-constraint clusters percolate at the critical temperature of the Ising model and their critical behavior is governed by the correlation-length exponent $\nu = 1$ found by Onsager. At the same time, they exhibit non-standard and distinct sets of exponents for the average cluster size and percolation strength.
		
\end{abstract}
	
\maketitle

\section{\label{sec:intoduction} Introduction}

In simple ferromagnets the definition of an order parameter is straightforward, and in the vast majority of cases one simply considers the magnetization~\cite{wj:chem}. For more intricate problems such as some frustrated systems with quenched disorder and certain quantum-spin models (easily measurable) order parameters are harder to come by~\cite{wegner:71,kitaev:03,castelnovo:07,johnston:12}. In spin glasses the free-energy landscape has many minima that are occupied at low temperatures, but which are not related to each other by simple symmetry transformations (such as, e.g., spin-flip symmetry)~\cite{binder:86a}. For such systems it was proposed to consider {\em self-consistent\/} definitions of ordering by constructing order parameters that capture the tendency of such systems to occupy the same set of metastable configurations. As Parisi showed~\cite{parisi:83}, such {\em overlap} definitions allow one to describe the transition to a short-range ordered spin-glass phase~\cite{mezard:book}. Different order parameters in general might also lead to different scaling and associated critical exponents, however. One of the simplest examples of this type is the ordering in the overlap space of the square-lattice ferromagnetic Ising model which is studied here. The overlap is a more general and conceptually robust order parameter also for this ferromagnetic system, and so it is worthwhile studying its behavior. In addition, such setups might have some more general relevance, for example for the description of layered Ising models with (asymptotically) vanishing coupling as present, for example, in multiplex networks~\cite{boccaletti:14}, where ordering might occur independently in the different levels of the graphs. More generally, a system composed of two or several layers of magnetic material with Ising-like anisotropy that are very weakly coupled would also be represented by such a model.
	
The study of ordering transitions of spin systems from a {\em geometrical\/} perspective has greatly enhanced our understanding of phase transitions. Such approaches naturally fall into the realm of percolation theory~\cite{stauffer_introduction_1994}, in which spin systems can be described by appropriately defined clusters, capable to encode the critical behavior of the system. Fortuin and Kasteleyn (FK)~\cite{fortuin_random-cluster_1972} showed that the $q$-state Potts model is equivalent to a site-bond correlated percolation problem, where clusters are defined as neighboring parallel spins, and bonds between them are deleted with a certain temperature-dependent probability. Such clusters percolate at the transition temperature and, even more importantly, they encode the critical behavior of the system, as suitably defined cluster exponents are found to be identical to the thermal ones (such clusters were independently also analyzed by Coniglio and Klein~\cite{coniglio:80a}). Apart from the conceptual importance of these results, they also allowed for the construction of powerful Monte Carlo algorithms by Swendsen and Wang~\cite{swendsen_nonuniversal_1987} and Wolff~\cite{wolff_collective_1989}, where whole FK clusters are flipped in contrast to local update schemes, such as the Metropolis algorithm~\cite{metropolis_equation_1953}. It is well established that the main advantage of this approach is the reduction of autocorrelation times in the vicinity of the critical point as compared to local update schemes.
	
While FK clusters encode the critical behavior of the system, this is generally not the case for the geometrical or spin clusters~\cite{binder:76b}. Instead, such clusters undergo a percolation transition that is normally distinct from the thermal one, with the related critical exponents also being different from the thermal ones (see Refs.~\cite{coniglio_correlated_2009,saberi_recent_2015} for a review). In two dimensions, however, the geometrical clusters percolate at the thermal transition point~\cite{coniglio:77} and it has been shown that they encode the tricritical behavior of the site-diluted $q=1$ Potts model~\cite{stella_scaling_1989}. In fact, analogous interrelations have been reported for the more general $q$-state Potts model and its diluted version for $0 \le q \le 4$ in both analytical and numerical terms; see Refs.~\cite{stella_scaling_1989,vanderzande_bulk_1989,vanderzande_fractal_1992,janke_geometrical_2004,janke_fractal_2005} and references therein. It is hence a natural question to investigate how such geometrical clusters defined in the overlap of the Ising model behave, and whether they percolate at the critical temperature of the ordering transition.

An additional motivation relates to the rather less clear connection between clusters and thermal phase transitions in spin-glass systems~\cite{arcangelis:91}. For such models the FK representation does not properly describe the phase transition, and the constructed clusters percolate at a temperature way above the spin-glass transition~\cite{machta_percolation_2008, fajen_percolation_2020}. Several types of clusters in the overlap of two copies, including geometric clusters, have been considered as potential candidates for the construction of cluster updates~\cite{houdayer_cluster_2001, joerg:05, machta_percolation_2008}. It is found there that the spin-glass transition is connected to the onset of a density difference of the two largest clusters of a suitable type, while percolation of such clusters occurs already above the spin-glass transition point \cite{machta_percolation_2008,munster_cluster_2023}. As we shall see below, the clusters considered for the Ising model in the present work are related to some of the cluster types discussed in the context of the spin-glass transition, cf.\ Ref.~\cite{munster_cluster_2023}.

The rest of this paper is organized as follows: In Sec.~\ref{sec:model} we introduce the replicated Ising model and the associated concept of soft- and hard-constraint clusters. We further outline the cluster-update Monte Carlo scheme used in the following to study the problem numerically. In Sec.~\ref{sec:observables} we elaborate on the relevant observables whose percolation properties are investigated in detail. In Sec.~\ref{sec:results} we report on a finite-size scaling analysis of the simulation data leading to estimates of the percolation temperature $T_{\text{p}}$ and the critical exponents $\nu$, $\beta/\nu$, and $\gamma/\nu$ characterizing the transition for the two cluster types. We also comment on the influence of corrections to scaling in the estimation of the exponent ratios $\beta/\nu$ and $\gamma /\nu$ when different definitions are used for the involved observables. Finally, in Sec.~\ref{sec:conclusions} we provide a summary of our work and an outlook.

%==================================================================================
%                              THE MODEL
%==================================================================================
	
%\section{The Model and observables}	\label{sec:the model and observables}
	
\section{Model and Simulation Details} 
\label{sec:model}

We study the nearest-neighbor, zero-field Ising model with Hamiltonian 
\begin{equation}\label{eq:ising_hamiltonian}
	\mathcal{H} = -J \sum_{\left\langle i,j\right\rangle} s_i s_j,
\end{equation}
where $J > 0$ indicates ferromagnetic interactions, $s_i = \pm 1$ denotes the spin on lattice site $i$, and $\left\langle \ldots \right\rangle$ refers to summation over nearest neighbors only. 
	
We now consider two identical copies of the system, each being described by the Hamiltonian~\eqref{eq:ising_hamiltonian}, resulting in two spin configurations, $s_i^{(1)}$ and $s_i^{(2)}$. Both systems are at the same temperature and do not interact with each other, thus being statistically independent. We then consider overlap variables $q_i$ at each site, i.e.,
\begin{equation}\label{eq:qi_prod}
	q_i = s^{(1)}_{i}\,s_i^{(2)}.
\end{equation}
The behavior of this overlap parameter has been discussed in depth for the case of spin-glass systems (see, e.g., Ref.~\cite{mezard:book}). Since $q_i = \pm 1$, the overlap configuration has the same configuration space $\{\pm 1\}^N$ as the spin lattices themselves (here, $N$ denotes the total number of lattice sites). As a consequence, all derived observables usually considered for $\{s_i\}$ can also be defined for $\{q_i\}$. While for spin glasses this approach leads to the spin-glass susceptibility and related quantities such as the spin-glass correlation length, and exploitation of the available gauge symmetry of the couplings results in a possible approach towards understanding the spin-glass phase \cite{haake:85}, the behavior of the overlap has hardly been studied for the case of ferromagnets.

\begin{figure}[]% pwrap soft constraint
	\begin{subfigure}[]{0.48\columnwidth}
	    \includegraphics[scale=0.44]{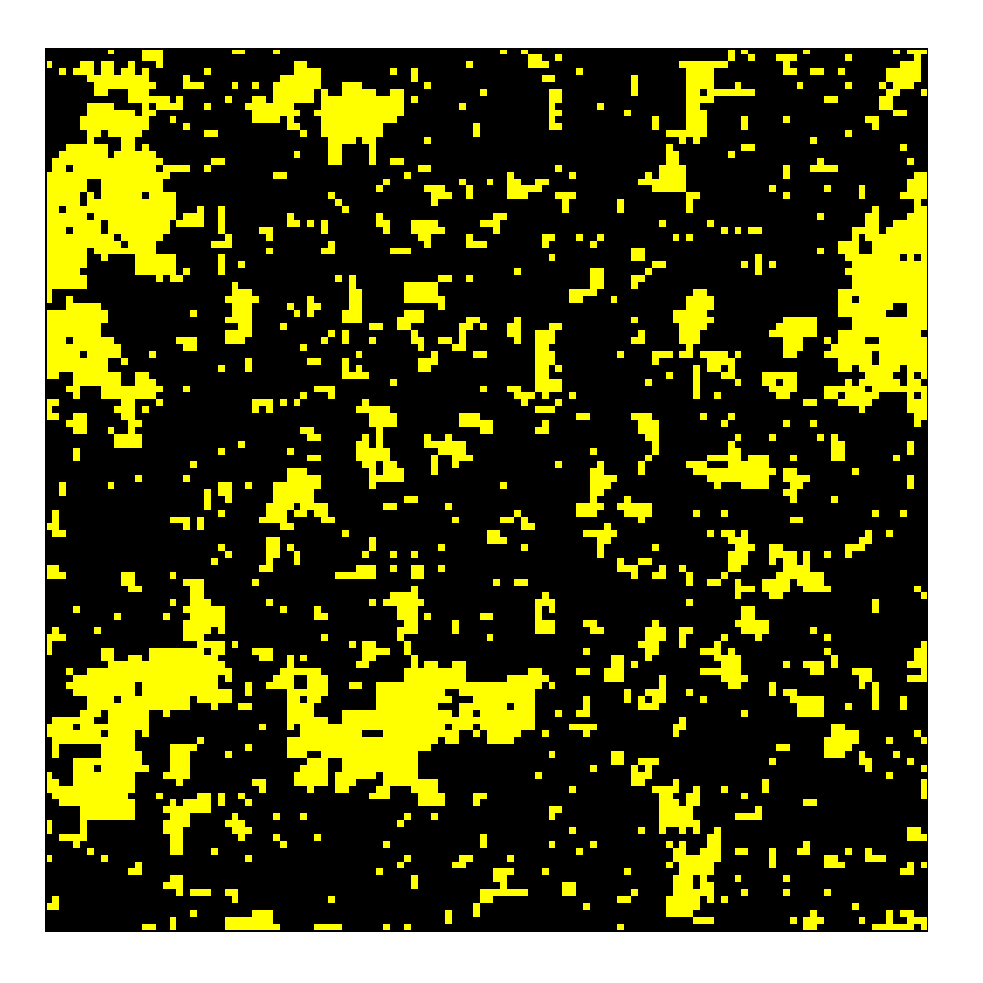}
        \caption{Replica 1.}
		\label{subfig:q2-replica1}
	\end{subfigure}
	\begin{subfigure}[]{0.48\columnwidth}
	    \includegraphics[scale=0.44]{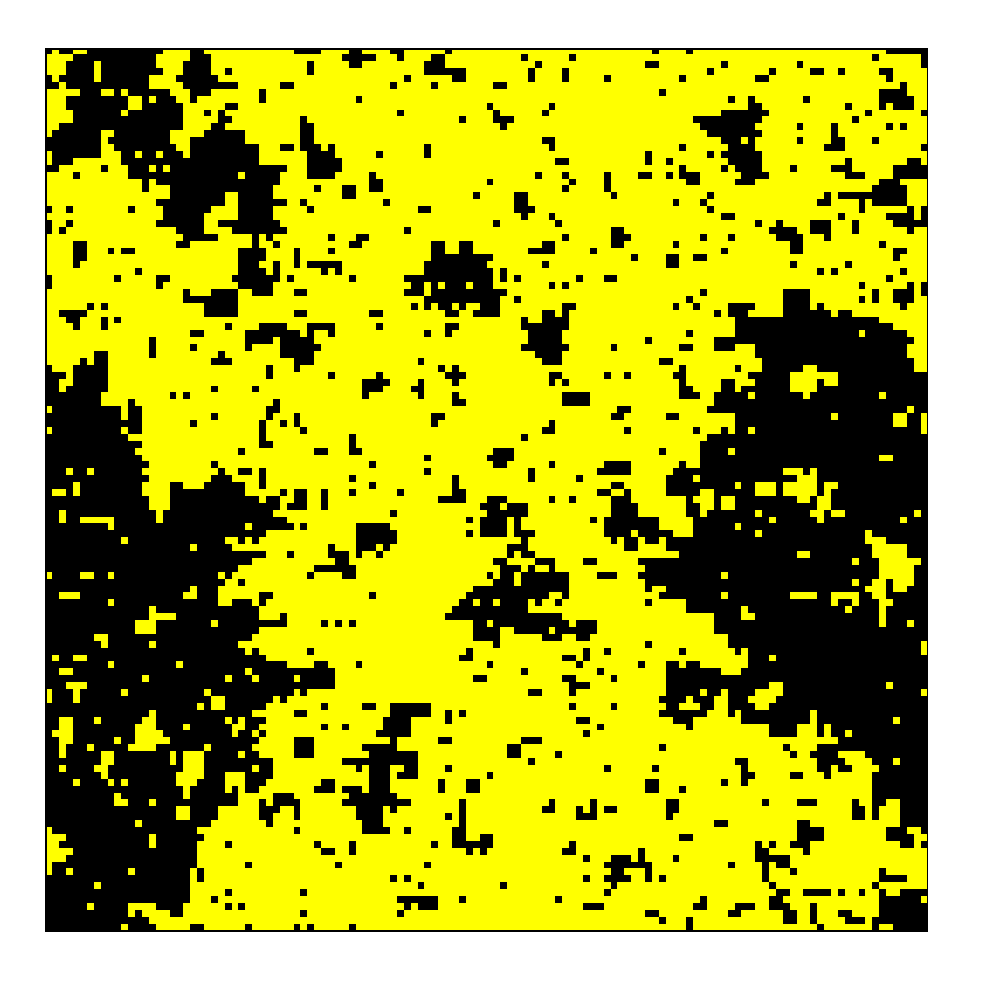}
        \caption{Replica 2.}
		\label{subfig:q2-replica2}
	\end{subfigure}
	
	\begin{subfigure}[]{0.48\columnwidth}
	    \includegraphics[scale=0.44]{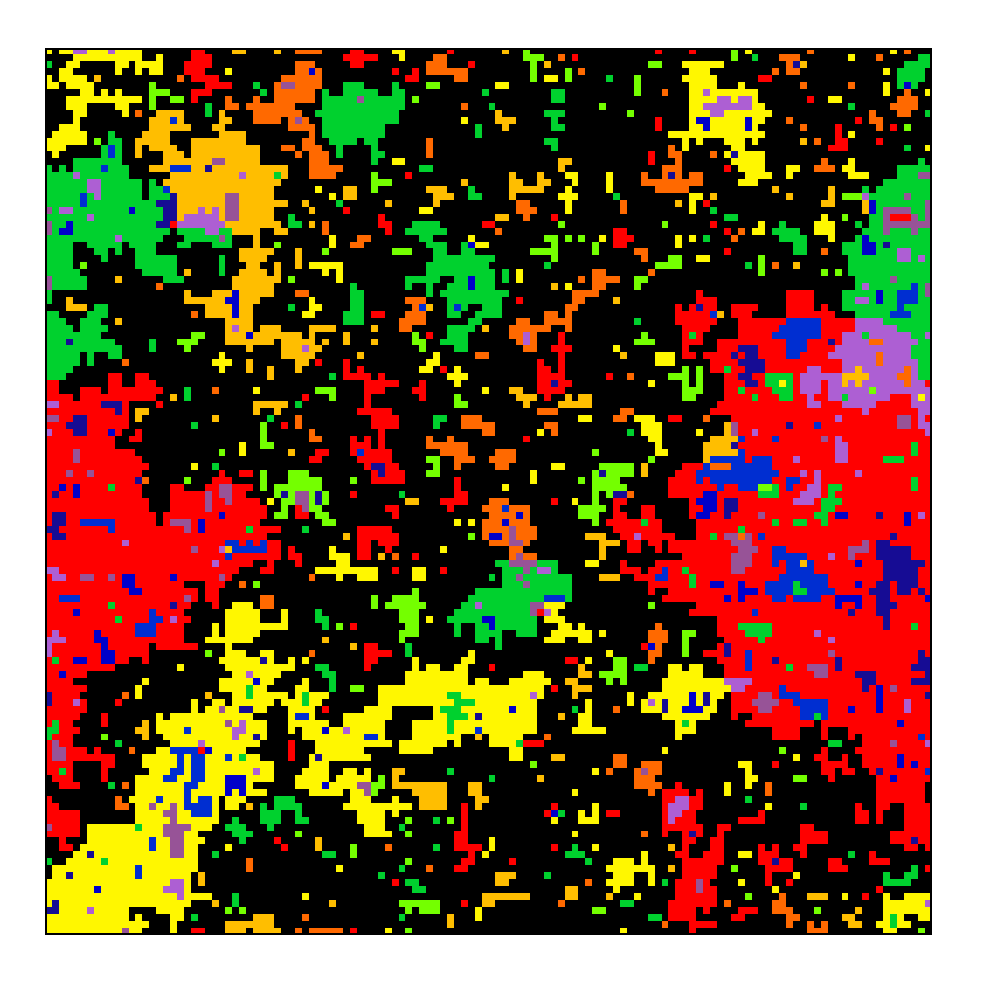}
		\caption{Soft-constraint clusters.}
        \label{subfig:q2-soft_constraint}
	\end{subfigure}
	\begin{subfigure}[]{0.48\columnwidth}
		\includegraphics[scale=0.44]{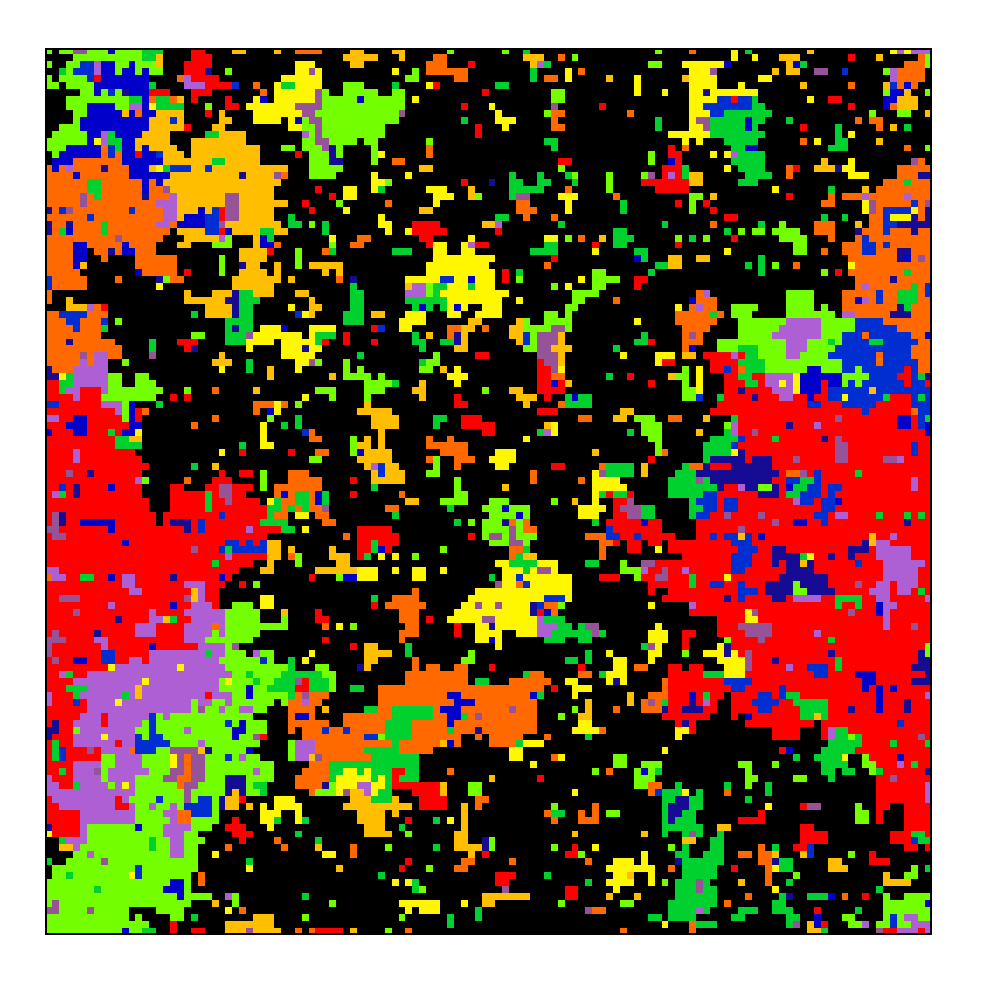}
        \caption{Hard-constraint clusters.}
		\label{subfig:q2-hard_constraint}
	\end{subfigure}
    \caption{Spin configurations as well as soft- and hard-constraint geometrical clusters of the two-dimensional Ising model at the Ising critical temperature $ T_{\text{c}} $. (\subref{subfig:q2-replica1}) Spin configuration of the first replica. (\subref{subfig:q2-replica2}) Spin configuration of the second replica.  (\subref{subfig:q2-soft_constraint}) The resulting soft-constraint clusters. (\subref{subfig:q2-hard_constraint}) The resulting hard-constraint clusters. In (\subref{subfig:q2-soft_constraint}) and (\subref{subfig:q2-hard_constraint}) all clusters, apart from the largest percolating one, are assigned colours at random. For the largest percolating cluster of both the soft- and hard-constraint definitions, the same colour (black) is assigned.
    %Note that in our setup we allow the black colour to be assigned \textit{only} to the largest percolating cluster.
	\label{fig:q2-snapshot_L128}}
\end{figure}

In overlap space we may then define geometrical clusters, i.e., sets of neighboring lattice sites with the same values of the overlap, that can be formed in two particular ways:
\begin{enumerate}[itemsep=2pt,parsep=2pt,leftmargin=20pt,label=(\arabic*)]
    \item \emph{Soft-constraint clusters} are created by joining spins on neighboring sites $(i,j)$ with $q_i = q_j$.
    \item \emph{Hard-constraint clusters} are formed by joining neighboring spins with  $s_i^{(1)} = s_j^{(1)}$ and $s_i^{(2)}=s_j^{(2)}$.
\end{enumerate}
From the above it is obvious that the hard-constraint clusters trivially satisfy $q_i=q_j$ and they are a subset of the soft-constraint clusters. The construction of these clusters is illustrated through the snapshots shown in Fig.~\ref{fig:q2-snapshot_L128}, where configurations of the two replicas are shown (top row) along with the soft- and hard-constraint clusters (bottom row) for a system of linear size $L = 128$ at the critical temperature of the square-lattice ferromagnetic Ising model. It is apparent that the typical clusters in the overlap are smaller than those in the two replicas, and that the hard constraint leads to smaller clusters than the soft constraint. We note that in the spin-glass setup, the prescription for soft-constrained clusters leads to what is there known as Houdayer clusters \cite{houdayer_cluster_2001}, while the hard constraint corresponds to a geometric-cluster version of the Chayes-Machta-Redner construction \cite{chayes:98,munster_cluster_2023}.

In the present work, we studied these clusters for the case of the Ising model on the square lattice.
%For the one replica case the soft and hard constraint clusters are identical. Here we study the percolation properties of such clusters for the case of two replicas. \myworries{Should I put a figure?}
In particular, we simulated the Ising model by considering two replicas on the square lattice with periodic boundary conditions for a range of temperatures including the exact Ising critical temperature, i.e.,  $ T_{\text{c}} = 2/ \left(1+\sqrt{2}\right) \approx 2.269185 $. Configurations were generated via the Swendsen-Wang algorithm~\cite{swendsen_nonuniversal_1987} applied to systems of linear sizes in the range $8\le L\le 2048$.
%$L = $  $8$, $10$, $16$, $20$, $32$, $ 40 $, $ 50 $, $ 64 $, $ 80 $, $ 100 $, $ 128 $, $ 160 $, $ 200 $, $ 256 $, $ 320 $, $ 400 $, $ 512 $, $ 640 $, $ 800 $, $ 1024 $, $ 1280 $, $ 1600 $ and $ 2048 $. 
For all system sizes and on each replica the total number of simulation steps was $1.1\times \tau_{\text{int, E}} \times 10^5$ sweeps, of which $\tau_{\text{int, E}} \times 10^4$ sweeps were discarded during equilibration. Here, $\tau_{\text{int, E}}$ denotes the integrated autocorrelation time of the energy~\cite{weigel_error_2010}. After every $\tau_{\text{int, E}}$ sweeps a measurement was taken, resulting in  $10^5$ measurements per run. The estimates of $\tau_{\text{int, E}}$, rounded up to the next largest integer, varied from $5$ sweeps for $L = 8 $ to $15$ sweeps for $L = 2048$. Note that in order to identify wrapping clusters we employed the method of Machta \textit{et al.}~\cite{machta_invaded_1996}. Finally, for all curve fitting performed throughout this paper we restricted ourselves to data with $L\geq L_{\rm min}$, adopting the standard $\chi^{2}$ test for goodness of the fit. Specifically, we considered a fit as being acceptable only if $Q > 0.01$, where $Q$ is the quality-of-fit parameter~\cite{numrec}. 
	
%==================================================================================
%                              OBSERVABLES
%==================================================================================
	
\section{Observables} 
\label{sec:observables}
	
\begin{figure}[]% pwrap soft constraint
	\begin{subfigure}[]{0.8\columnwidth}
	    %{\Large\resizebox{1.0\columnwidth}{!}{{\Large\input{Figures/q2-soft_pwrap_xory-zoomed.tex}}}}
        %q12-soft_pwrap_xory-zoomed.pdf}
        %\includegraphics[scale=0.23]{./figures/test1.png}
        \includegraphics[scale=0.42]{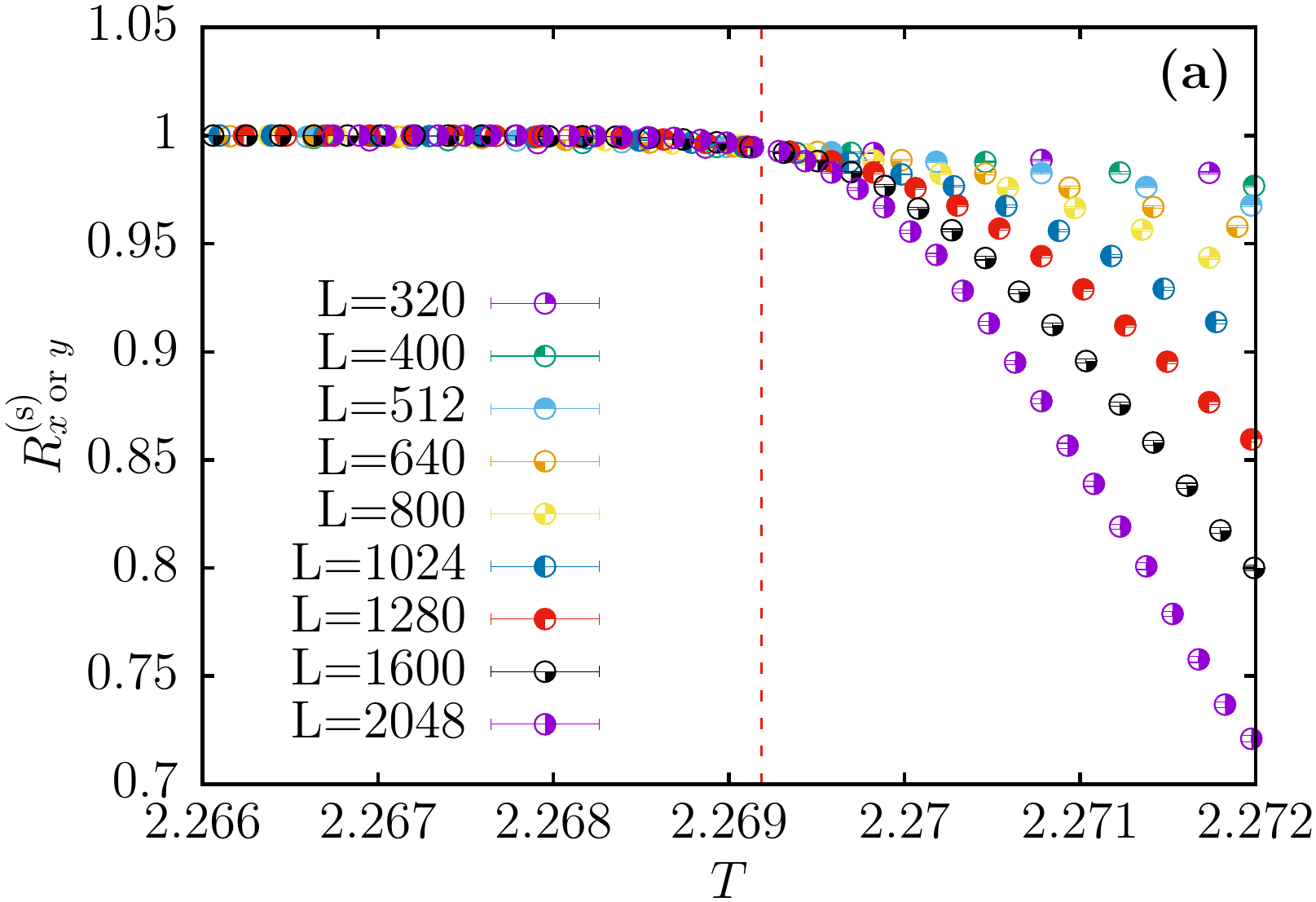}
		\label{subfig:q2_pwrap_xory_soft-zoomed}
	\end{subfigure}
 
	\begin{subfigure}[]{0.8\columnwidth}
	    %{\Large\resizebox{1.0\columnwidth}{!}{\input{Figures/q2-soft_pwrap_xandy-zoomed.tex}}}
        %\includegraphics[]{./figures/q2-soft_pwrap_xandy-zoomed.pdf}
        \includegraphics[scale=0.42]{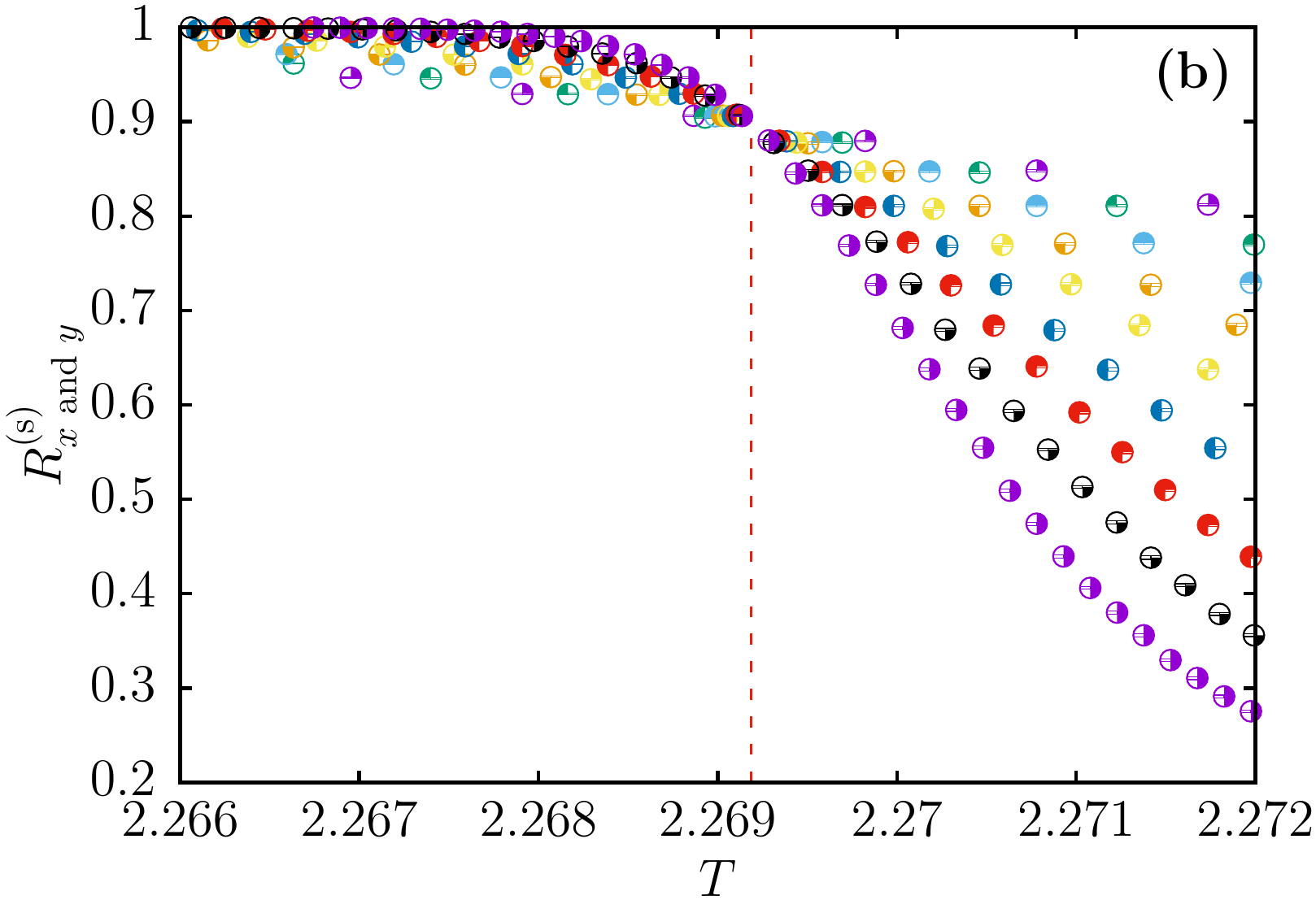}
		\label{subfig:q2_pwrap_xandy_soft-zoomed}
	\end{subfigure}
	
	\begin{subfigure}[]{0.8\columnwidth}
	    %{\Large\resizebox{1.0\columnwidth}{!}{\input{Figures/q2-soft_pwrap_x-zoomed.tex}}}
        %\includegraphics[width=1.0\textwidth]{./figures/q2-soft_pwrap_x-zoomed.pdf}
        \includegraphics[scale=0.42]{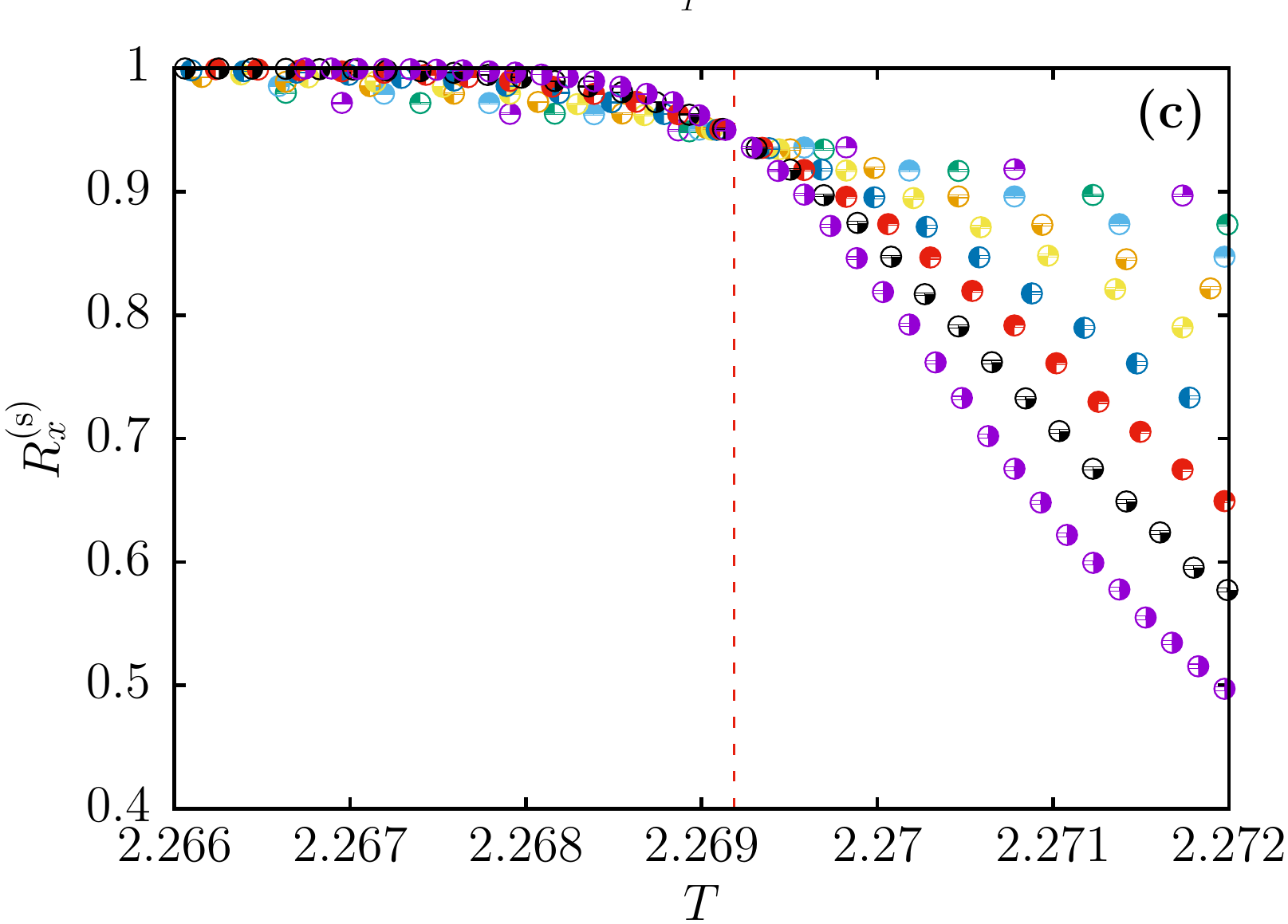}
		\label{subfig:q2_pwrap_x_soft-zoomed}
	\end{subfigure}
 
	\begin{subfigure}[]{0.8\columnwidth}
		%{\Large\resizebox{1.0\columnwidth}{!}{\input{Figures/q2-soft_pwrap_xnoty-zoomed.tex}}}
        %\includegraphics[width=1.0\textwidth]{./figures/q2-soft_pwrap_xnoty-zoomed.pdf}
        \includegraphics[scale=0.42]{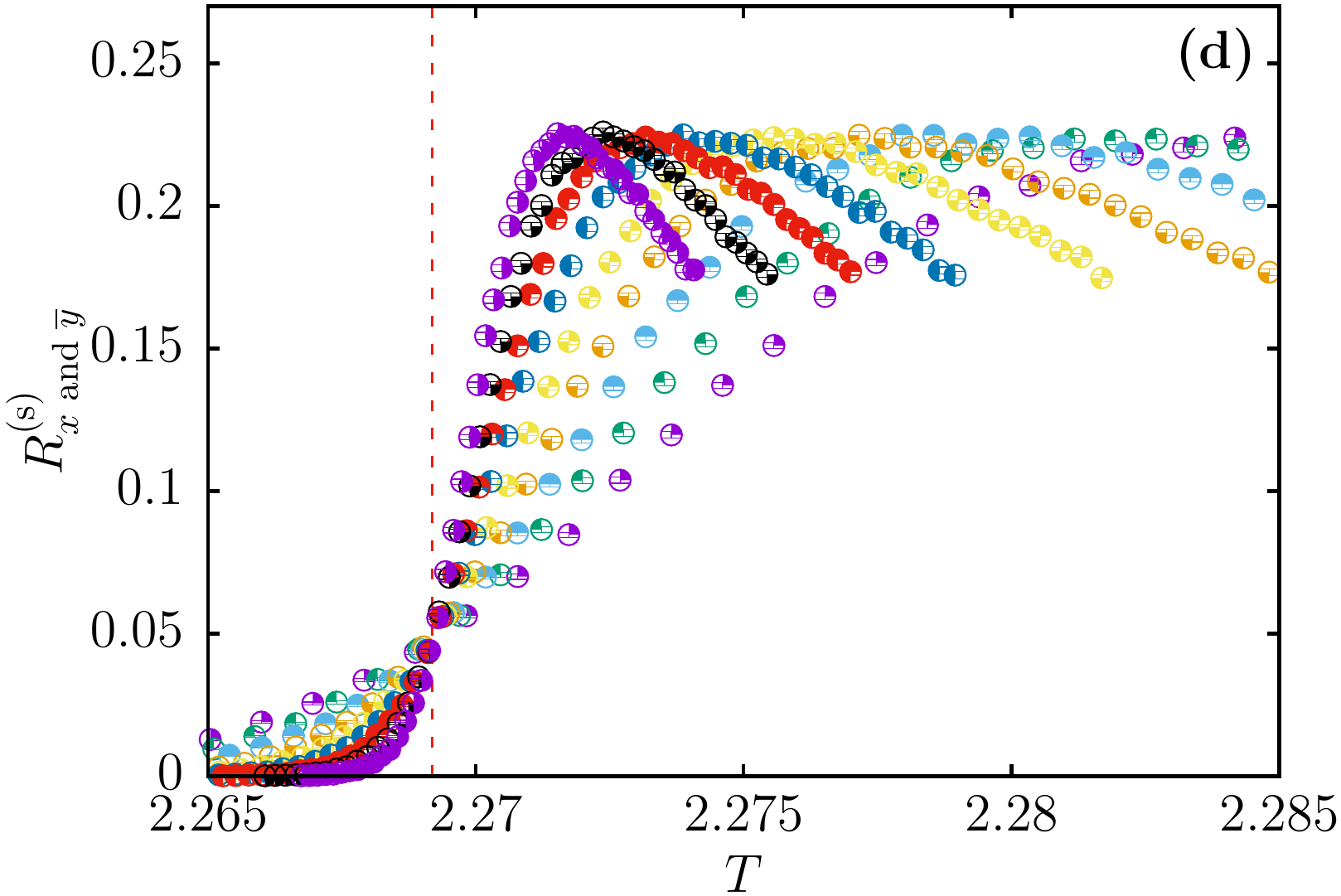}
		\label{subfig:q2_pwrap_xnoty_soft-zoomed}
	\end{subfigure}
	
	%\caption{Wrapping probabilities of the soft constraint clusters $R^{\text{(s)}}$  as function of temperature $ T $, for the largest system sizes $L$ considered of \subref{subfig:q2_pwrap_xory_soft-zoomed} $  R_{\text{x or y}}^{\text{(s)}}  $ \subref{subfig:q2_pwrap_xandy_soft-zoomed} $  R_{\text{x and y}}^{\text{(s)}}  $ \subref{subfig:q2_pwrap_x_soft-zoomed} $  R_{\text{x}}^{\text{(s)}}  $ \subref{subfig:q2_pwrap_xnoty_soft-zoomed} $ R_{\text{x and } \overline{\text{y}}}^{\text{(s)}} $. The dashed vertical line marks the transition temperature of the 1-replica Ising model.
    \caption{Wrapping probabilities $R^{\text{(s)}}$ of the soft-constraint clusters as a function of temperature $T$ for (a) $R_{\text{x or y}}^{\text{(s)}}$, (b) $R_{\text{x and y}}^{\text{(s)}}$, (c) $R_{\text{x}}^{\text{(s)}}$, and (d) $R_{\text{x and } \overline{\text{y}}}^{\text{(s)}}$. Only results for the larger system sizes are shown ($L \geq 320$). Here and in the following, the dashed vertical lines mark the critical temperature $T_\mathrm{c}$ of the Ising model.
	\label{fig:pwrap_soft}}
\end{figure}

\begin{figure}[] % pwrap hard constraint
	\begin{subfigure}[]{0.8\columnwidth}
	    %{\Large\resizebox{1.0\columnwidth}{!}{{\Large\input{Figures/q2-hard_pwrap_xory-zoomed.tex}}}}
        %\includegraphics[width=1.0\textwidth]{./figures/q2-hard_pwrap_xory-zoomed.pdf}
        \includegraphics[scale=0.42]{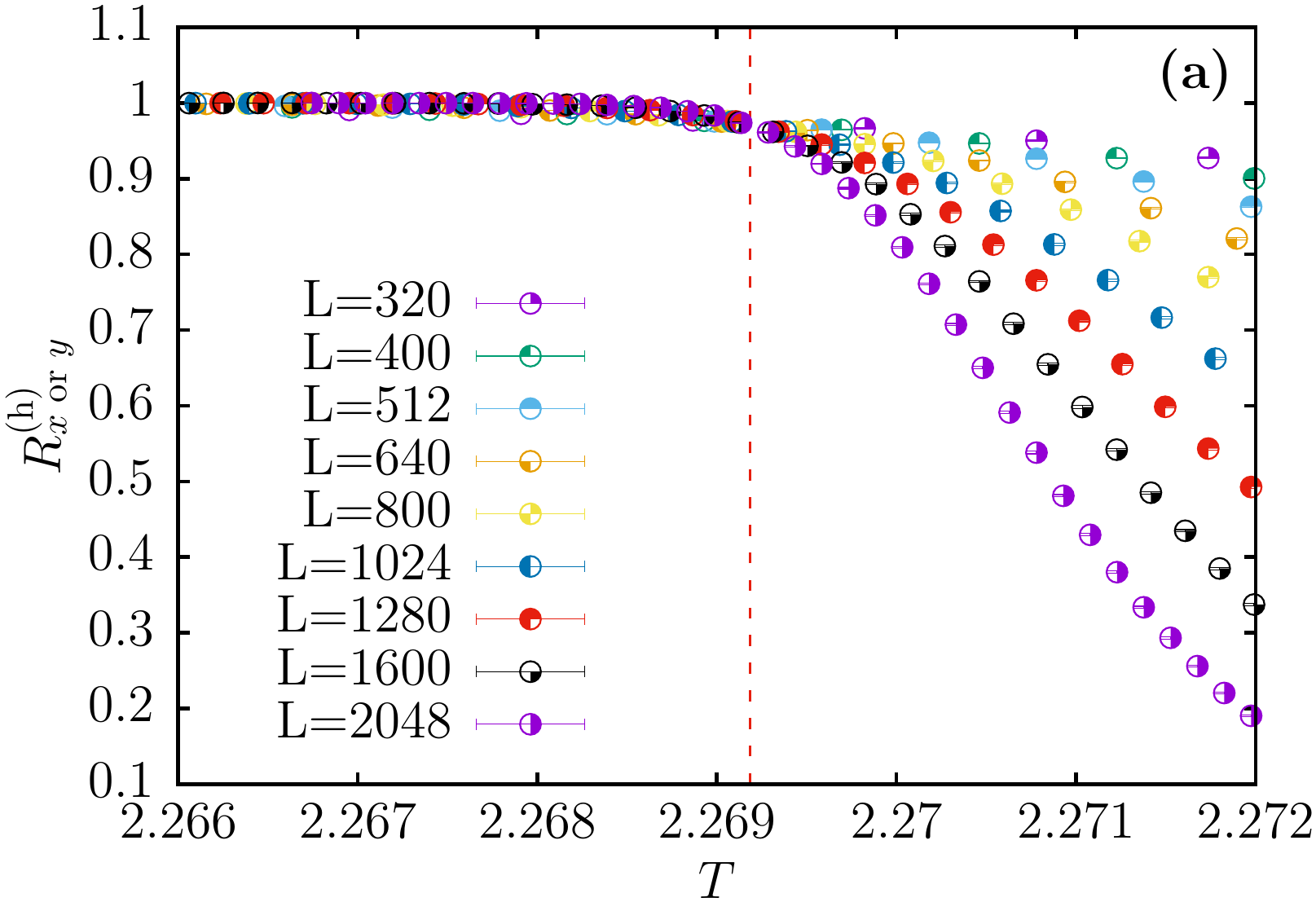}
		\label{subfig:q2_pwrap_xory_hard-zoomed}
	\end{subfigure}
	\begin{subfigure}[]{0.8\columnwidth}
	    %{\Large\resizebox{1.0\columnwidth}{!}{\input{Figures/q2-hard_pwrap_xandy-zoomed.tex}}}
        %\includegraphics[width=1.0\textwidth]
        \includegraphics[scale=0.42]{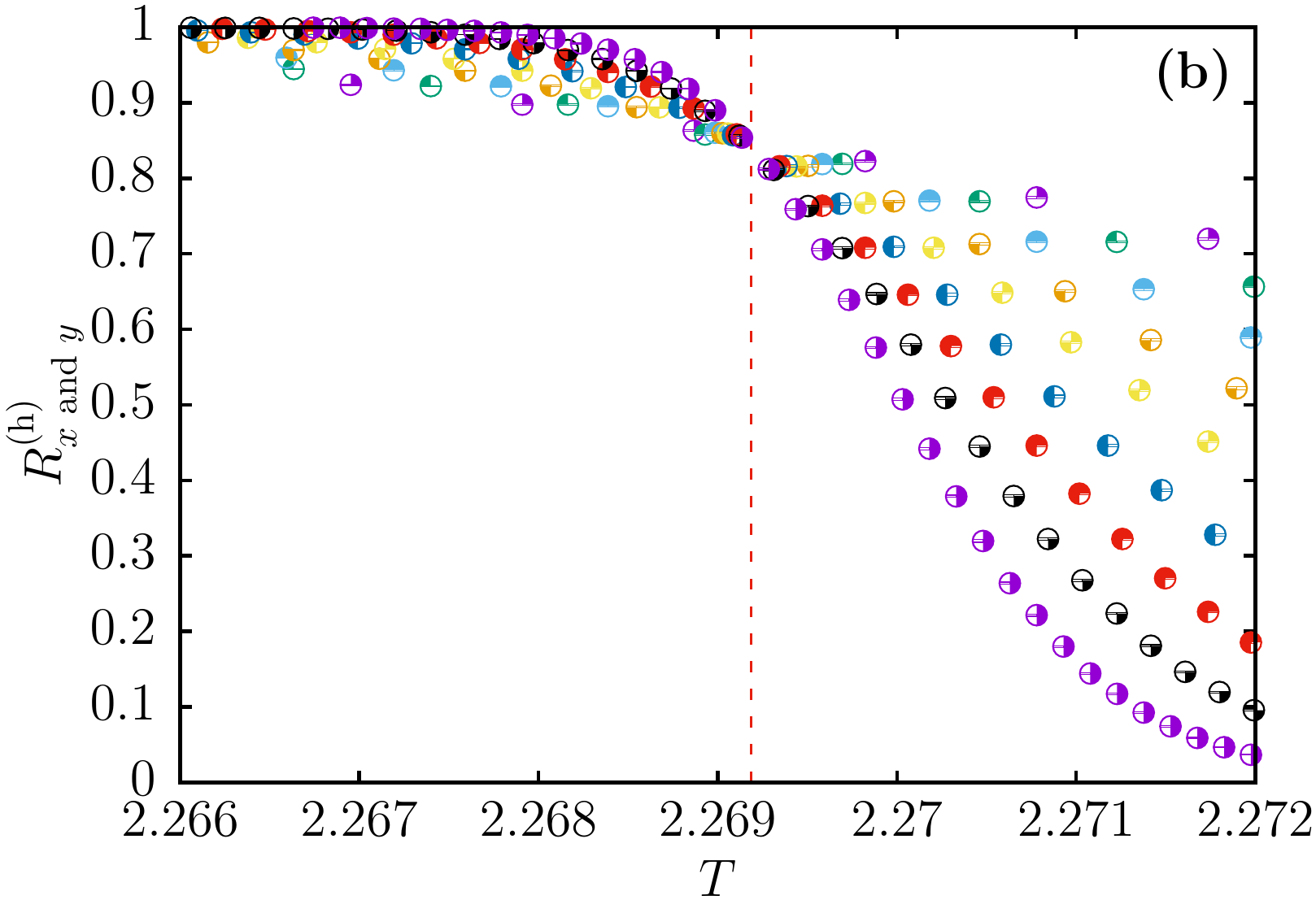}
		\label{subfig:q2_pwrap_xandy_hard-zoomed}
	\end{subfigure}
	
	\begin{subfigure}[]{0.8\columnwidth}
	    %{\Large\resizebox{1.0\columnwidth}{!}{\input{Figures/q2-hard_pwrap_x-zoomed.tex}}}
        %\includegraphics[width=1.0\textwidth]{./figures/q2-hard_pwrap_x-zoomed.pdf}
        \includegraphics[scale=0.42]{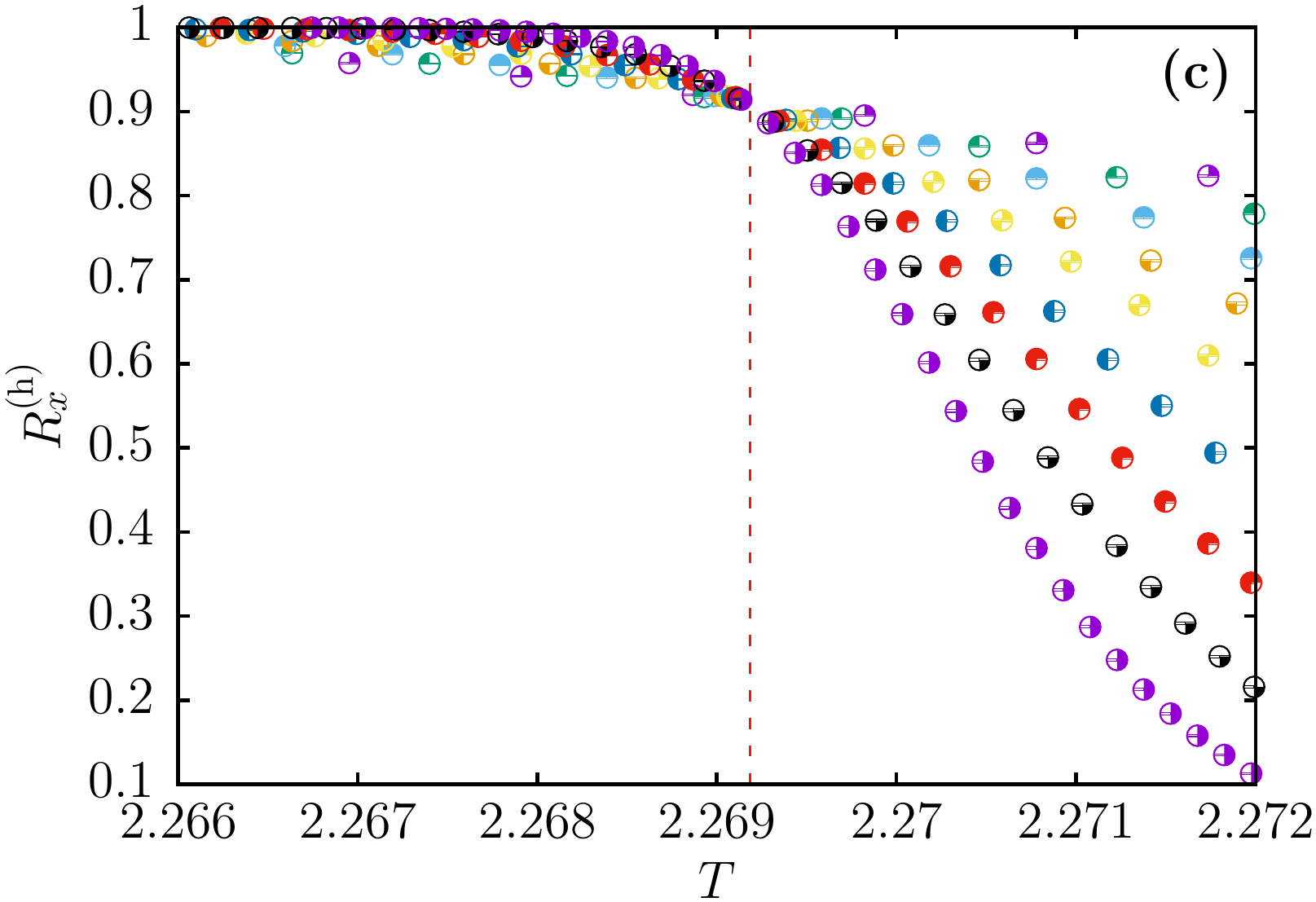}
		\label{subfig:q2_pwrap_x_hard-zoomed}
	\end{subfigure}
	\begin{subfigure}[]{0.8\columnwidth}
    	%{\Large\resizebox{1.0\columnwidth}{!}{\input{Figures/q2-hard_pwrap_xnoty-zoomed.tex}}}
        %\includegraphics[width=1.0\textwidth]{./figures/q2-hard_pwrap_xnoty-zoomed.pdf}
        \includegraphics[scale=0.42]{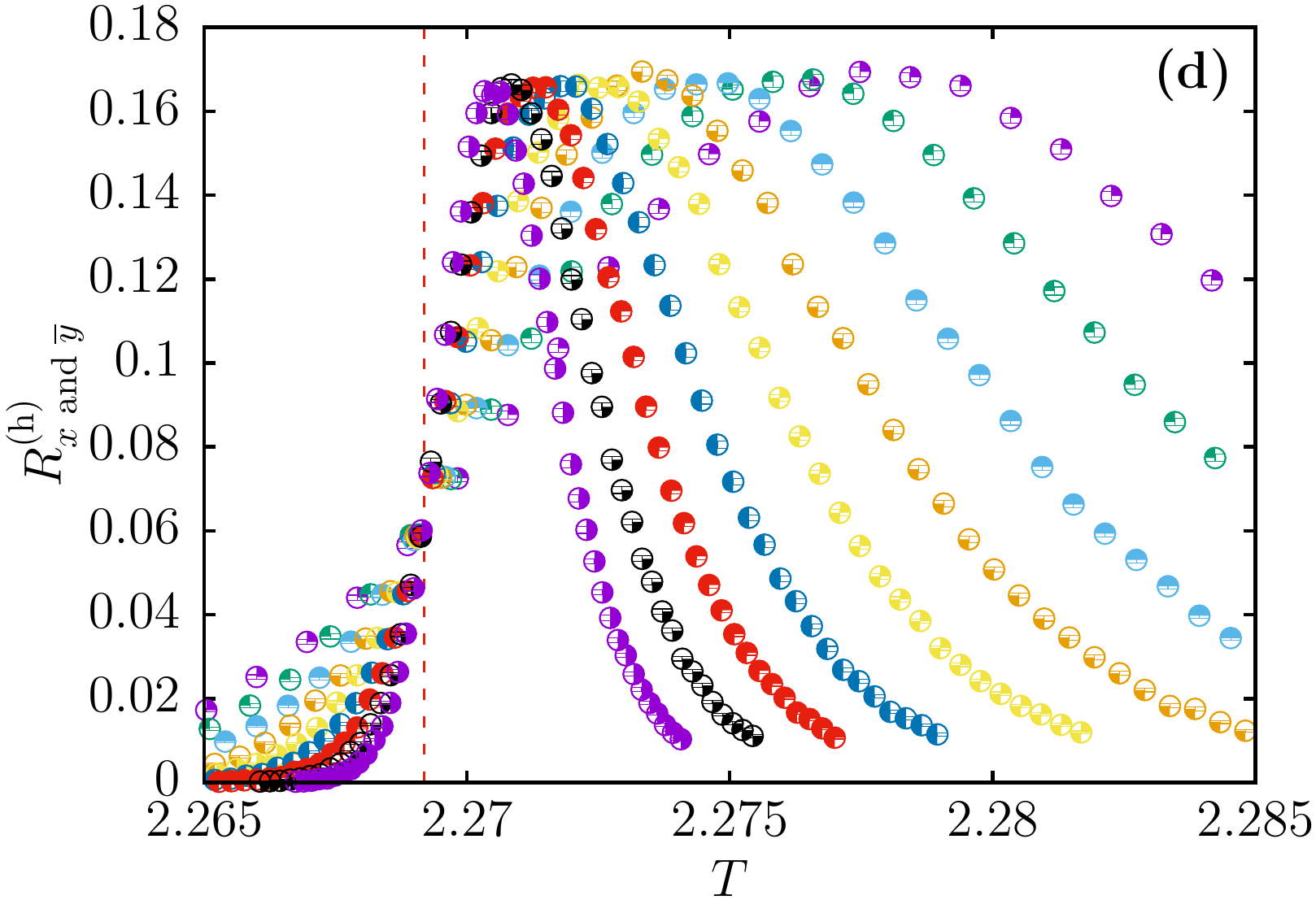}
		\label{subfig:q2_pwrap_xnoty_hard-zoomed}
	\end{subfigure}
	
	%\caption{Wrapping probabilities of the hard constraint clusters $R^{\text{(h)}}$ as function of temperature $ T $, for the largest system sizes $L$ considered of \subref{subfig:q2_pwrap_xory_hard-zoomed} $  R_{\text{x or y}}^{\text{(h)}}  $ \subref{subfig:q2_pwrap_xandy_hard-zoomed} $  R_{\text{x and y}}^{\text{(h)}}  $ \subref{subfig:q2_pwrap_x_hard-zoomed} $  R_{\text{x}}^{\text{(h)}}  $ \subref{subfig:q2_pwrap_xnoty_hard-zoomed} $ R_{\text{x and } \overline{\text{y}}}^{\text{(h)}} $. The dashed vertical line marks the transition temperature of the 1-replica Ising model.}
	%\label{fig:pwrap_hard}
    \caption{Wrapping probabilities $R^{\text{(h)}}$ of the hard-constraint clusters as a function of temperature $T$, analogous to the data for soft-constraint clusters shown in Fig.~\ref{fig:pwrap_soft}.}
	\label{fig:pwrap_hard}
\end{figure}

To investigate the percolation transition of geometrical overlap clusters, the main relevant quantities are the percolation strength $P_{\infty}$, the average cluster size $S$, and the wrapping probability $R$~\cite{stauffer_introduction_1994}. The latter is defined as the probability that, given a spin configuration, at least one cluster wraps around the periodic boundaries of a finite lattice and is connected back to itself. In the thermodynamic limit, one expects that $R = 1$ for temperatures below the percolation transition at $T_p$, and $R = 0$ at temperatures above $T_p$. The wrapping of a cluster can occur in various ways, and here we consider the following cases, in analogy to Ref.~\cite{newman_fast_2001}:
\begin{enumerate}[itemsep=2pt,parsep=2pt,leftmargin=20pt,label=(\arabic*)]
\item $R_{\text{x or y}}$ is the probability that a cluster wraps around the lattice in horizontal or vertical (or both) direction(s).
\item $R_{\text{x and y}}$ is the probability that a cluster wraps in horizontal and in vertical direction.
\item  $R_{\text{x}}$ is the probability that a cluster wraps in horizontal direction. Obviously, on the square lattice the probability that a cluster wraps in vertical direction $R_{\text{y}} \equiv R_{\text{x}}$.
\item  $R_{\text{x and } \overline{\text{y}}}$ is the probability that a cluster wraps around one but not the other direction. Here we choose the probability that a cluster wraps around the horizontal and not the vertical direction. The symmetry of the square lattice indicates that $R_{\text{x and } \overline{\text{y}}} \equiv R_{\text{y and } \overline{\text{x}}}$.
\end{enumerate}
The wrapping probability $R$ is a dimensionless quantity, and so one expects finite-size scaling of the form~\cite{stauffer_introduction_1994,privman:privman}
\begin{equation}
\label{eq:wrapping_scaled}
  R = \Tilde{R}\left[\left(T-T_{\text{p}}\right) L^{1/ \nu}\right].
\end{equation}    
Hence, the $R$ curves for systems of different sizes are expected to cross, up to finite-size corrections, at the same point, marking the transition temperature $T_\mathrm{p}$. In addition, since the scaling function $\tilde{R}$ is expected to be universal, so is the value of $R = \tilde{R}(0)$ at the crossing point~\cite{stauffer_introduction_1994}. This behavior is nicely verified in Figs.~\ref{fig:pwrap_soft} and \ref{fig:pwrap_hard}, where the various wrapping probabilities are plotted as a function of temperature $T$ for the larger system sizes studied and for both soft- and hard-constraint clusters. Except for $R_{\text{x and } \overline{\text{y}}}$, all wrapping probabilities increase with decreasing temperature, indicative of the onset of the percolating phase. The crossing of data sets for different system sizes is found to occur very close to the critical temperature $T_\mathrm{c}$ of the Ising model shown in Figs.~\ref{fig:pwrap_soft} and \ref{fig:pwrap_hard} as a dashed vertical line. This latter observation also holds for $R_{\text{x and } \overline{\text{y}}}$ with the essential difference that this observable exhibits a maximum, the position of which is expected to shift to its asymptotic value as $L\rightarrow \infty$~\cite{newman_fast_2001}. In numerical studies of ordinary, uncorrelated site or bond percolation, however, it was shown that $ R_{\text{x and } \overline{\text{y}}} $ exhibits both crossing points and maxima only in three dimensions~\cite{martins_percolation_2003}, whereas in two dimensions the crossing region is absent~\cite{newman_fast_2001,martins_percolation_2003}. Thus, the existence of both maxima and crossing points in $R_{\text{x and } \overline{\text{y}}}$ in two dimensions marks an interesting feature of the percolation signature of geometrical clusters in the overlap of the Ising model.

%For ordinary percolation in two dimensions, $ R_{\text{x and } \overline{\text{y}}} $ is a non-monotonic function of temperature and estimation of $ T_{\text{c}} $ can be obtained from the location of the maxima \cite{newman_fast_2001,martins_percolation_2003}, in three dimensions though,  $ R_{\text{x and } \overline{\text{y}}} $ exhibit both maximum and a crossing region \cite{martins_percolation_2003}. Here we will see that for the two-dimensional Ising model the behaviour of $ R_{\text{x and } \overline{\text{y}}} $ is similar to ordinary percolation in three dimensions. \myworries{Should I put a figure for the $ R^{\left(\text{1}\right)} $?}

%scrap from caption of Fig:2:  as function of temperature $ T $, for the larger system sizes $L$ considered of (a)~$ R_{\text{x or y}}^{\text{(h)}} $ (b)~$ R_{\text{x and y}}^{\text{(h)}} $ (c)~$ R_{\text{x}}^{\text{(h)}} $ (d)~$ R_{\text{x and } \overline{\text{y}}}^{\text{(h)}} $. The dashed vertical line marks the transition temperature of the Ising model

Of central importance in percolation theory is the \emph{cluster number} $n_s$, denoting the expected number of clusters of $s$ sites per lattice site~\cite{stauffer_introduction_1994}. Thus the average cluster size can be expressed as
\begin{equation}\label{eq:average_cluster_size}
	S = \frac{\sum_{s'} s^2 n_s}{\sum_{s'} s  n_s},	
\end{equation}
where $s n_s$ corresponds to the probability of a randomly picked site to belong to a cluster of size $s$. The notation $s'$ indicates that the sums are restricted to certain subsets of clusters. Denoting the set of clusters in a configuration as $C$ and letting $P$ be a subset of $C$ containing the percolating clusters, we can introduce the following definitions for $S$:
\begin{enumerate}[itemsep=2pt,parsep=2pt,leftmargin=20pt,label=(\arabic*)]
\item All clusters are included: $C$.
\item Exclude the largest cluster in each measurement: $C \; \backslash \; \text{max} \; C$.
\item Exclude all percolating clusters: $C \; \backslash \; P$.
\item Exclude all clusters percolating in horizontal and in vertical direction: $C \; \backslash \; P_{\text{x and y}}$.
\item Exclude all clusters percolating in one specific direction, e.g., horizontal: $C \; \backslash \; P_{\text{x}}$.
\item Exclude all clusters percolating in one but not the other direction, e.g., horizontal and not vertical: $C \; \backslash \; P_{\text{x and } \overline{\text{y}}}$.
\end{enumerate}

\begin{figure} % s soft hard (23 points 4 plots)
	\begin{subfigure}[]{0.5\textwidth}
	    %\hspace{-0.8cm}
		%\resizebox{0.96\columnwidth}{!}{\large\input{Figures/q2-soft_s-zoomed.tex}}
        \includegraphics[width=0.95\textwidth]{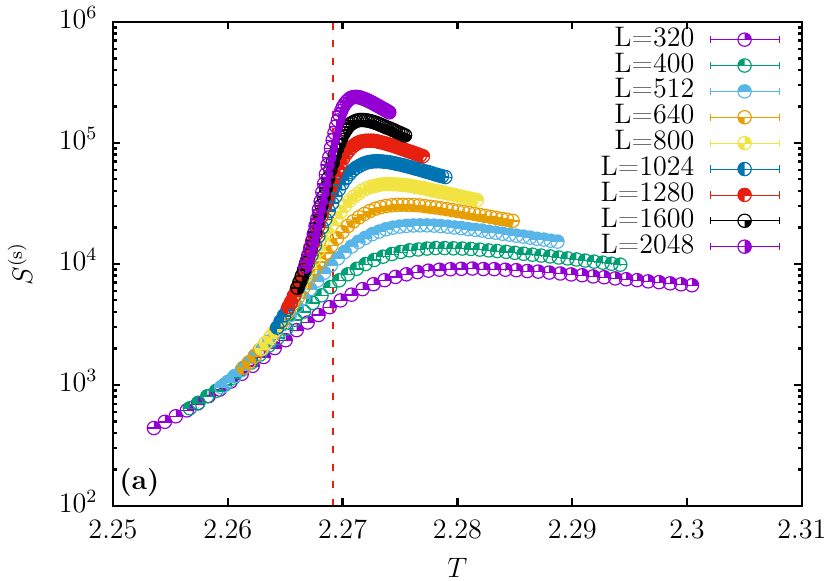}
		\label{subfig:q2_s_soft-zoomed}
	\end{subfigure}
 
	\begin{subfigure}[]{0.5\textwidth}
	    %\hspace{-0.8cm}
	    %\resizebox{0.96\columnwidth}{!}{\large\input{Figures/q2-hard_s-zoomed.tex}}
        \includegraphics[width=0.95\textwidth]{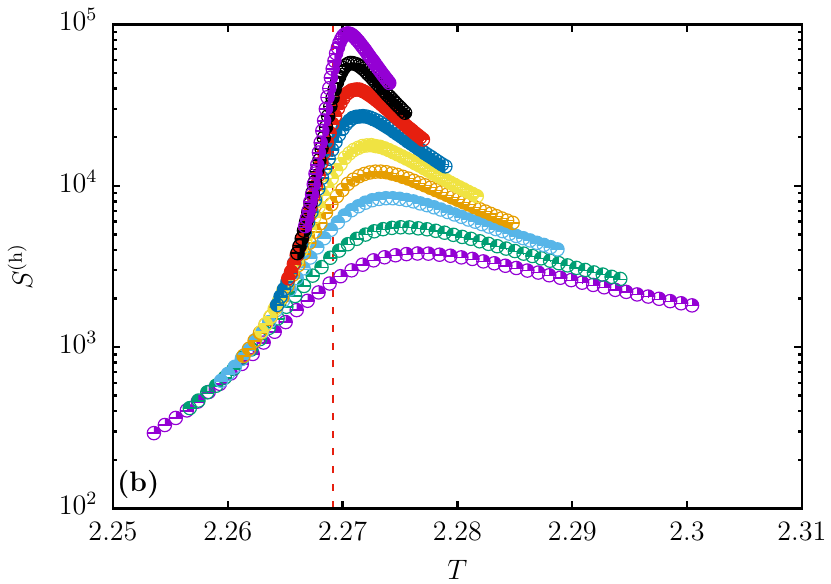}
		\label{subfig:q2_s_hard-zoomed}
	\end{subfigure}
	
	%\caption{Average cluster size $S$ on a semi-log axis as a function of temperature $ T $ for the largest system sizes $ L $ considered for the \subref{subfig:q2_s_soft-zoomed} soft and  \subref{subfig:q2_s_hard-zoomed} hard constraint clusters. The dashed vertical line marks the transition temperature of the 1-replica Ising model.}

    \caption{Average cluster size $S$ according to the definition $C \; \backslash \; \text{max} \; C$ as a function of temperature on a semi-logarithmic scale. Results for both (a) soft-constraint and (b) hard-constraint clusters are presented for the larger system sizes studied ($L \geq 320$).}
	\label{fig:q2_s_soft_hard}
\end{figure}

\begin{figure}[]% pwrap soft constraint
    %\begin{subfigure}[]{0.5\textwidth}
	    %\resizebox{0.96\linewidth}{!}{\large\input{Figures/q2-L1024-soft_s.tex}}\\
        %\hspace{-0.8cm}
        %\includegraphics[width=0.475\textwidth]{./figures/q2-L1024-soft_s.pdf}
        \includegraphics[width=0.475\textwidth]{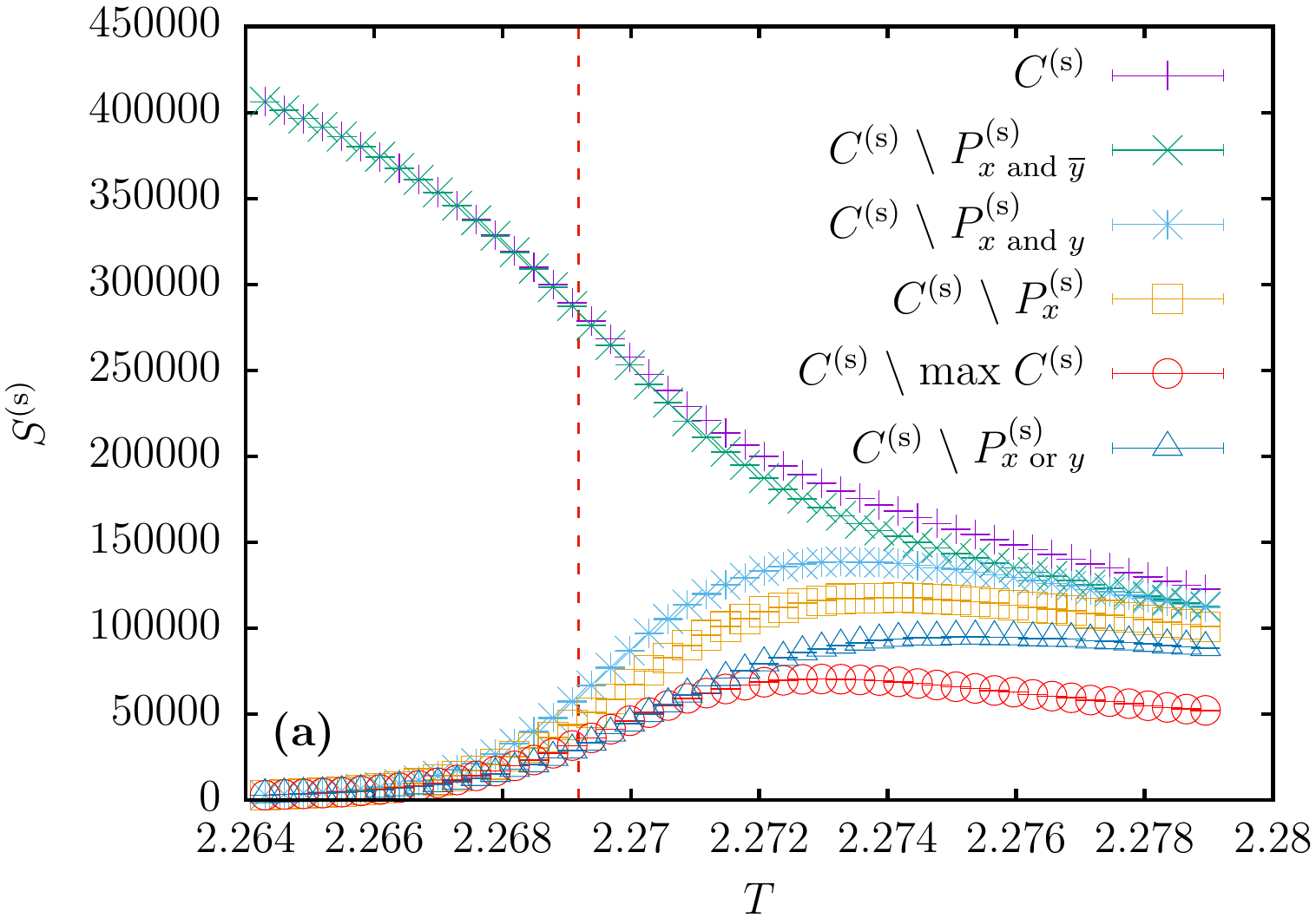}
     %   \label{subfig:q2-L1024-soft_s}
    %\end{subfigure}

    %\begin{subfigure}[]{0.5\textwidth}
        %\resizebox{0.96\linewidth}{!}{\large\input{Figures/q2-L1024-hard_s.tex}}
        %\hspace{-0.8cm}
        %\includegraphics[width=0.475\textwidth]{./figures/q2-L1024-hard_s.pdf}
        \includegraphics[width=0.475\textwidth]{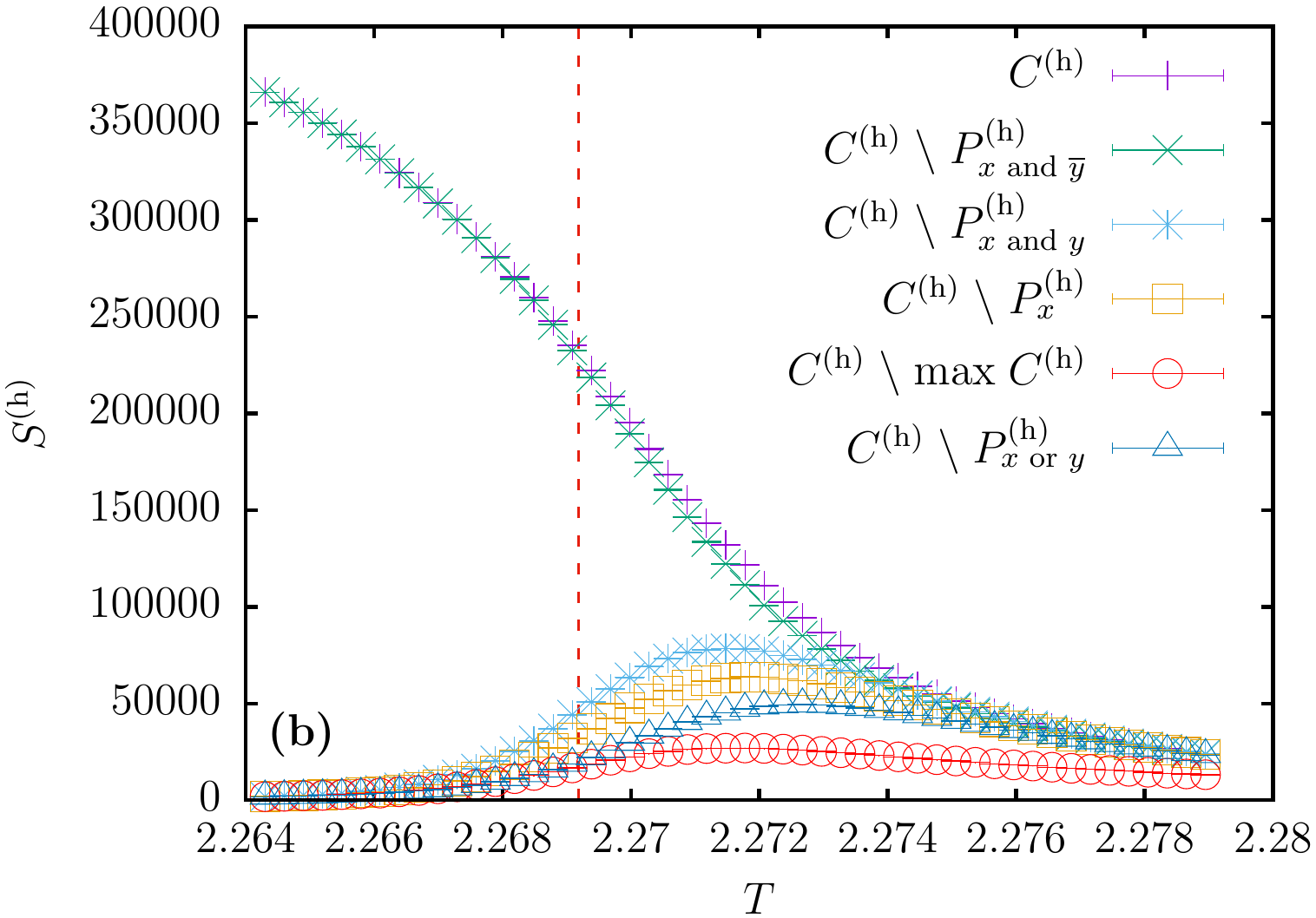}
    %\end{subfigure}
    \caption{Average cluster size $S$ vs.\ $T$ of (a) the soft-constraint and (b) the hard-constraint clusters for $L=1024$ using the different cluster sets defined in the text.}
		\label{fig:L1024-soft-hard-s}
\end{figure}

In most numerical studies of percolation the employed definition of the average cluster size excludes the largest cluster in each measurement, corresponding to our case (2)~\cite{stauffer_introduction_1994}. With this convention,  $S$ has a maximum around the percolation point, since in the non-percolating phase the size of many contributing clusters increases, while in the percolating regime most spins belong to the largest cluster which is not counted towards the sum. This estimate is shown in Fig.~\ref{fig:q2_s_soft_hard} where $S$ is plotted as a function of $T$ for the larger system sizes and for both soft- and hard-constraint clusters. In the vicinity of the percolation point, the average cluster size is expected to follow a scaling form according to~\cite{stauffer_introduction_1994}
\begin{equation}
\label{eq:average_cluster_size_scaled}
  S\left(L,T\right) = L^{\gamma / \nu} \Tilde{S}\left[\left(T-T_{\text{p}}\right) L^{1/ \nu}\right], 
\end{equation}
which can be used for determining the critical exponent ratio $\gamma/\nu$ conventionally associated to the scaling of the magnetic susceptibility. The behaviors of the remaining definitions of $S$ are shown for the soft-constraint and hard-constraint clusters in Fig.~\ref{fig:L1024-soft-hard-s} (a) and (b), respectively. Most definitions show a maximum of $S$, the exceptions being the full cluster set $C$ as well as $C \; \backslash \; P_{\text{x and } \overline{\text{y}}}$.

\begin{figure} % p soft hard
	\begin{subfigure}[]{0.5\textwidth}
	    %\hspace{-0.8cm}
		%\resizebox{0.96\linewidth}{!}{\large\input{Figures/q2-soft_p.tex}}
        \includegraphics[width=0.95\textwidth]{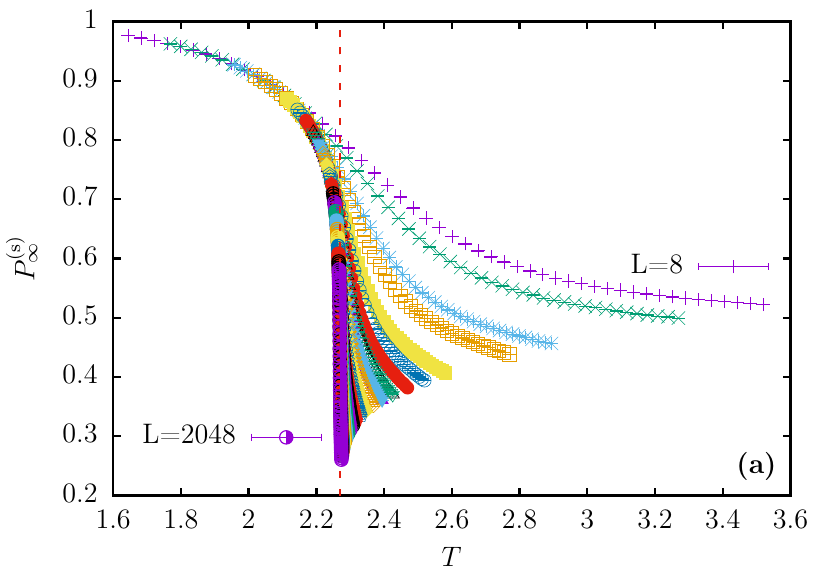}
		\label{subfig:q2_p_soft-zoomed}
	\end{subfigure}
	\begin{subfigure}[]{0.5\textwidth}
	    %\hspace{-0.8cm}
		%\resizebox{0.96\linewidth}{!}{\large\input{Figures/q2-hard_p.tex}}
        \includegraphics[width=0.95\textwidth]{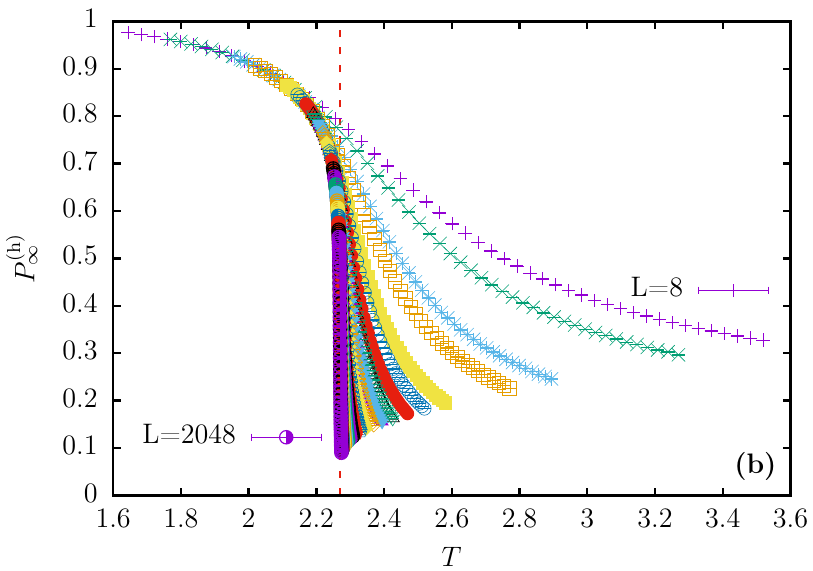}
		\label{subfig:q2_p_hard-zoomed}
	\end{subfigure}
	
	%\caption{Percolation strength $P_{\infty}$  as a function of temperature $ T $ for the different system sizes $L$ considered for the \subref{subfig:q2_s_soft-zoomed} soft and  \subref{subfig:q2_s_hard-zoomed} hard constraint clusters. The dashed vertical line marks the transition temperature of the 1-replica Ising model.}
    \caption{Percolation strength $P_{\infty}$ as a function of temperature for both (a) soft-constraint  and (b) hard-constraint clusters. For clarity, only the legends of the smallest and largest system sizes are highlighted.}
	\label{fig:q2_p_soft_hard}
\end{figure}

The percolation strength $P_{\infty}$ corresponds to the fraction of sites belonging to the infinite cluster in the thermodynamic limit. For finite-size systems it is usually estimated from the fraction of sites belonging to the largest cluster. In Fig.~\ref{fig:q2_p_soft_hard}, $P_{\infty}$ is plotted against $T$ for the full range of system sizes studied and for both soft- and hard-constraint clusters. Note that (i) as the temperature decreases $P_{\infty}$ increases, indicating the appearance of a percolating cluster, and (ii) for $T=0$ we have $P_{\infty} = 1$ as all spins belong to the percolating cluster. We remind that when studying the FK clusters in magnetic systems, the percolation strength corresponds to the magnetization of the system \cite{fortuin_random-cluster_1972}. Finite-size scaling theory suggests a scaling form
\begin{equation}
\label{eq:strength_scaled}
  P_{\infty}\left(L,T\right) = L^{-\beta / \nu} \Tilde{P}_{\infty}\left[\left(T-T_{\text{p}}\right) L^{1/ \nu}\right],
\end{equation}
where $\beta/\nu$ denotes the corresponding critical exponent ratio.

\begin{figure}[]% pwrap soft constraint
	   %\resizebox{0.96\linewidth}{!}{\large\input{Figures/q2-L1024-soft_p.tex}}\\
        %\hspace{-0.8cm}
       %\includegraphics[width=0.48\textwidth]{./figures/q2-L1024-soft_p.pdf}
       \includegraphics[width=0.48\textwidth]{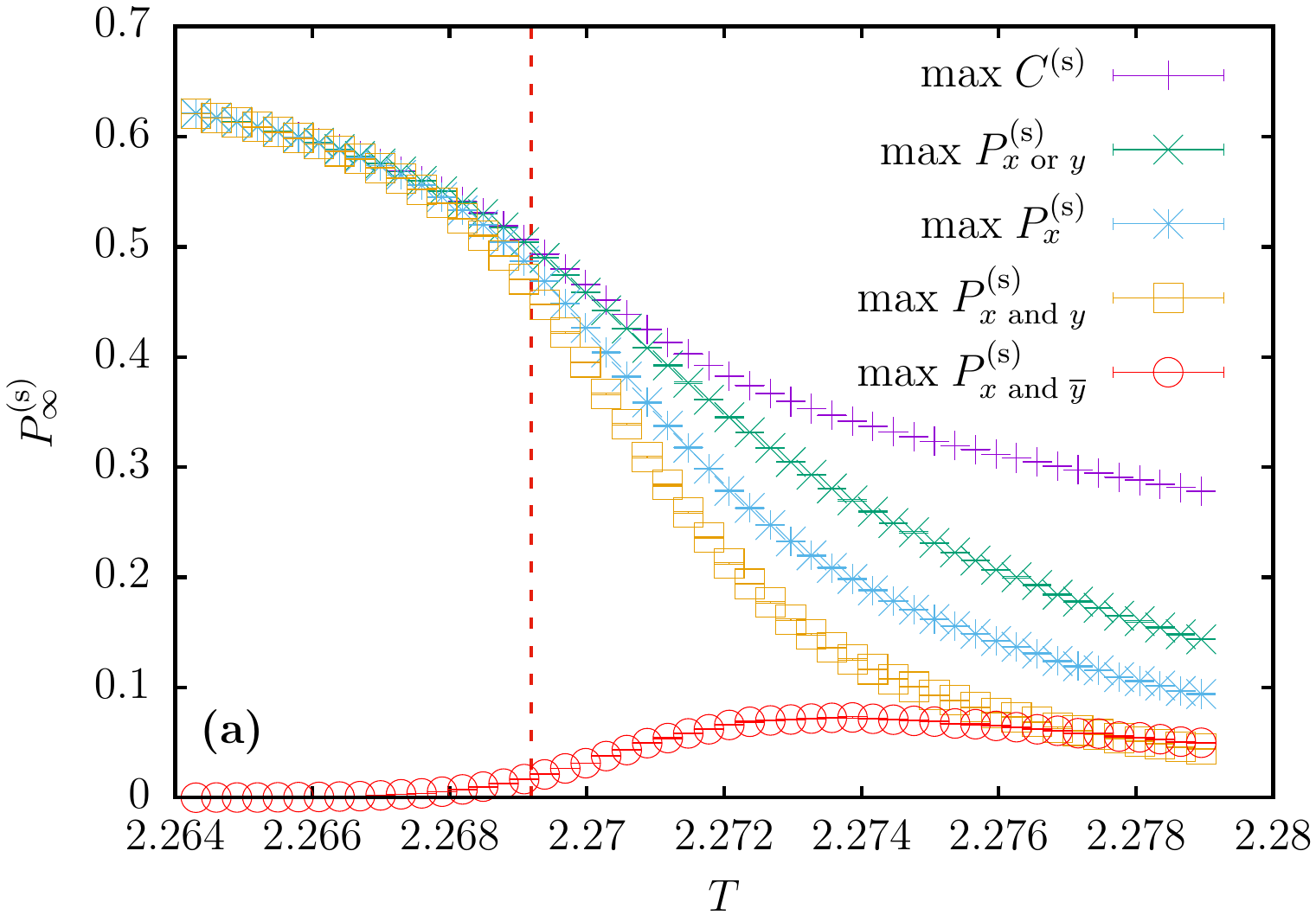}
        %\resizebox{0.96\linewidth}{!}{\large\input{Figures/q2-L1024-hard_p.tex}}

        %\hspace{-0.8cm}
       %\includegraphics[width=0.48\textwidth]{./figures/q2-L1024-hard_p.pdf}
       \includegraphics[width=0.48\textwidth]{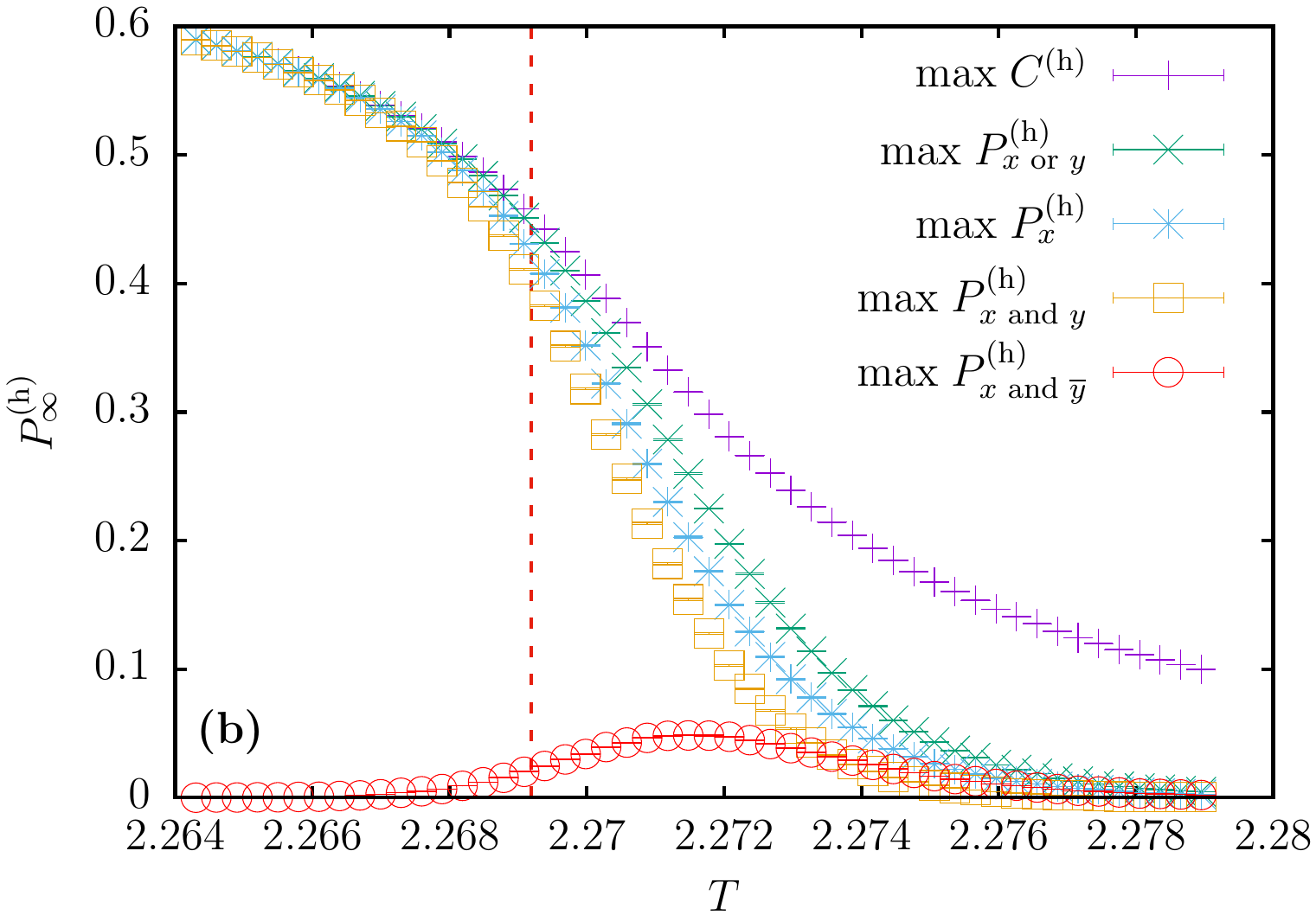}
        \caption{Percolation strength $P_{\infty}$ vs.\ $T$ of (a) the soft-constraint and (b) the hard-constraint clusters for $L=1024$ and the cluster types discussed in the main text.}
		\label{fig:L1024-soft-hard-p}
\end{figure}

Analogous to the treatment of $R$ and $S$, it is natural to also consider modified percolation strengths $P_\infty$ by studying the fractions of sites occupied by the following subsets:
\begin{enumerate}[itemsep=2pt,parsep=2pt,leftmargin=20pt,label=(\arabic*)]
\item Largest cluster: $\text{max} \; C$.
\item Largest percolating cluster: $ \text{max} \; P $.
\item Largest cluster that percolates in horizontal and in vertical direction: $\text{max} \; P_{\text{x and y}} $.
\item Largest cluster that percolates in one specific direction, e.g., horizontal: $\text{max} \; P_{\text{x}}$.
\item Largest cluster that percolates in one but not the other direction, e.g., horizontal and not vertical: $\text{max} \; P_{\text{x and } \overline{\text{y}}}$.
\end{enumerate}
The temperature dependence of the percolation strength according to these definitions is shown for an $L=1024$ system and the soft-constraint and hard-constraint clusters in Fig.~\ref{fig:L1024-soft-hard-p} (a) and (b), respectively. With the exception of the unlikely $P_{\text{x and } \overline{\text{y}}}$ clusters, the different definitions lead to the same behavior for temperatures below the percolation point, but visible differences above.

In the following, we will use the scaling forms \eqref{eq:wrapping_scaled}, \eqref{eq:average_cluster_size_scaled} and \eqref{eq:strength_scaled} with the universal scaling functions $\Tilde{R}$, $\Tilde{P}_{\infty}$, and $\Tilde{S}$ in order to determine the correlation-length critical exponent $\nu$ as well as the exponent ratios $\gamma/\nu$ and $\beta/\nu$ of the average cluster size and the percolation strength, respectively, as well as the percolation temperature $T_\mathrm{p}$.

\begin{figure} % maximum derivative of pwrap soft hard  with reweighting (23points 4 plots)
	\begin{subfigure}[]{0.5\textwidth}
        %\hspace{-0.8cm}
		%\resizebox{0.96\linewidth}{!}{\large\input{Figures/q2_max_derivative_soft_from_derivative_reweighting.tex}}
        %\includegraphics[width=0.93\textwidth]{./figures/q2_max_derivative_soft_from_derivative_reweighting.pdf}
		\includegraphics[width=0.91\textwidth]{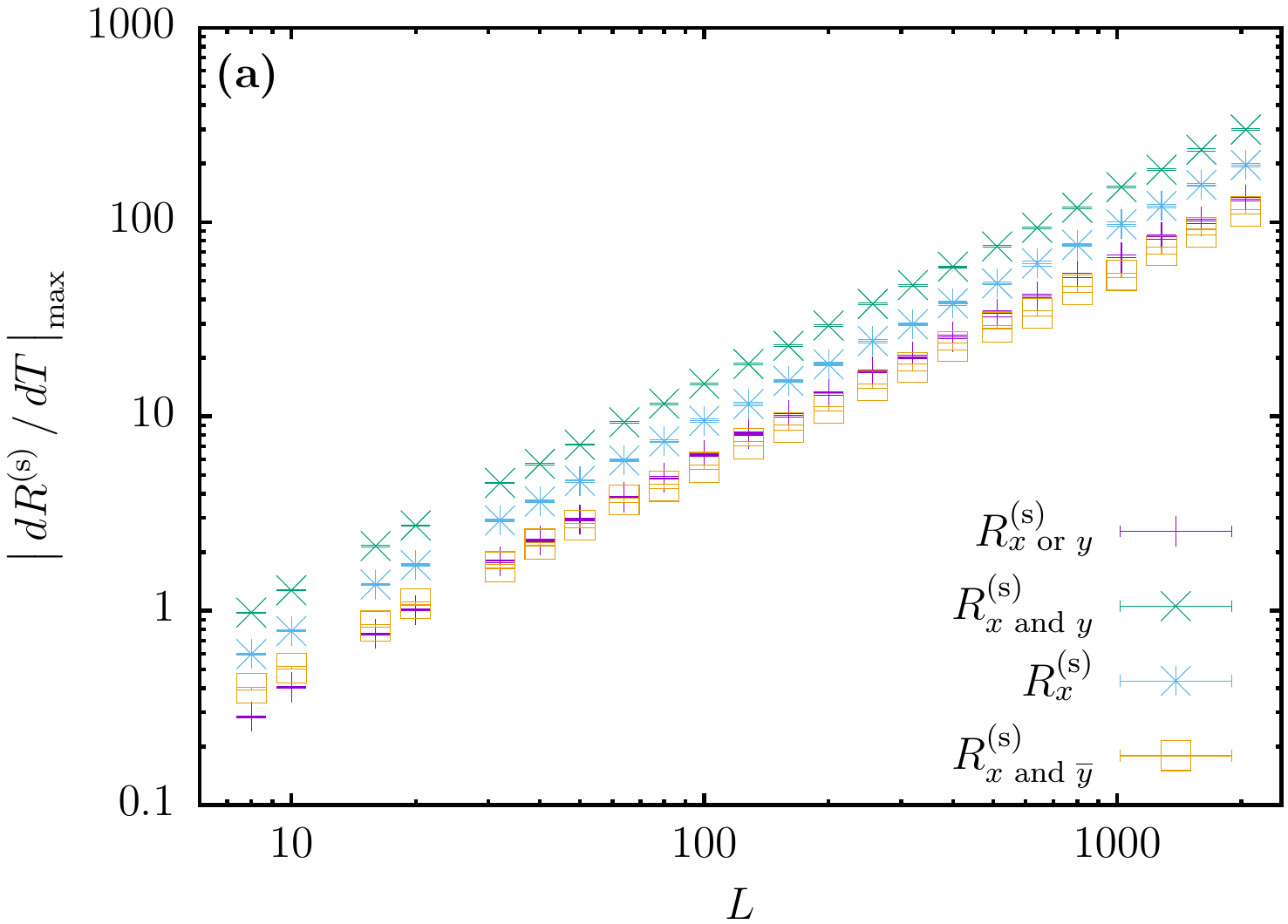}\label{fig:q2_max_derivative_soft_from_derivative_reweighting}
	\end{subfigure}
	\begin{subfigure}[]{0.5\textwidth}
        %\hspace{-0.8cm}
		%\resizebox{0.96\linewidth}{!}{\large\input{Figures/q2_max_derivative_hard_from_derivative_reweighting.tex}}
        %\includegraphics[width=0.93\textwidth]{./figures/q2_max_derivative_hard_from_derivative_reweighting.pdf}
        \includegraphics[width=0.91\textwidth]{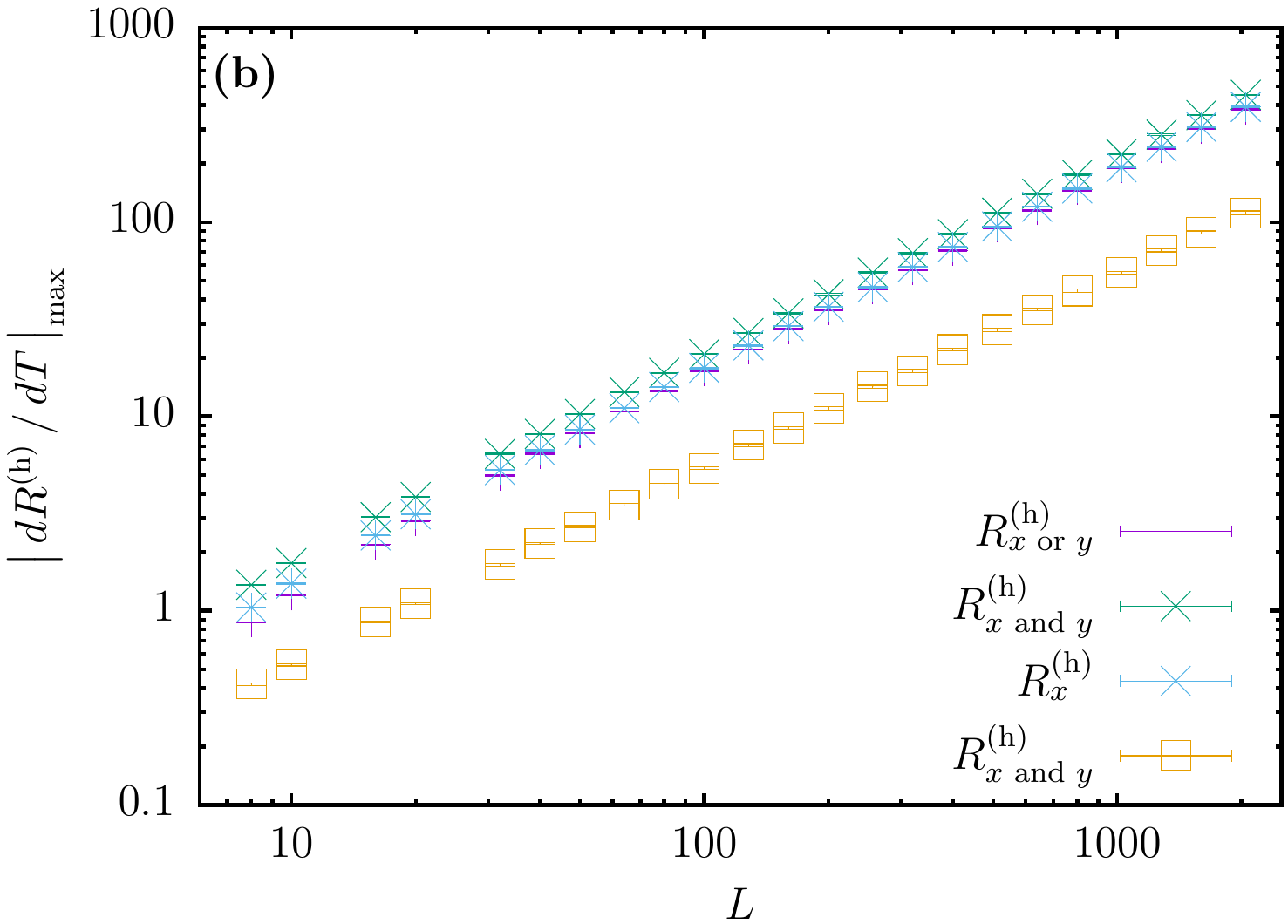}\label{fig:q2_max_derivative_hard_from_derivative_reweighting}
	\end{subfigure}
		
	%\caption{Log-log plot of $ \left|dR/dT\right|_{\text{max}} $ for the different definitions of the wrapping probabilities as a function of system size $ L $ for the \subref{fig:q2_max_derivative_soft_from_derivative_reweighting} soft and  \subref{fig:q2_max_derivative_hard_from_derivative_reweighting} hard constraint clusters.}
    \caption{Maximum slope of the wrapping probability, $\left|\d R/\d T\right|_{\text{max}}$, as a function of system size $L$ on a double logarithmic scale for all variants of the wrapping probabilities and for (a) soft-constraint and (b) hard-constraint clusters.}
	\label{fig:q2_max_derivative_from_derivative_reweighting}
\end{figure}

\section{Main Results} 
\label{sec:results}

\subsection{Correlation-length exponent} 
\label{subsec:nu_exponent}
	
In Monte Carlo studies of phase transitions the temperature-derivative of the Binder cumulant~\cite{binder_finite_1981} provides a reliable estimate for the critical exponent $\nu$~\cite{ferrenberg_critical_1991}. In random percolation, a similar behavior is expected for the derivative of the wrapping probability with respect to the bond occupation probability \cite{martins_percolation_2003}. For the present problem of clusters in a thermal problem it is natural, in contrast, to consider the temperature-derivative of the wrapping probabilities. As the latter are monotonic functions of the temperature --- except for $R_{\text{x and } \overline{\text{y}}}$ --- one expects that the maximum of the absolute value of its first derivative should 
scale as
\begin{equation}\label{eq:max_derivative}
	\left| \frac{\d R}{\d T} \right|_{\text{max}} \sim L^{1 / \nu}.
\end{equation}  
As shown in Sec.~\ref{sec:observables}, the exception to this rule is $R_{\text{x and } \overline{\text{y}}}$ which is a non-monotonic function of $T$, showing both a maximum and a crossing region. Nevertheless, if we restrict ourselves to the vicinity of the crossing regime, this observable is also expected to follow the scaling behaviour of Eq.~\eqref{eq:max_derivative}. 
Importantly,  Eq.~\eqref{eq:max_derivative} allows one to obtain estimates of $\nu$ without prior knowledge of the 
percolation temperature $T_\mathrm{p}$. 
To determine the maximum of $ \left| \d R / \d T \right|$, both the first and second derivatives are computed using the symmetric-finite-difference definition, and the root is located using the bisection method~\cite{numrec}. The required estimates at nearby temperatures are extracted from the simulation data by means of single-histogram reweighting~\cite{ferrenberg_new_1988}, using a step size $\Delta T = 10^{-7}$.

\begin{figure} % nu soft hard from derivative maxima with reweighting (21 points 4 plots)
	\begin{subfigure}[]{0.5\textwidth}
        %\hspace{-0.5cm}  
		%\resizebox{0.96\linewidth}{!}{\large\input{Figures/q2_nu_soft_from_derivative_reweighting.tex}}
        %\includegraphics[width=0.92\textwidth]{./figures/q2_nu_soft_from_derivative_reweighting.pdf}
        \includegraphics[width=0.92\textwidth]{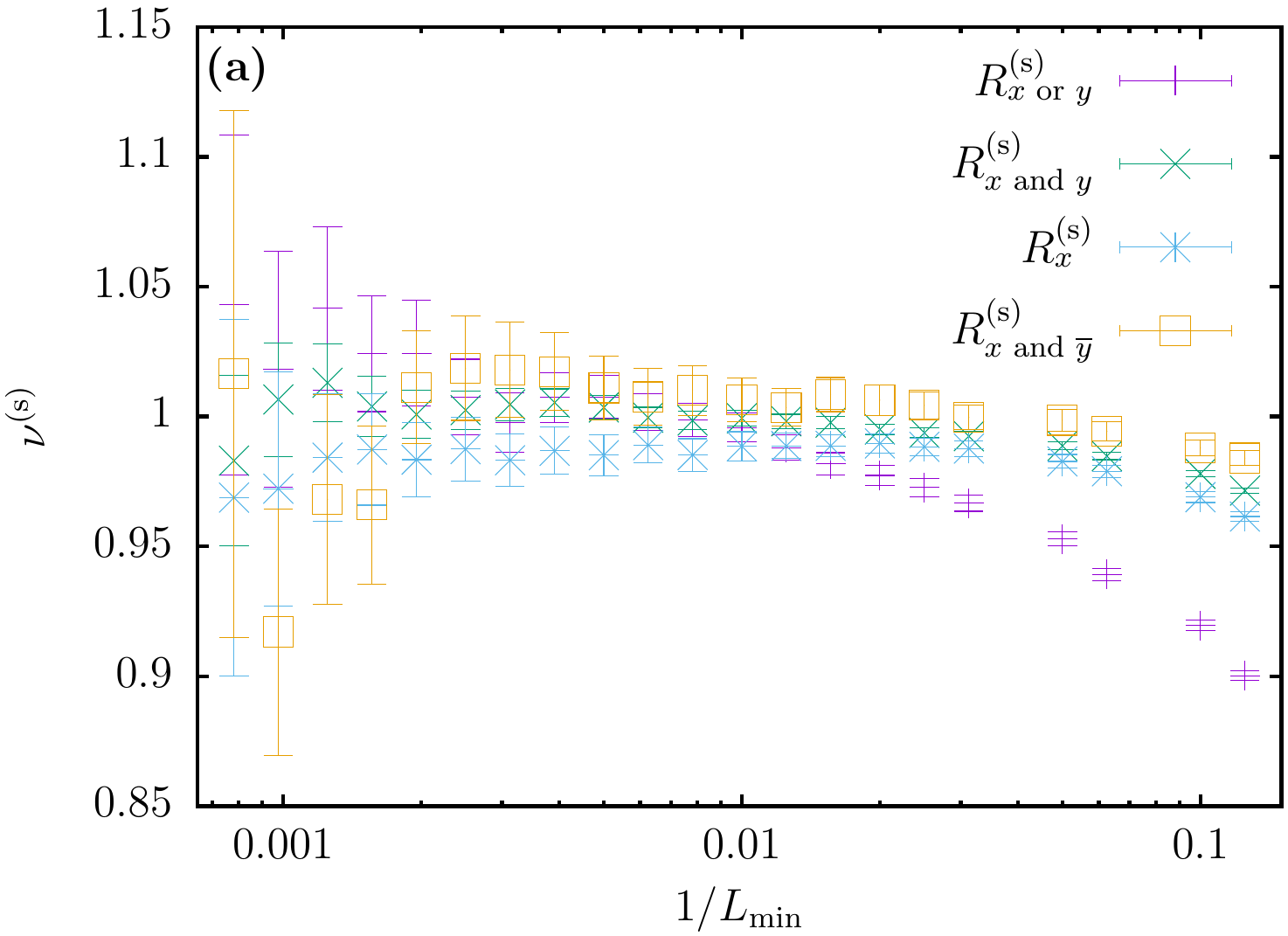}
		\label{fig:q2_nu_soft_from_derivative_reweighting}
	\end{subfigure}
	\begin{subfigure}[]{0.5\textwidth}
        %\hspace{-0.5cm}
		%\resizebox{0.96\linewidth}{!}{\large\input{Figures/q2_nu_hard_from_derivative_reweighting.tex}}
		%\includegraphics[width=0.92\textwidth]{./figures/q2_nu_hard_from_derivative_reweighting.pdf}
        \includegraphics[width=0.92\textwidth]{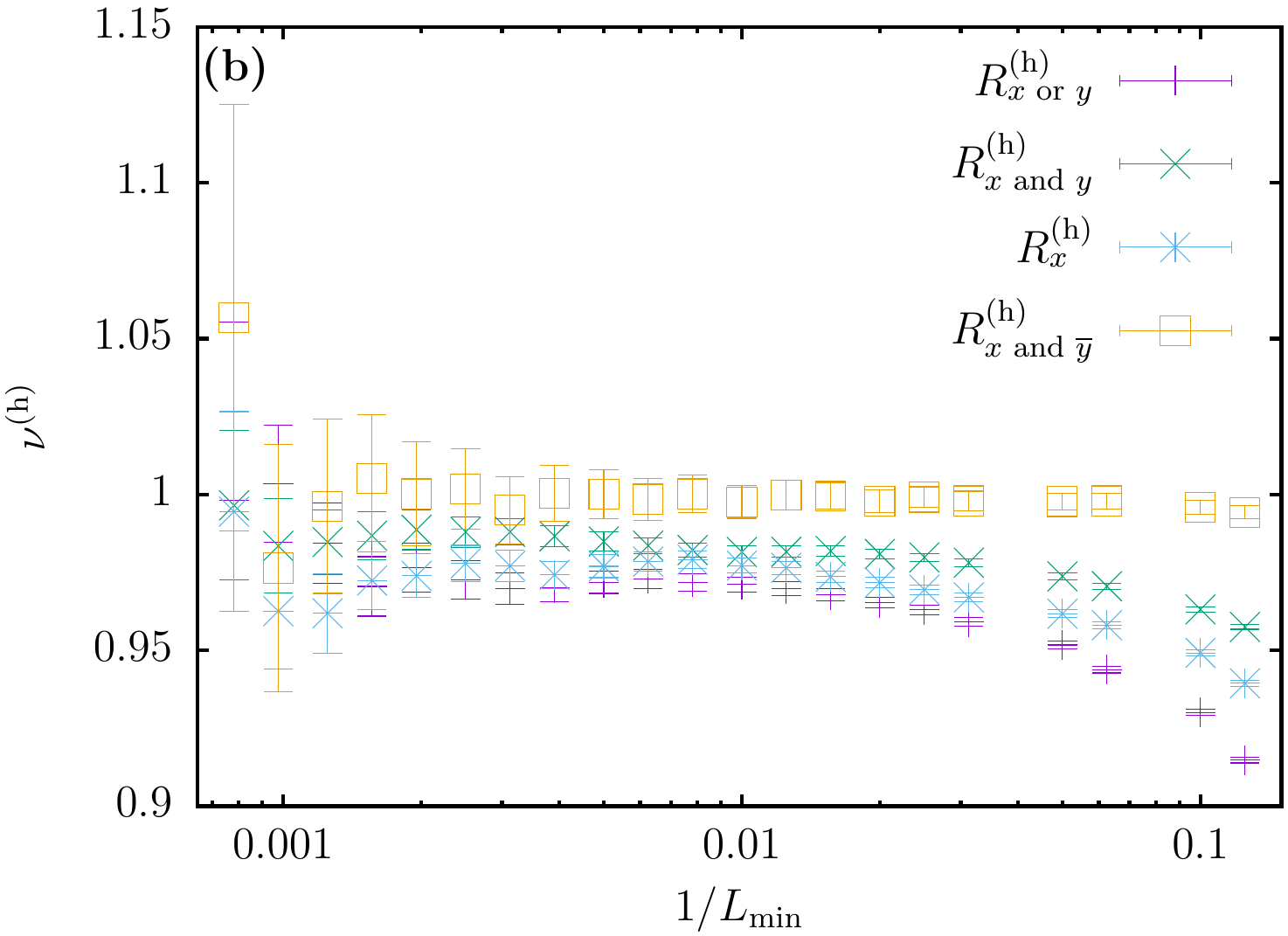}
        \label{fig:q2_nu_hard_from_derivative_reweighting}
	\end{subfigure}
	
	%\caption{Estimates of the $ \nu $ exponent for the different definitions of the wrapping probabilities as function of $ 1/L_{\text{min}} $ for the \subref{fig:q2_nu_soft_from_derivative_reweighting} soft and \subref{fig:q2_nu_hard_from_derivative_reweighting} hard constraint clusters. Estimates extracted from linear  fits (on a log-log scale) of $ \left|dR/dT\right|_{\text{max}} $ as a function of system size $ L $.}
    \caption{Effective values of the exponent $\nu$ vs.\ the inverse lower cutoff size $ 1/L_{\text{min}}$ on a semi-logarithmic scale. The estimates were obtained from fits of the form~\eqref{eq:max_derivative} to the data of Fig.~\ref{fig:q2_max_derivative_from_derivative_reweighting}.}
	\label{fig:q2_nu_soft_hard_from_derivative_reweighting}
\end{figure}

In Fig.~\ref{fig:q2_max_derivative_from_derivative_reweighting}, $ \left| \d R / \d T \right|_{\text{max}}$ is shown as a function of $L$ for the different wrapping probabilities of the soft- and hard-constraint clusters. Fits of the form~\eqref{eq:max_derivative} were 
performed for system sizes on intervals $L_{\text{min}} \le L \le L_{\text{max}}$ by systematically increasing the lower cut-off $L_{\text{min}}$, while keeping the upper cut-off fixed at $L_{\text{max}} = 2048$. The resulting effective values of $\nu$ are shown in
Fig.~\ref{fig:q2_nu_soft_hard_from_derivative_reweighting}. The final estimates we quote for both the soft- and hard-constraint clusters using the $R_{\text{x and y}}^{(\text{s})}$ 
and $R_{\text{x and } \overline{\text{y}}}^{(\text{h})}$ definitions, respectively, are%\myworries{correction to scaling?}
% \begin{align}
% 	\label{eq:q2-soft_estimate_nu}
% 	\nu^{\text{(s)}} &= 1.005(5), &  \chi^2/\text{d.o.f.} &\approx 0.39,& L_{\text{min}}&=256, & & R_{\text{x and y}}^{(\text{s})}, \\
% 	\label{eq:q2-hard_estimate_nu}
% 	\nu^{\text{(h)}} &= 1.00(3), &  \chi^2/\text{d.o.f.} &\approx 0.96,& L_{\text{min}}&=800, & & R_{\text{x and } \overline{\text{y}}}^{(\text{h})}.
% \end{align}

\begin{subequations}
\begin{equation}
    \label{eq:q2-soft_estimate_nu}
	\nu^{\text{(s)}} = 1.005(5) \;\; (L_{\text{min}}=256), \\
\end{equation}
\begin{equation}
\label{eq:q2-hard_estimate_nu}
	\nu^{\text{(h)}} = 1.00(3) \;\; (L_{\text{min}}=800).
\end{equation}
\end{subequations}
These results suggest that the critical exponent of the correlation length is the same for both cluster types and consistent with that of the Ising model, i.e., $\nu = 1$.

%Additionally, as a precaution against unavoidable corrections to %scaling we neglect estimates corresponding to small values of %$L_{\text{min}}$ (say $L_{\text{min}}$=100). 

%In Tables \ref{table:table-q2_nu_soft_from_derivative_reweighting}-\ref{table:table-q2_nu_hard_from_derivative_reweighting} we report the estimates of $ \nu $ for different fit intervals,  the deviation of the estimates from the value $\nu=1$ of the 1-replica Ising model in multiples of their estimated statistical errors $\Delta_{\sigma}$, the degrees of freedom  (d.o.f.), the \textit{chi-square} per d.o.f.  $\left( x^2/ \text{d.o.f.}\right) $, and the respective quality-of-fit parameter $ Q $ for the sc and hc clusters respectively. 
	
%Both for sc and hc clusters, estimates are in good agreement with $\nu=1$, with reasonable values of $x^2/ \text{d.o.f.} $ and $Q$ for certain values of $L_\text{min}$ and above, and for all wrapping probabilities considered. The above is showed in Fig. \ref{fig:q2_nu_soft_hard_drom_derivative_reweighting}, where the estimates of the exponent $\nu$ are plotted as a functions of $1 / L_\text{min}$
	
\subsection{Percolation temperature} 
\label{subsec:tc}

For the estimation of the percolation temperature $T_{\text{p}}$ we 
considered the intersection of the wrapping probabilities of pairs of system sizes $(L, L')$ as a function of $T$, following the original prescription by Binder for the magnetization cumulant~\cite{binder_finite_1981,ferrenberg_critical_1991}. The points where these wrapping probabilities 
cross scale as~\cite{binder_finite_1981}
\begin{equation}\label{eq:tc_crossing}
	T_{\text{cross}}\left(L,b\right) = T_{\text{p}} + aL^{-\left( 1/\nu + \omega \right)} \left(\frac{b^{-\omega}-1}{b^{1 / \nu}-1}\right),
\end{equation}
where $a$ is a non-universal scaling parameter, $\omega$ is the corrections-to-scaling exponent, and $b = L'/L$ is the quotients ratio,  fixed hereafter to $b = 2$. Crossings were determined using the bisection method~\cite{numrec} alongside the single-histogram reweighting technique~\cite{ferrenberg_new_1988}.

\begin{figure} % Tc cross soft hard from reweighting (20 points 4 plots)
    \begin{subfigure}[]{0.5\textwidth}
        %\hspace{-0.8cm}
		%\resizebox{0.96\linewidth}{!}{\large\input{Figures/q2_Tc_crossings_soft_from_reweighting.tex}}
		%\includegraphics[width=0.92\textwidth]{./figures/q2_Tc_crossings_soft_from_reweighting.pdf}
        \includegraphics[width=0.92\textwidth]{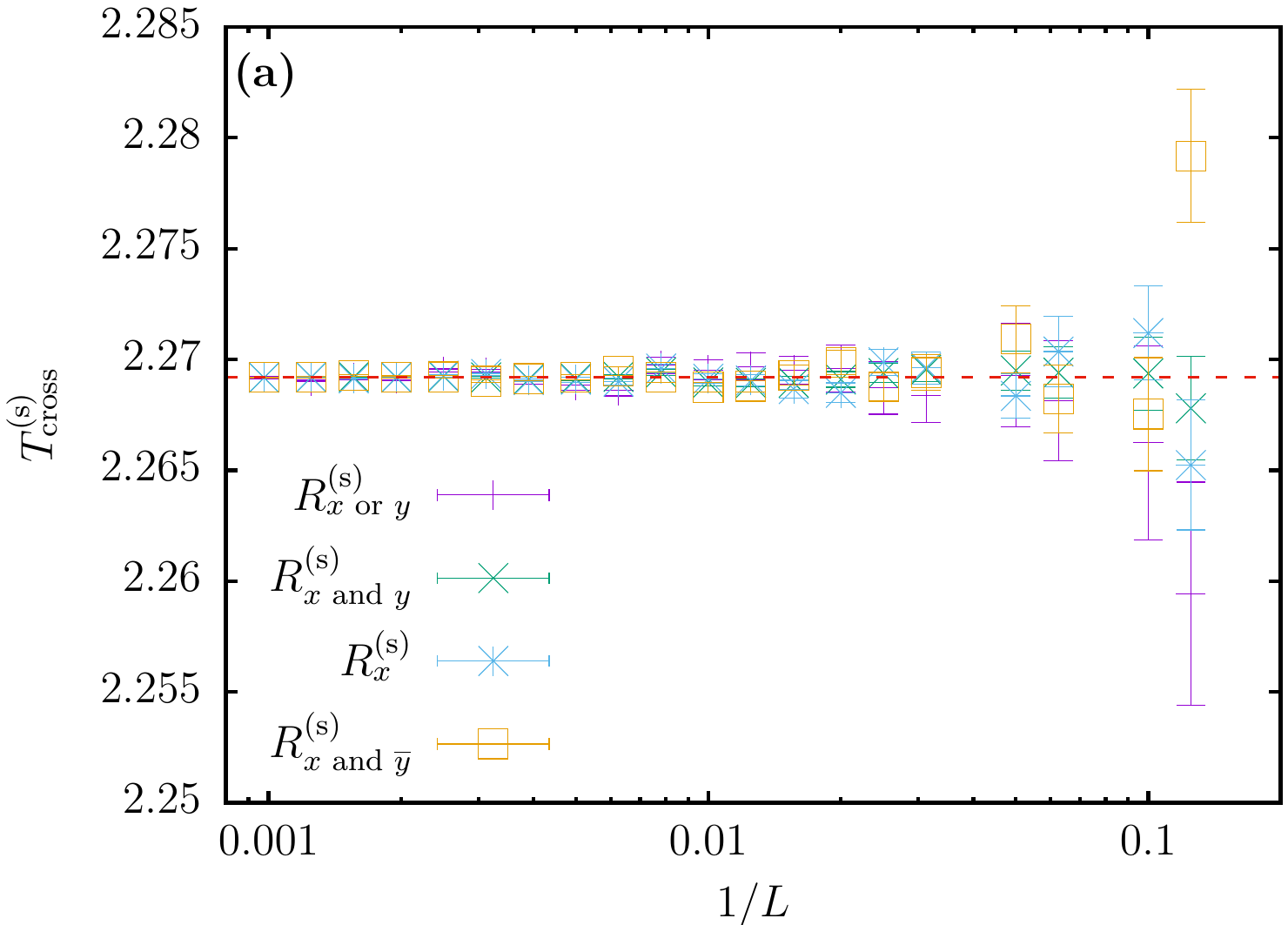}
        \label{fig:q2_Tc_crossings_soft_from_reweighting}
    \end{subfigure}
    
	\begin{subfigure}[t]{0.5\textwidth}
       %\hspace{-0.8cm}
	      %\resizebox{0.96\linewidth}{!}{\large\input{Figures/q2_Tc_crossings_hard_from_reweighting.tex}}
		%\includegraphics[width=0.92\textwidth]{./figures/q2_Tc_crossings_hard_from_reweighting.pdf}
        \includegraphics[width=0.92\textwidth]{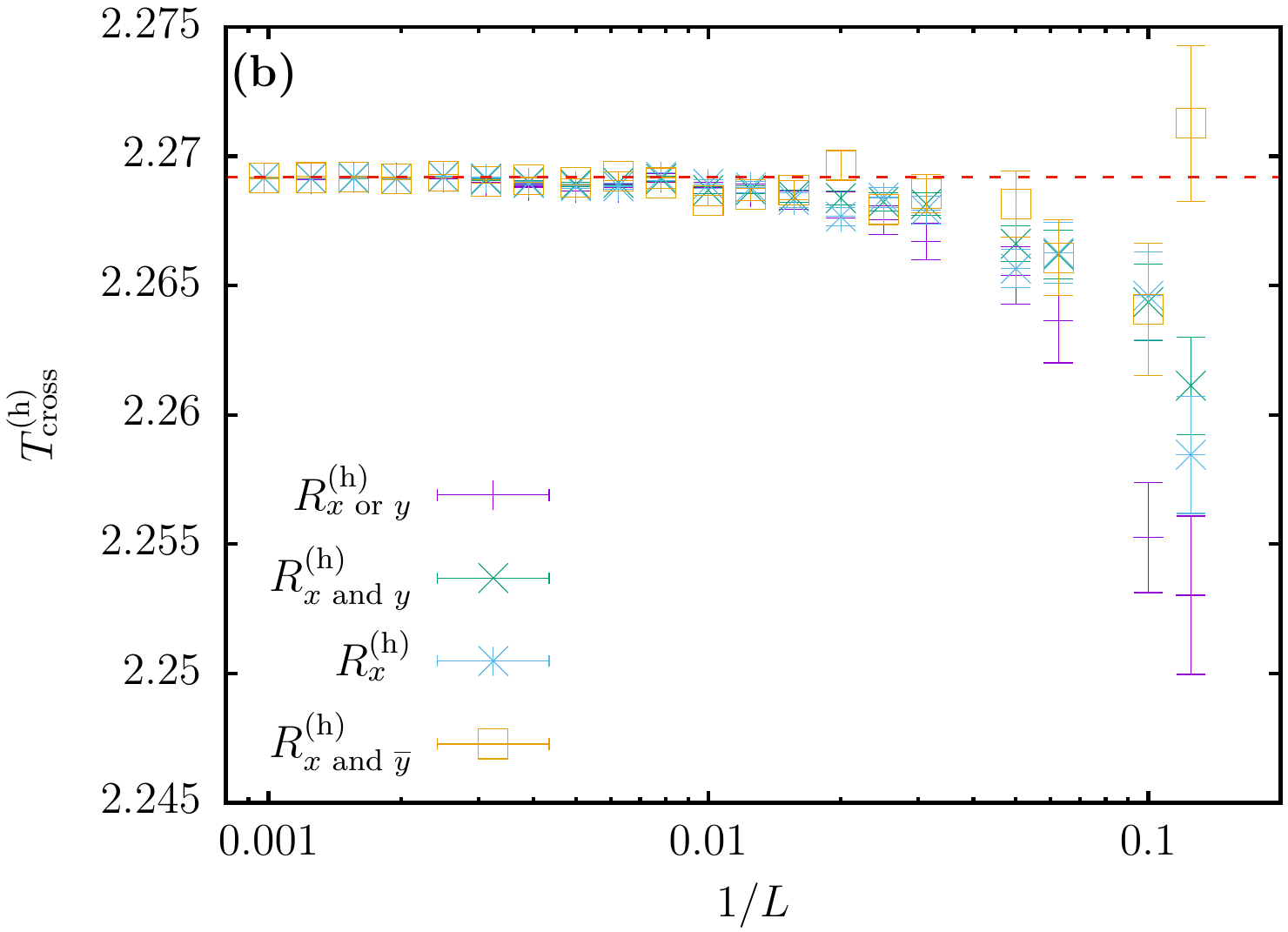}
        \label{fig:q2_Tc_crossings_hard_from_reweighting}
	\end{subfigure}
	%\caption{Estimates of the crossing temperatures of pairs $ (L,2L) $ for the different definitions of the wrapping probabilities as a function of $ 1/L $ for the \subref{fig:q2_Tc_crossings_soft_from_reweighting} soft and  \subref{fig:q2_Tc_crossings_hard_from_reweighting} hard constraint clusters. The dashed horizontal line marks the transition temperature of the 1-replica Ising model.}
    \caption{Estimates of crossing temperatures for (a) soft-constraint and (b) hard-constraint clusters on a semi-logarithmic scale. The dashed horizontal line marks the transition temperature of the Ising model.}
	\label{fig:q2_Tc_crossings_soft_hard_from_reweighting}
\end{figure}

Figure~\ref{fig:q2_Tc_crossings_soft_hard_from_reweighting} showcases the scaling of the crossing points of the various wrapping probabilities for both the soft- and hard-constrained clusters, which appear to be consistent with the critical temperature of the Ising model, up to surprisingly small finite-size effects. In order to obtain more accurate estimates of $T_{\text{p}}$  we performed fits using Eq.~\eqref{eq:tc_crossing}. Due to the smallness of the corrections, however, the accuracy of our data did not allow us to resolve their detailed form, resulting in fits of poor quality and consequently in unreliable estimates of the involved parameters. As an alternative, we fixed $\nu$ to the expected value $\nu = 1$, so that
%$ = 1$$ + \omega = 1$ in 
Eq.~\eqref{eq:tc_crossing} (ignoring scaling corrections) simplifies to
\begin{equation}
\label{eq:tc_crossing-linear}
    T_{\text{cross}} = T_{\text{p}} + a / L.
\end{equation}

%which is a reasonable choice since $\nu$ is consistent with a value around 1 and assuming that the correction-to-scaling exponent is small. 

%Under this construction Eq.~\eqref{eq:tc_crossing} becomes

%straight-line fits; results are shown in Tables \ref{table:table-q2_Tc_crossings_soft_reweighting}, \ref{table:table-q2_Tc_crossings_hard_reweighting} for the sc and hc clusters respectively. For all wrapping probabilities and for the two types of clusters, results are consistent with  straight lines parallel to the horizontal axis, with reasonable $ x^2 / \text{d.o.f.} $ and $ Q $ for all $ L_{\text{min}} \ge 320 $, where we see  that in between error bars the critical temperature of the two-replica Ising model in two dimensions, coincides with the critical temperature of the one-replica (standard) Ising model i.e., $ T_{\text{c}} \approx 2.269185 $. Finally, we note that the accuracy of our data do not allow us to perform fits using Eq. \ref{eq:tc_crossing}, which could provide, additionally to $ T_{\text{c}} $, estimates of the correction exponent $ \omega $.  
	
\begin{figure} % Tc cross soft hard from reweighting (20 points 4 plots)
	\begin{subfigure}[]{0.5\textwidth}
        %\hspace{-0.5cm}
		%\resizebox{0.96\linewidth}{!}{\large\input{Figures/q2_Tc_estimates_soft-fit_from_reweighting.tex}}
		%\includegraphics[width=0.93\textwidth]{./figures/q2_Tc_estimates_soft-fit_from_reweighting.pdf}
        \includegraphics[width=0.93\textwidth]{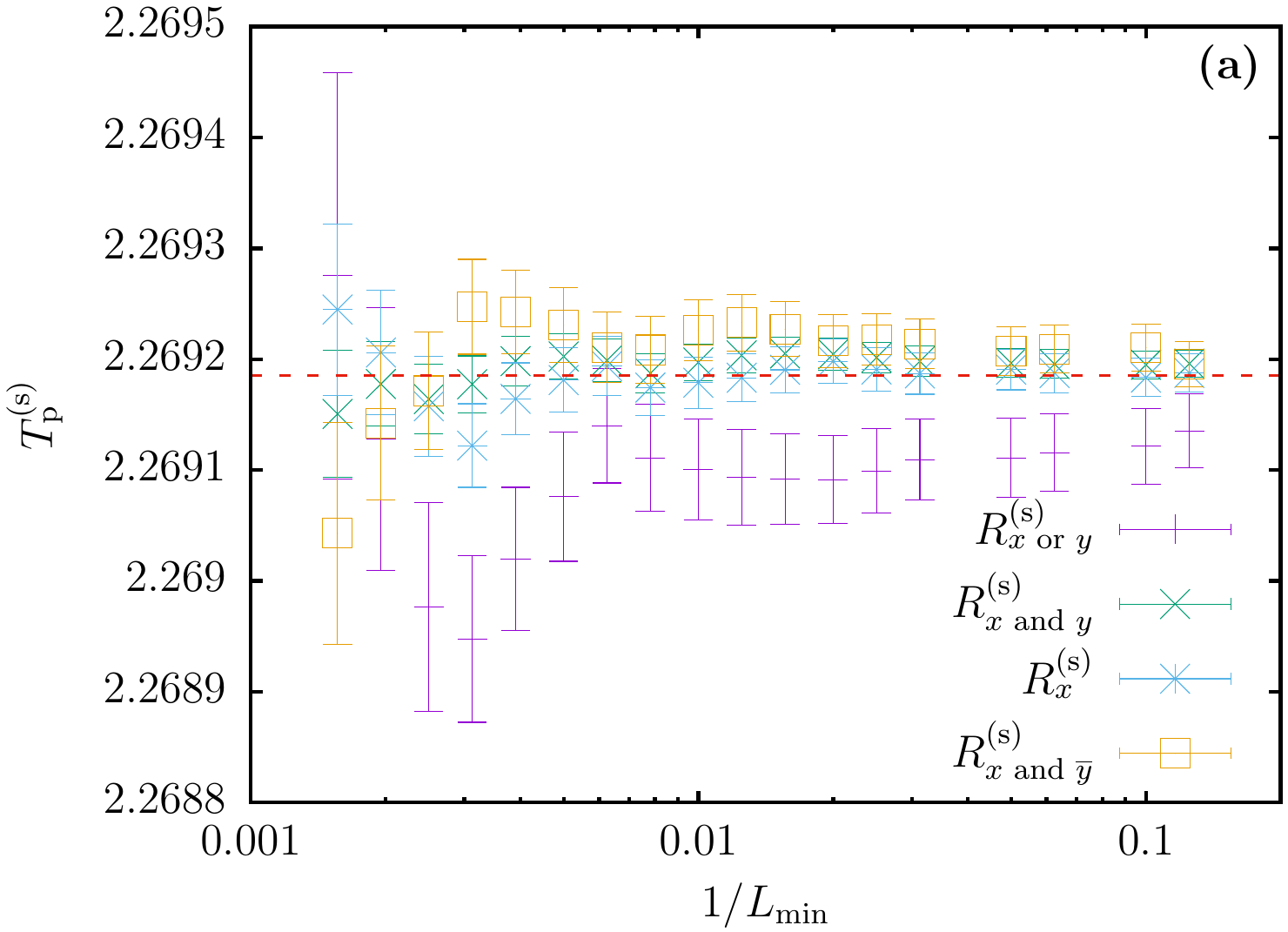}
        \label{fig:q2_Tc_estimates_soft-fit_from_reweighting}
	\end{subfigure}
	\begin{subfigure}[]{0.5\textwidth}
        %\hspace{-0.5cm}
		%\resizebox{0.96\linewidth}{!}{\large\input{Figures/q2_Tc_estimates_hard-fit_from_reweighting}}
        %\includegraphics[width=0.93\textwidth]{./figures/q2_Tc_estimates_hard-fit_from_reweighting.pdf}
        \includegraphics[width=0.93\textwidth]{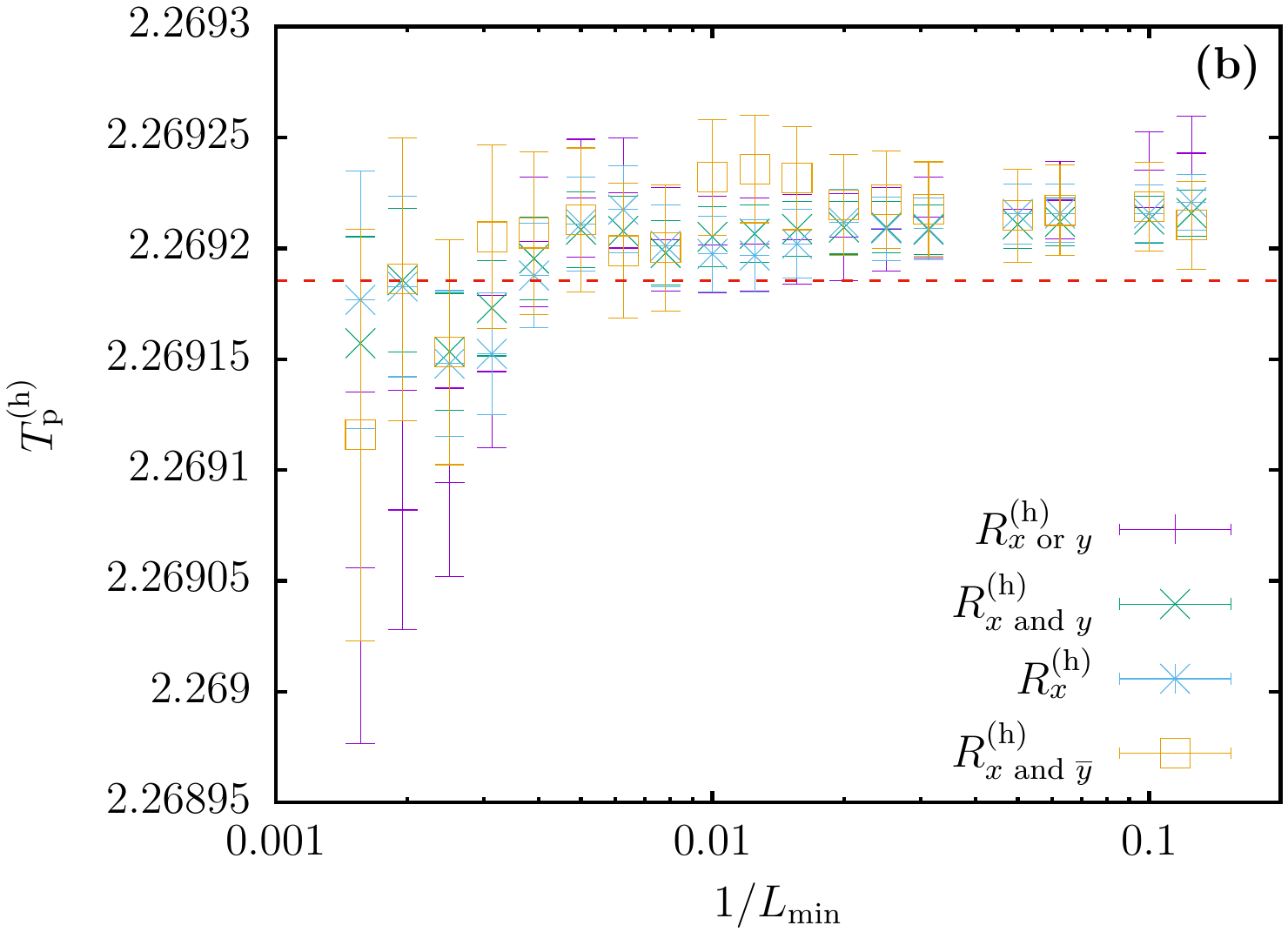}
		\label{fig:q2_Tc_estimates_hard-fit_from_reweighting}
	\end{subfigure}
	
	%\caption{Estimates of the critical temperature for the different definitions of the wrapping probabilities as a function of $ 1/L_{\text{min}} $ for the \subref{fig:q2_Tc_estimates_soft-fit_from_reweighting} soft and \subref{fig:q2_Tc_estimates_hard-fit_from_reweighting} hard constraint clusters. Estimates extracted from the linear fits of Eq.~\eqref{eq:tc_crossing-linear}. The dashed horizontal line marks the transition temperature of the 1-replica Ising model.}

    \caption{Estimates of the percolation temperature $T_\mathrm{p}$ vs. $1/L_{\rm min}$ on a semi-logarithmic scale. Results obtained from fits of the form~\eqref{eq:tc_crossing-linear} on the data of Fig.~\ref{fig:q2_Tc_crossings_soft_hard_from_reweighting}. The dashed horizontal line marks here as well the transition temperature of the Ising model.}
	\label{fig:q2_Tc_estimates_soft_hard-fit_from_reweighting}
\end{figure}		

Estimates of the percolation temperature resulting from the linear fits of Eq.~\eqref{eq:tc_crossing-linear} are shown in Fig.~\ref{fig:q2_Tc_estimates_soft_hard-fit_from_reweighting} for both soft- and hard-constraint clusters; they are found to be consistent with the critical temperature of the Ising ferromagnet, i.e., $T_{\rm c} = 2.269185\ldots$. 
We note that the results for $R_{\text{x or y}}$ have slightly elevated statistical errors as compared to the other definitions, a feature that can be traced back to the fact that $R_{\text{x or y}}$ is very close to one at $T_c$ such that the curves cross at a smaller angle, cf.\ Figs.~\ref{fig:pwrap_soft}(a) and \ref{fig:pwrap_hard}(a).
In addition, the parameter $a$ of Eq.~\ref{eq:tc_crossing-linear} for all fits is consistent with zero within error bars, indicating that the data can also be described by a constant $T_\mathrm{cross}(L) = T_\mathrm{p} = T_\mathrm{c}$, independent of $L$.
For random percolation it has been observed that most estimates of pseudo-critical points on the square lattice converge even more quickly to the thermodynamic limit than expected from the $L^{-1/\nu}$ scaling \cite{ziff:02}, namely proportional to $L^{-(1/\nu+1)}$. Performing fits with a corresponding variant of Eq.~\eqref{eq:tc_crossing-linear} with the $1/L$ term replaced by $1/L^2$ yield perfectly compatible results, indicating that our data are in fact not able to resolve these very small deviations from the infinite-size behavior.
At this point we may safely conclude that soft- and hard-constraint clusters undergo a percolation transition at the critical temperature $T_\mathrm{c}$ of the Ising model.

%We note here, that the reasonable estimates that we  obtain from fits using Eq.~\ref{eq:tc_crossing-linear} do not imply that corrections to scaling are absent but rather not visible given the accuracy of our data
	
% \input{./Tables/table-q2_Tc_crossings_soft_reweighting.tex}
% \input{./Tables/table-q2_Tc_crossings_hard_reweighting.tex}

\subsection{Scaling at criticality} 
\label{subsec:at tc}
	
\begin{figure} % p soft hard (23 points 4 plots)
	\begin{subfigure}[]{0.5\textwidth}
        %\hspace{-0.55cm}
		%\resizebox{0.96\linewidth}{!}{\large\input{Figures/q2_p_soft_Tc.tex}}
        %\includegraphics[width=0.93\textwidth]{./figures/q2_p_soft_Tc.pdf}
        \includegraphics[width=0.93\textwidth]{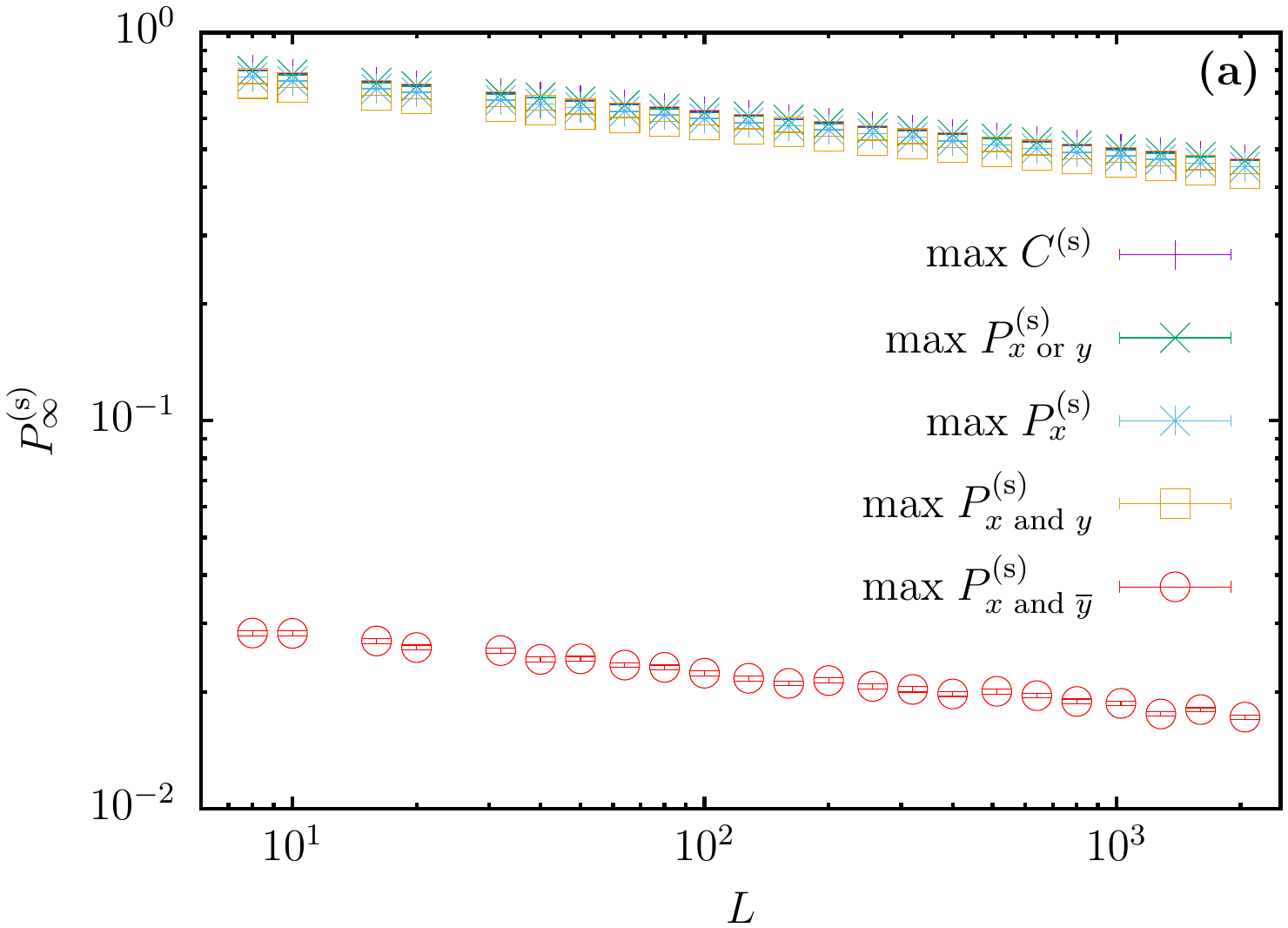}
		\label{fig:q2_p_soft_Tc}
	\end{subfigure}
 
	\begin{subfigure}[]{0.5\textwidth}
        %\hspace{-0.55cm}
		%\resizebox{0.96\linewidth}{!}{\large\input{Figures/q2_p_hard_Tc.tex}}
        %\includegraphics[width=0.93\textwidth]{./figures/q2_p_hard_Tc.pdf}
        \includegraphics[width=0.93\textwidth]{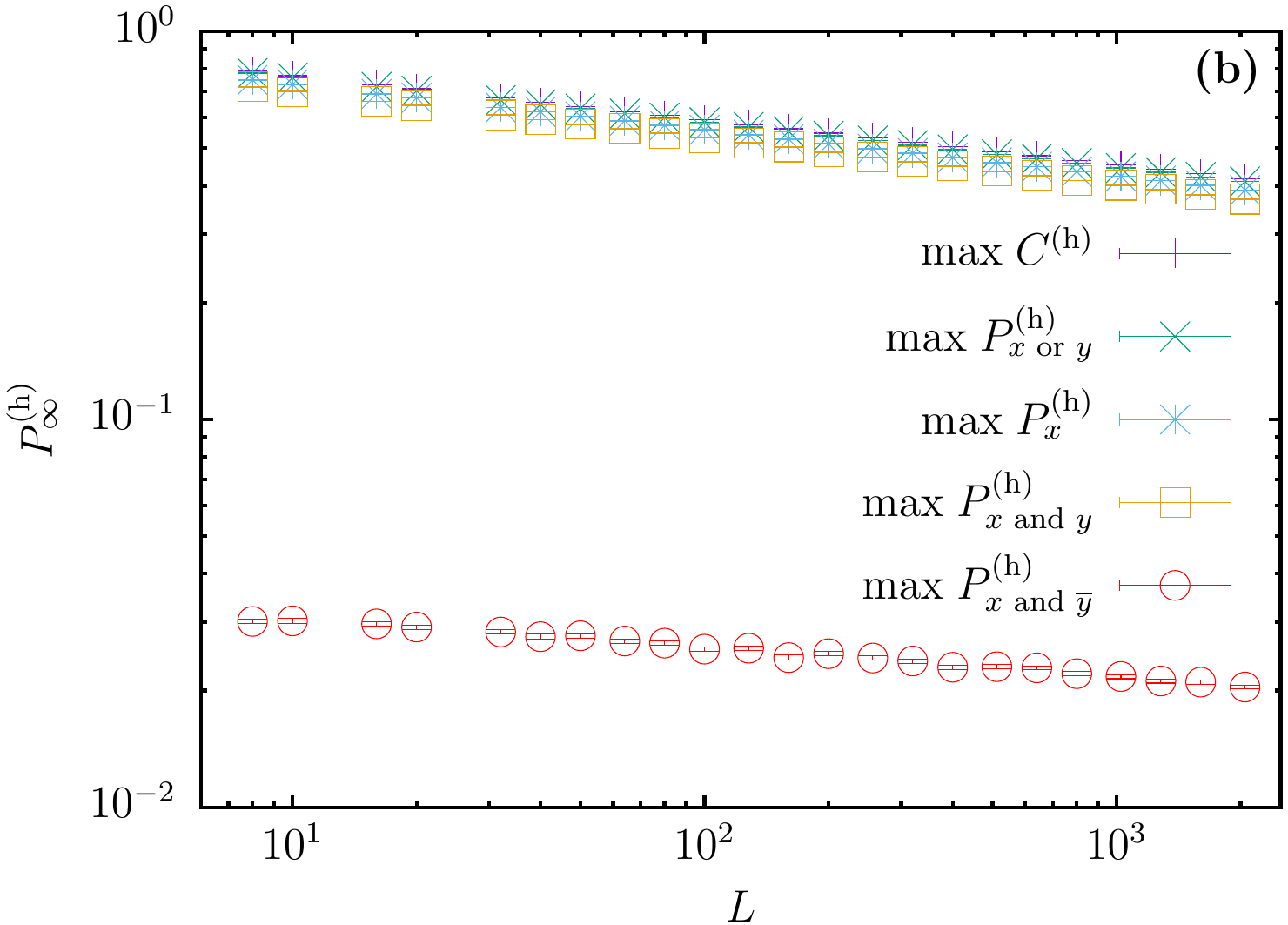}
		\label{fig:q2_p_hard_Tc}
	\end{subfigure}
	
	%\caption{Log-log plot of the percolation strength for the different definitions as a function of the system size $ L $ for the \subref{fig:q2_p_soft_Tc} soft and \subref{fig:q2_p_hard_Tc} hard constraint clusters, at the critical temperature of the 1-replica Ising model.}

    \caption{Percolation strength $P_\infty$ vs.\ $L$ at the critical temperature of the Ising model on a double logarithmic scale for (a) soft-constraint and (b) hard-constraint clusters.}
	\label{fig:q2_p_soft_hard_Tc}
\end{figure}

\begin{figure} % s soft hard (23 points 4 plots)
	\begin{subfigure}[]{0.5\textwidth}
        %\hspace{-0.8cm}
		%\resizebox{0.96\linewidth}{!}{\large\input{Figures/q2_s_soft_Tc.tex}}
        %\includegraphics[width=0.93\textwidth]{./figures/q2_s_soft_Tc.pdf}
        \includegraphics[width=0.93\textwidth]{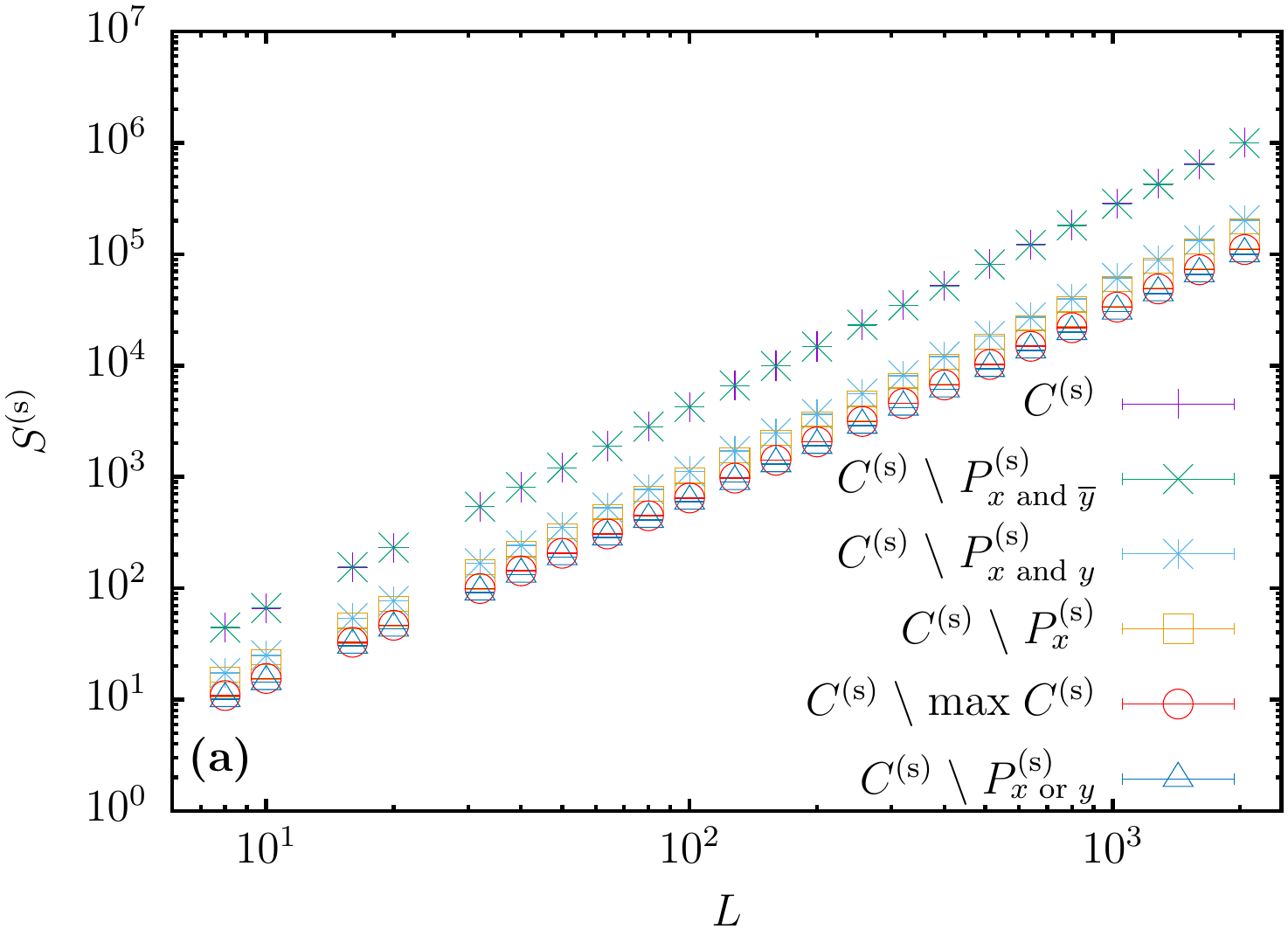}
		\label{fig:q2_s_soft_Tc}
	\end{subfigure}
	\begin{subfigure}[]{0.5\textwidth}
        %\hspace{-0.8cm}
		%\resizebox{0.96\linewidth}{!}{\large\input{Figures/q2_s_hard_Tc.tex}}
        %\includegraphics[width=0.93\textwidth]{./figures/q2_s_hard_Tc.pdf}
        \includegraphics[width=0.93\textwidth]{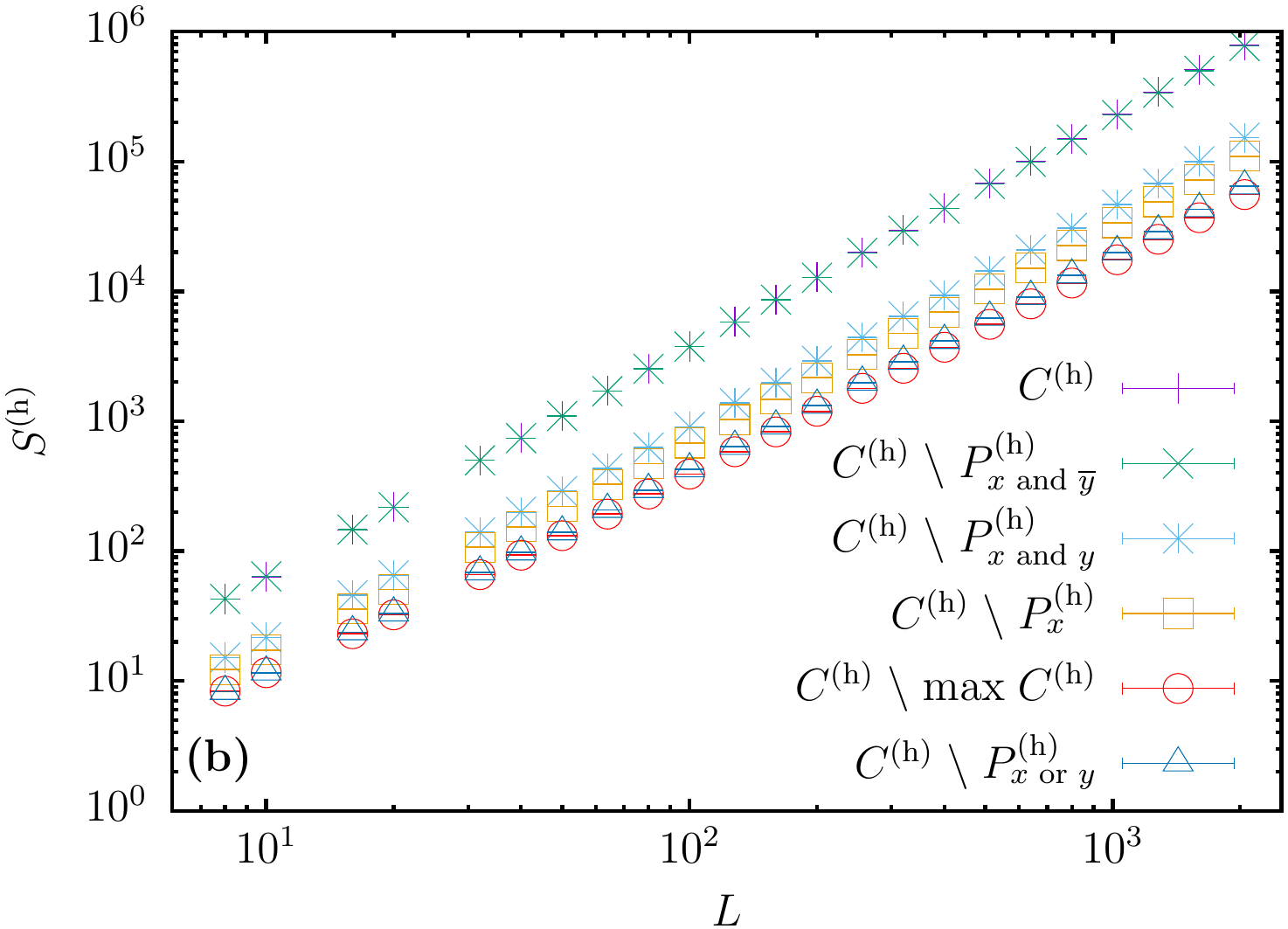}
		\label{fig:q2_s_hard_Tc}
	\end{subfigure}
	
	%\caption{Log-log plot of the average cluster size for the different definitions as a function of the system size $ L $ for the \subref{fig:q2_s_soft_Tc} soft and \subref{fig:q2_s_hard_Tc} hard constraint clusters, at the critical temperature of the 1-replica Ising model.}

    \caption{Average cluster size $S$ vs.\ $L$ at the critical temperature of the Ising model on a double logarithmic scale for (a) soft-constraint and (b) hard-constraint clusters.}
	\label{fig:q2_s_soft_hard_Tc}
\end{figure}

We proceed with the computation of critical exponents related to the percolation strength, $P_{\infty}$, and the average cluster size, $S$, for the two types of 
clusters at criticality. Since we found convincing evidence of the identity of the percolation and critical exponents, $T_\mathrm{p} = T_\mathrm{c} = 2.269185\ldots$ in the previous section, we considered the scaling of $P_{\infty}$ and $S$ at the fixed temperature $T = T_\mathrm{c}$; the result is shown for the soft- and hard-constraint clusters in Figs.~\ref{fig:q2_p_soft_hard_Tc} and \ref{fig:q2_s_soft_hard_Tc}, respectively. For both observables, all data appear to follow straight parallel lines,
suggesting minor corrections to scaling and universal exponents, independent of the definition used.
	
At the percolation point, according to Eqs.~\eqref{eq:average_cluster_size_scaled} and \eqref{eq:strength_scaled} the scaling functions become constant, thus allowing for the estimation of the involved exponents from finite-size scaling. In the following, we present the results of systematically fitting those functional forms to the data for $P_{\infty}$ and $S$ for all sets of definitions, following the established protocol of varying $L_{\rm min}$ as described above, allowing us to monitor the influence of corrections to scaling.

\begin{figure}[] % beta/nu 1 soft hard (21 points 4 plots)
	\begin{subfigure}[]{0.5\textwidth}
        %\hspace{-0.8cm}
		%\resizebox{0.96\linewidth}{!}{\large\input{Figures/q2_beta_soft_Tc-no_corrections-1.tex}}
        %\includegraphics[width=0.93\textwidth]{./figures/q2_beta_soft_Tc-no_corrections-1.pdf}
        \includegraphics[width=0.93\textwidth]{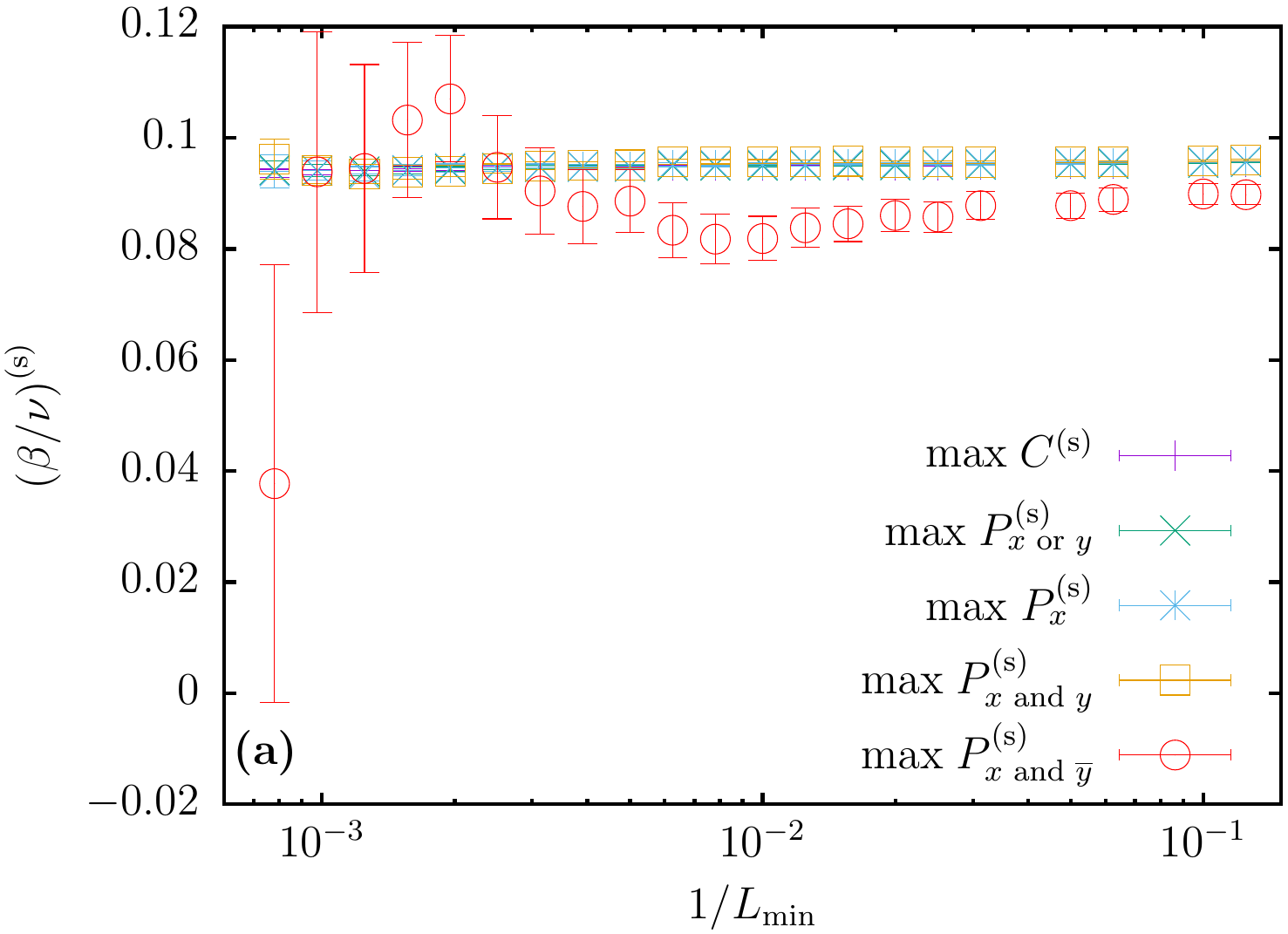}
		\label{fig:q2_beta_soft_Tc-no_corrections-1}
	\end{subfigure}
	\begin{subfigure}[]{0.5\textwidth}
        %\hspace{-0.8cm}
		%\resizebox{0.96\linewidth}{!}{\large\input{Figures/q2_beta_hard_Tc-no_corr0ections-1.tex}}
        %\includegraphics[width=0.93\textwidth]{./figures/q2_beta_hard_Tc-no_corrections-1.pdf}
        \includegraphics[width=0.93\textwidth]{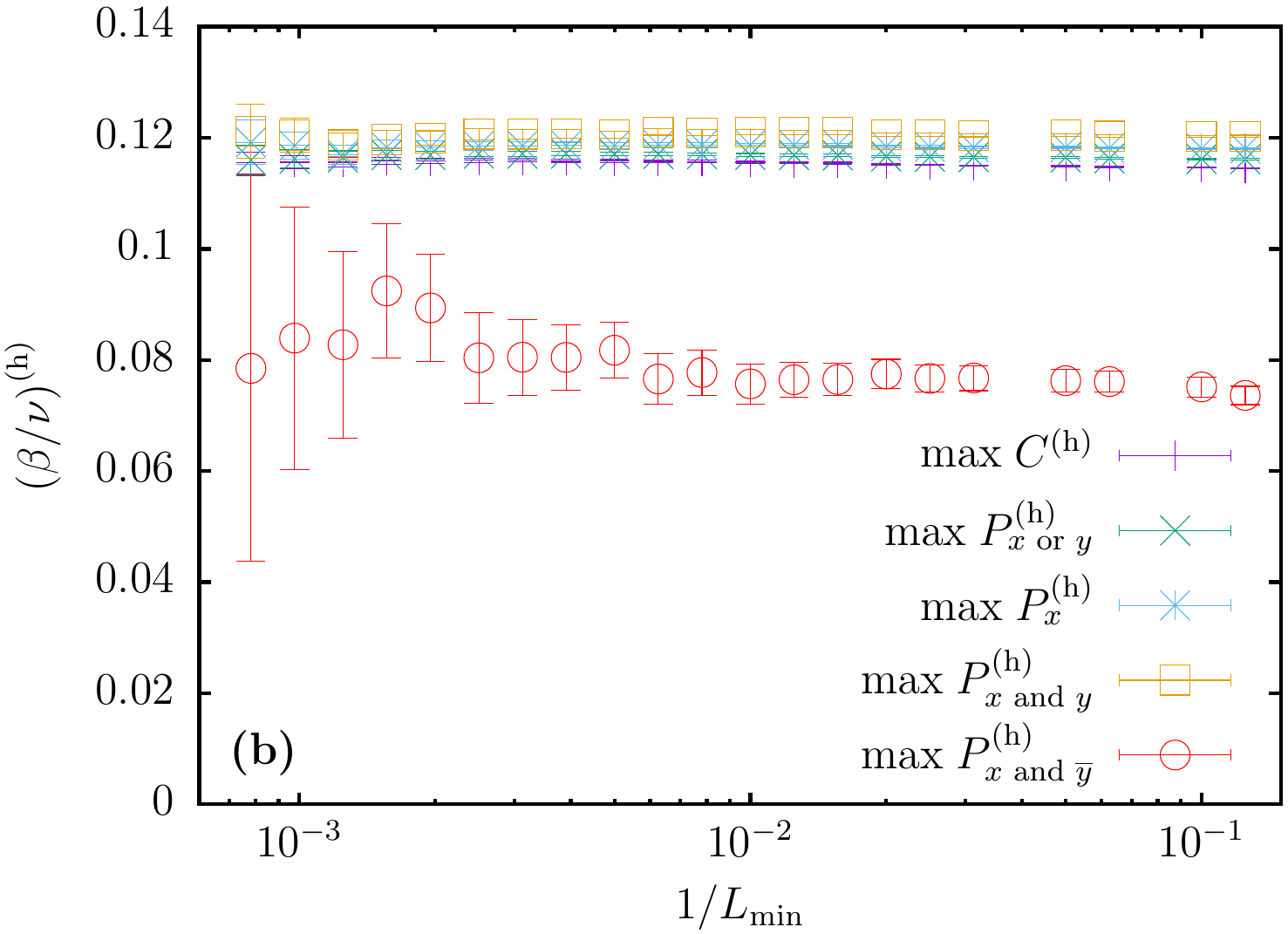}
		\label{fig:q2_beta_hard_Tc-no_corrections-1}
	\end{subfigure}
	
	%\caption{Exponent ratio $ \beta/\nu$ for the different definitions as a function of $1/L_{\text{min}}$ for the \subref{fig:q2_beta_soft_Tc-no_corrections-1} soft and  \subref{fig:q2_beta_hard_Tc-no_corrections-1} hard constraint clusters.}

    \caption{Effective exponent ratios $\beta/\nu$ extracted from fits of the functional forms \eqref{eq:average_cluster_size_scaled} and \eqref{eq:strength_scaled} as a function of the inverse lower size cut-off $1/L_{\text{min}}$ for (a) soft-constraint and (b) hard-constraint clusters.}
	\label{fig:q2_beta_soft_hard_Tc-no_corrections-1}
\end{figure}

Figure~\ref{fig:q2_beta_soft_hard_Tc-no_corrections-1} shows our estimates of the effective exponent ratios $\left(\beta / \nu\right)^{\text{(s)}} $ and $ \left(\beta / \nu\right)^{\text{(h)}}$ while varying the cutoff $L_{\rm min}$. Besides the $\text{max} \; P_{\text{x and } \overline{\text{y}}}$ data, it is evident that the exponents converge relatively quickly to the values $\left(\beta / \nu\right)^{\text{(s)}} \approx 0.095$  and $\left(\beta / \nu\right)^{\text{(h)}} \approx 0.12$ for the soft- and hard-constraint clusters, respectively. The fact that the numerical data for the $\text{max} \; P_{\text{x and } \overline{\text{y}}}$ definition provide unreliable estimates of the involved exponent can be readily understood: clusters percolating in one but not the other direction are sparse, leading to poor statistics in the estimation of the percolation strength and the associated exponent. This observation has also been reported in Ref.~\cite{akritidis_corrections_2022} for the Ising model. We note, however, that it is also conceivable that the rather special constraint ``\text{x} and $\overline{\text{y}}$'' leads to an asymptotically different fractal dimension for such clusters as compared to the unconstrained case.
	
%All of the above observations are supported by Tables \ref{table:q2_beta_soft_different_definitions-no_corrections}-\ref{table:q2_beta_hard_different_definitions-no_corrections}, where the estimates of $ \beta / \nu $ for different fit intervals, the degrees of freedom (d.o.f.), the  $ x^2 / \text{d.o.f.} $, and the respective quality-of-fit parameter Q are reported for the sc and hc clusters respectively. 

As there is no systematic trend visible in our data that could possible reveal the existence of corrections to scaling for $\left(\beta / \nu \right)^{\text{(s)}}$ and $\left(\beta / \nu \right)^{\text{(h)}}$, we did not attempt to perform fits including correction terms. Instead, as a trade-off between unavoidable corrections to scaling and reasonable values of $\chi^2$, we choose sufficiently large $L_\mathrm{min}$ for our final estimates of $ \left(\beta / \nu \right)^{\text{(s)}}$ and $\left(\beta / \nu \right)^{\text{(h)}}$,
%Tables \ref{table:q2_beta_soft_different_definitions-no_corrections} and \ref{table:q2_beta_hard_different_definitions-no_corrections}, are:
\begin{subequations}
\begin{equation}
    \label{eq:beta_soft_estimation}
	\left(\frac{\beta}{\nu} \right)^{\text{(s)}} = 0.0950(7) \;\; (L_{\text{min}}=320),\\
\end{equation}
\begin{equation}
\label{eq:beta_hard_estimation}
	\left(\frac{\beta}{\nu} \right)^{\text{(h)}} = 0.1184(11) \;\; (L_{\text{min}}=512).
\end{equation}
\end{subequations}

% 	\begin{figure}[] % beta/nu 2 soft hard (21 points 4 plots)
% 		\begin{subfigure}[]{0.5\textwidth}
% 			\resizebox{1.0\linewidth}{!}{\large\input{Figures/q2_beta_soft_Tc-no_corrections-2.tex}}
% 			\caption{}
% 			\label{fig:q2_beta_soft_Tc-no_corrections-2}
% 		\end{subfigure}
% 		\begin{subfigure}[]{0.5\textwidth}
% 			\resizebox{1.0\linewidth}{!}{\large\input{Figures/q2_beta_hard_Tc-no_corrections-2.tex}}
% 			\caption{}
% 			\label{fig:q2_beta_hard_Tc-no_corrections-2}
% 		\end{subfigure}
% 		\captionsetup{ justification=raggedright}
% 		\caption{Same as Fig. \ref{fig:q2_beta_soft_hard_Tc-no_corrections-1}, without the $ \text{max} \; P^{\left(1\right)} $ definition.}
% 		\label{fig:q2_beta_soft_hard_Tc-no_corrections-2}
% 	\end{figure}
	
% \input{./Tables/table-q2_beta_soft_different_definitions-no_corrections.tex}
% \input{./Tables/table-q2_beta_hard_different_definitions-no_corrections.tex}
	
\begin{figure}[]% gamma/nu  soft hard (21 points 4 plots)
	\begin{subfigure}[]{0.5\textwidth}
        %\hspace{-0.5cm}
		%\resizebox{0.96\linewidth}{!}{\large\input{Figures/q2_gamma_soft_Tc-no_corrections.tex}}
        %\includegraphics[width=0.93\textwidth]{./figures/q2_gamma_soft_Tc-no_corrections.pdf}
        \includegraphics[width=0.93\textwidth]{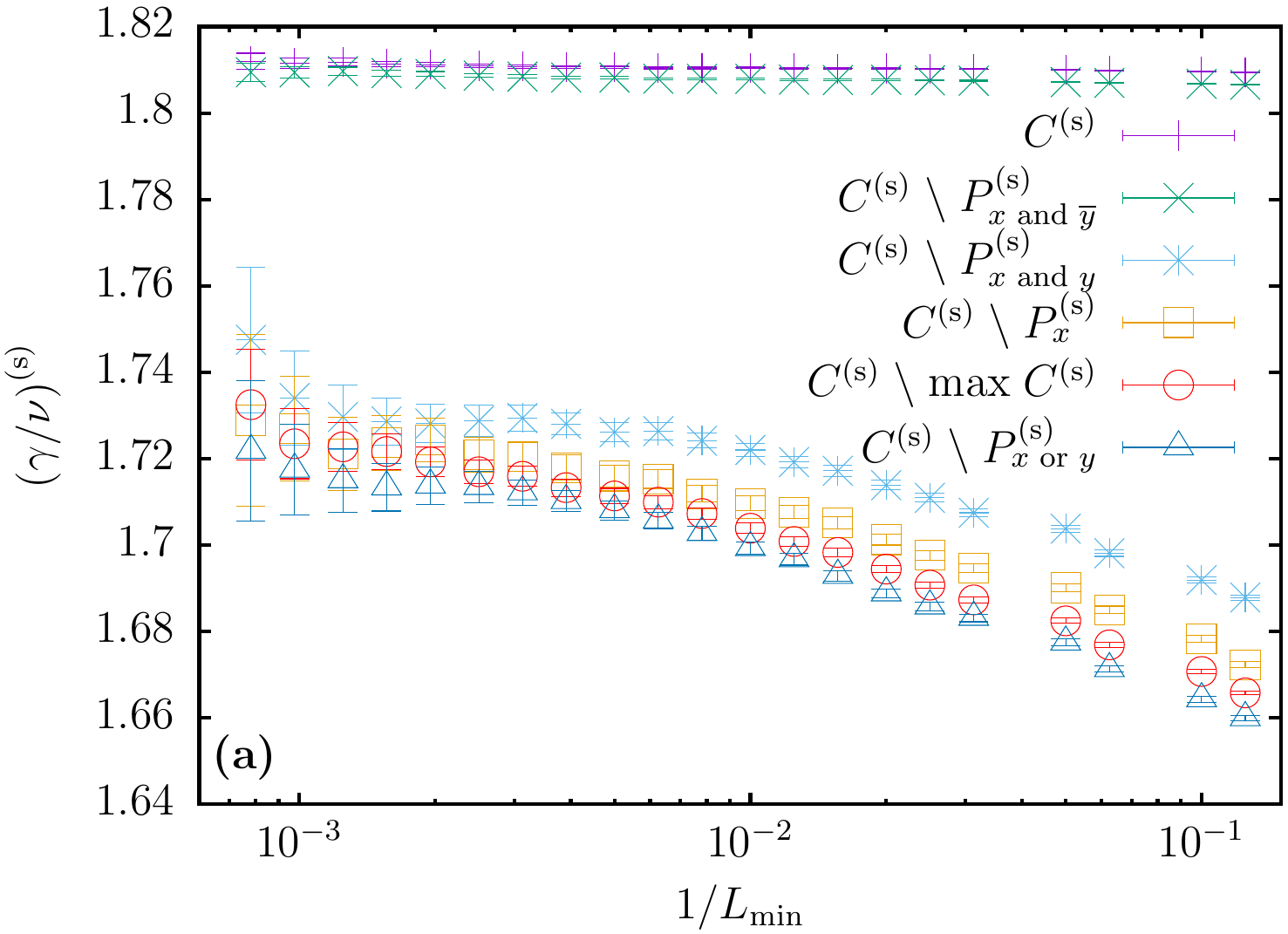}
		\label{fig:q2_gamma_soft_Tc-no_corrections}
	\end{subfigure}
	\begin{subfigure}[]{0.5\textwidth}
        %\hspace{-0.5cm}
		%\resizebox{0.96\linewidth}{!}{\large\input{Figures/q2_gamma_hard_Tc-no_corrections.tex}}
        %\includegraphics[width=0.93\textwidth]{./figures/q2_gamma_hard_Tc-no_corrections.pdf}
        \includegraphics[width=0.93\textwidth]{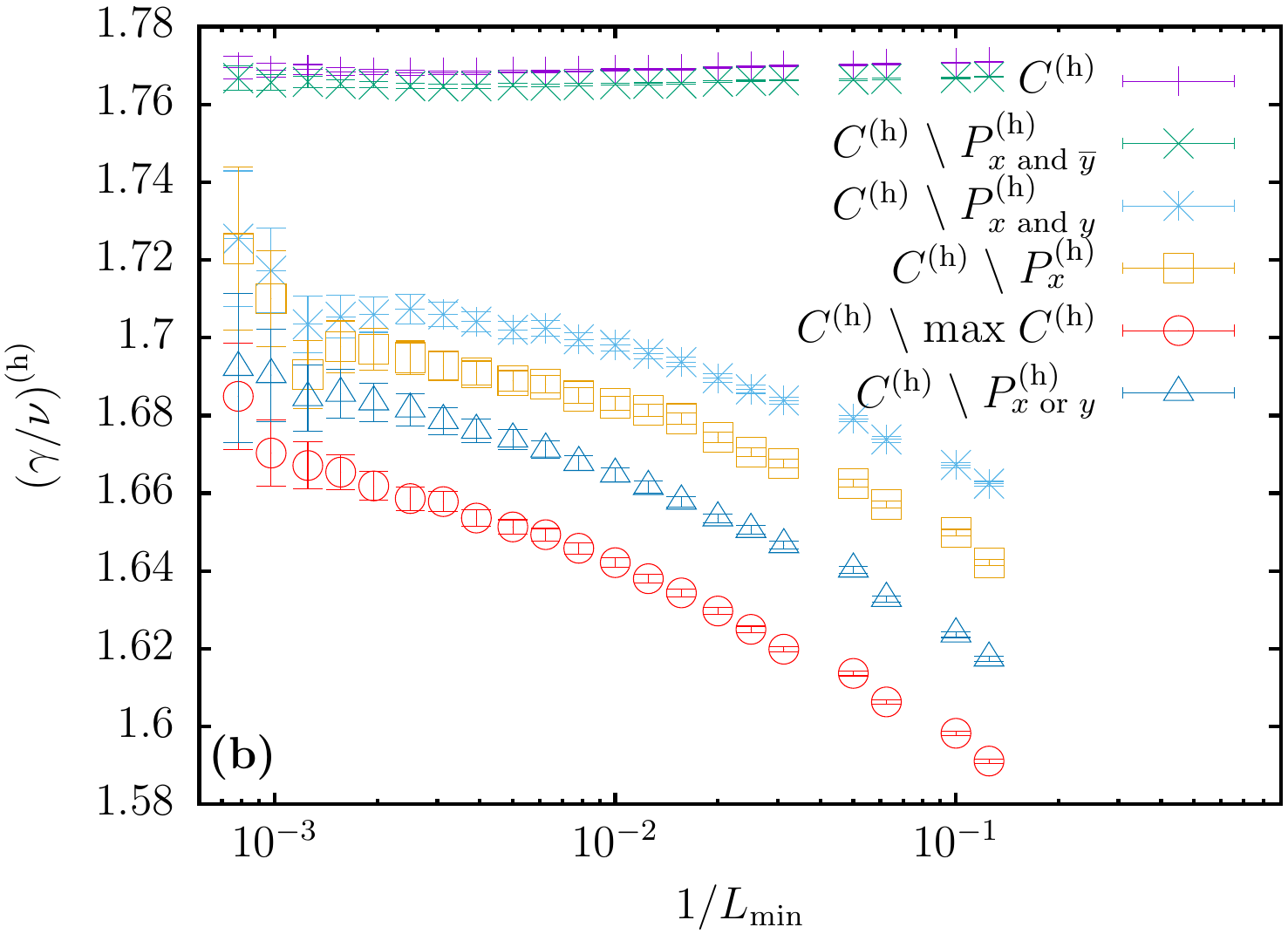}
		\label{fig:q2_gamma_hard_Tc-no_corrections}
	\end{subfigure}
	
	%\caption{Exponent ratio $ \gamma/\nu$ for the different definitions as a function of $1/L_{\text{min}}$ for the \subref{fig:q2_gamma_soft_Tc-no_corrections} soft and  \subref{fig:q2_gamma_hard_Tc-no_corrections} hard constraint clusters.}

    \caption{Effective exponent ratios $\gamma/\nu$ vs. $1/L_{\text{min}}$ for (a) soft-constraint and (b) hard-constraint clusters.}
	\label{fig:q2_gamma_soft_hard_Tc-no_corrections}
\end{figure}

Estimates for the effective exponent ratios  $\left(\gamma / \nu\right)^{\text{(s)}}$ and $ \left(\gamma / \nu\right)^{\text{(h)}} $ are shown in Fig. \ref{fig:q2_gamma_soft_hard_Tc-no_corrections}. For the $C$ and $C \; \backslash \; P_{\text{x and } \overline{\text{y}}}$ definitions, the exponent ratio  converges relatively quickly to the value $\left(\gamma / \nu\right)^{\text{(s)}} \approx 1.81$  and $\left(\gamma / \nu\right)^{\text{(h)}} \approx 1.77$ for the soft- and hard-constraint clusters respectively, indicating that corrections to scaling are not substantial. On the other hand, for the rest of the definitions, corrections to scaling become important and the convergence to an asymptotic value is rather slow. The fact that $C$ and $C \; \backslash \; P_{\text{x and } \overline{\text{y}}}$ give similar results is to be expected, as the latter definition excludes clusters that rarely appear, thus not altering significantly the sums in Eq.~\eqref{eq:average_cluster_size}. Note that the reduction of scaling corrections for the average cluster size through the choice of the $C$ definition was also reported in Ref.~\cite{janke_fractal_2005} for the Ising model, while a systematic study of the behavior for the Ising model from all of the above definitions was presented in Ref.~\cite{akritidis_corrections_2022}.
	
% \input{./Tables/table-q2_gamma_soft_different_definitions-no_corrections.tex}
% \input{./Tables/table-q2_gamma_hard_different_definitions-no_corrections.tex}

% Joint fit ===================================================================
	
\begin{figure}[] % gamma soft hard joint (19 points 1 plot)
	\begin{subfigure}[]{0.5\textwidth}
        %\hspace{-0.8cm}
		%\resizebox{0.96\linewidth}{!}{\large\input{Figures/q2_gamma_soft_Tc-joint-with_corrections.tex}}
        \includegraphics[width=0.93\textwidth]{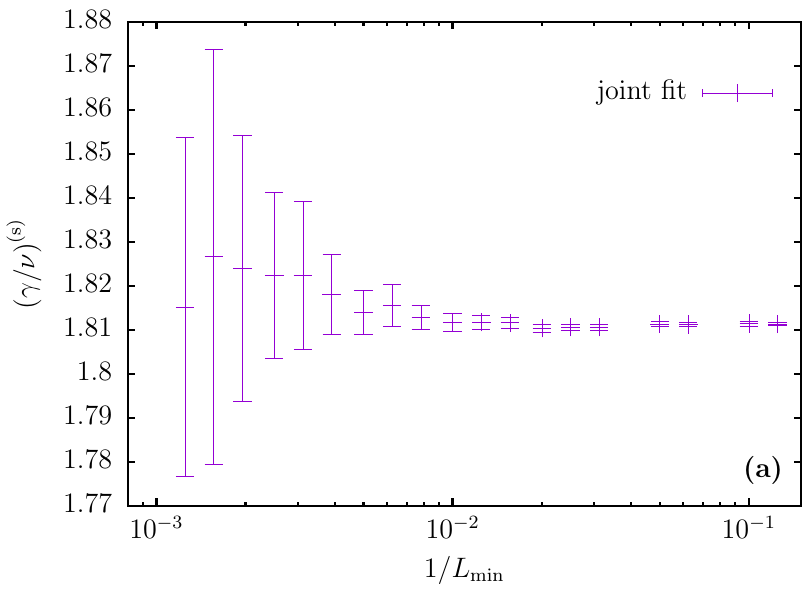}
		\label{fig:q2_gamma_soft_Tc-joint-with_corrections}
	\end{subfigure}
	\begin{subfigure}[]{0.5\textwidth}
        %\hspace{-0.8cm}
		%\resizebox{0.96\linewidth}{!}{\large\input{Figures/q2_gamma_hard_Tc-joint-with_corrections.tex}}
		\includegraphics[width=0.93\textwidth]{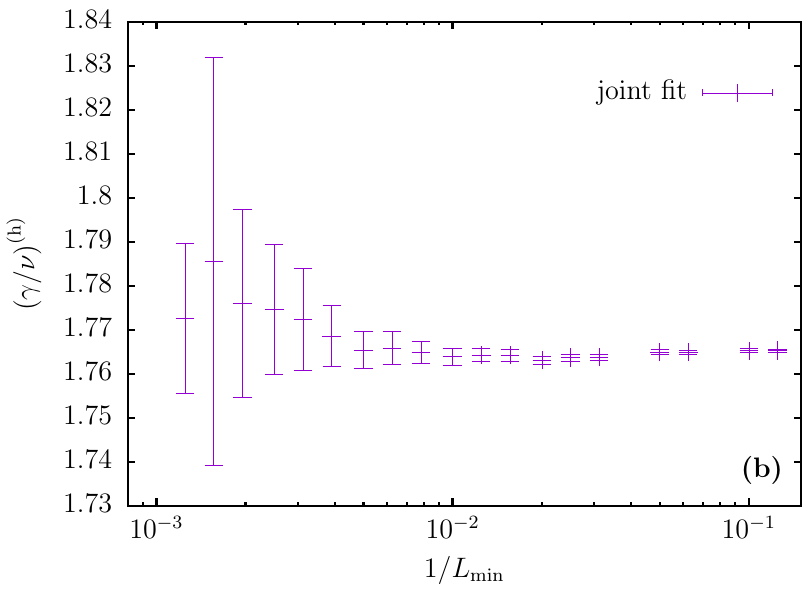}
        \label{fig:q2_gamma_hard_Tc-joint-with_corrections}
	\end{subfigure}
	
	%\caption{Exponent ratio $\gamma/ \nu $ resulting from the joint fit of all definitions as a function of $1/L_{\text{min}}$ for the \subref{fig:q2_gamma_soft_Tc-joint-with_corrections} soft and \subref{fig:q2_gamma_hard_Tc-joint-with_corrections} hard constraint clusters.}

    \caption{Exponent ratio $\gamma/ \nu $ resulting from the joint fit of all definitions as a function of $1/L_{\text{min}}$ for (a) soft-constraint and (b) hard-constraint clusters.}
	\label{fig:q2_gamma_soft_hard_Tc-joint-with_corrections}
\end{figure}

\begin{figure}[] % omega soft hard joint (19 points 1 plot)
	\begin{subfigure}[]{0.5\textwidth}
        %\hspace{-0.5cm}
		%\resizebox{0.96\linewidth}{!}{\large\input{Figures/q2_omega_soft_Tc-joint-with_corrections.tex}}
        \includegraphics[width=0.93\textwidth]{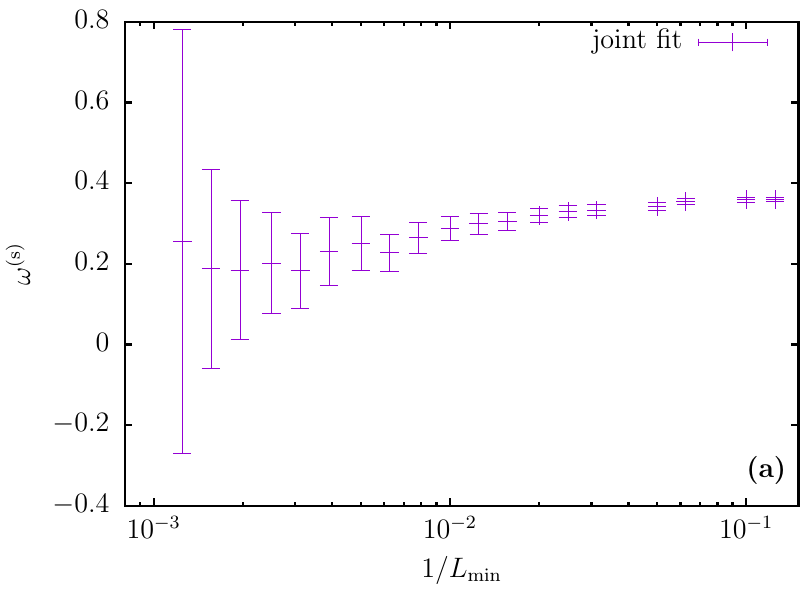}
		\label{fig:q2_omega_soft_Tc-joint-with_corrections}
	\end{subfigure}
	\begin{subfigure}[]{0.5\textwidth}
        %\hspace{-0.5cm}
		%\resizebox{0.96\linewidth}{!}{\large\input{Figures/q2_omega_hard_Tc-joint-with_corrections.tex}}
        \includegraphics[width=0.93\textwidth]{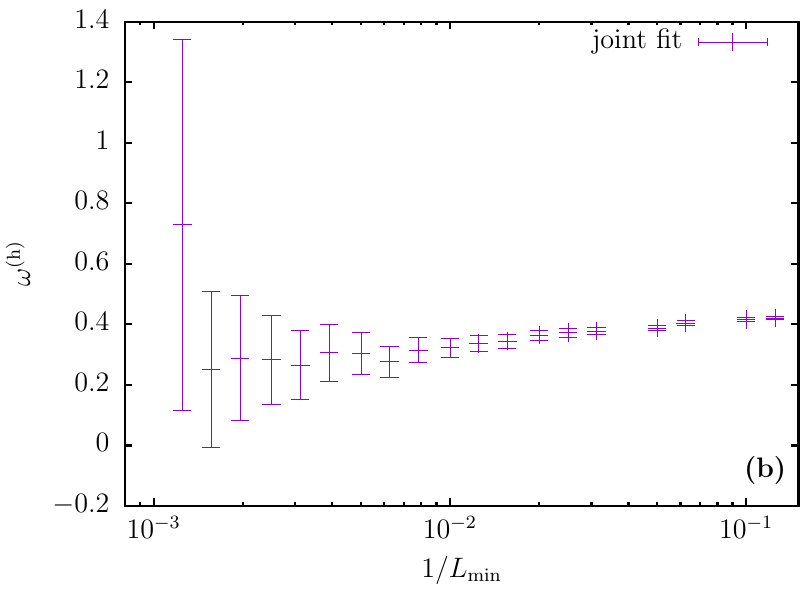}
		\label{fig:q2_omega_hard_Tc-joint-with_corrections}
	\end{subfigure}
	
	%\caption{Correction-to-scaling exponent $\omega$ resulting from the joint fit of all definitions as a function of $1/L_{\text{min}}$ for the \subref{fig:q2_omega_soft_Tc-joint-with_corrections} soft and \subref{fig:q2_omega_hard_Tc-joint-with_corrections} hard constraint clusters.}
 
    \caption{Corrections-to-scaling exponent $\omega$ resulting from the joint fit of all definitions as a function of $1/L_{\text{min}}$ for (a) soft-constraint and (b) hard-constraint clusters.}
	\label{fig:q2_omega_soft_hard_Tc-joint-with_corrections}
\end{figure}

More reliable estimates for the exponent ratio $\gamma/\nu$ can be retrieved by taking into account corrections to scaling for both soft- and hard-constraint clusters. To arrive at somewhat stable estimates for the correction-to-scaling exponent, we performed joint fits to the data for the different definitions of $S$ including one correction term using the \textit{Ansatz}
\begin{equation}
\label{eq:s_scaling_at_Tc_omega}
	S = aL^{\gamma / \nu} \left( 1+bL^{-\omega} \right),
\end{equation}
where $a$ and $b$ are the non-common fitting parameters and $\gamma/\nu$, and $\omega$ the shared parameters. Since data from different definitions are not statistically independent, as they result from the same Monte Carlo series, a naive implementation of the above fitting procedure will result in erroneous error estimation of the fit parameters. Thus, in order to provide reliable estimates for the errors of the involved parameters, we employed the jackknife method~\cite{efron}.
For all values of $L_{\text{min}}$, estimates of $\gamma/\nu$ agree within error bars as is shown in Fig.~\ref{fig:q2_gamma_soft_hard_Tc-joint-with_corrections}. Additionally, in Fig.~\ref{fig:q2_omega_soft_hard_Tc-joint-with_corrections} the exponent $\omega$ is plotted as a function of $L_{\text{min}}$. The fact that the error bars in the estimates of $\gamma/\nu$ and $\omega$  are increasing with $L_{\text{min}}$ is of course a consequence of the decreasing number of degrees of freedom. However, the errors in $\omega$ increase rapidly, and for $L_{\text{min}} \ge 320$ they are comparable with their absolute values for both cluster types. Our final estimates of the involved exponent ratio $\gamma/\nu$ of the soft- and hard-constraint clusters are: \myworries{What values to chose as final answer?}

\begin{subequations}
\begin{equation}
    \label{eq:gamma_soft_estimation}
	\left(\frac{\gamma}{\nu} \right)^{\text{(s)}} = 1.814(5)\;\; (L_{\text{min}}=200),\\
\end{equation}
\begin{equation}
\label{eq:gamma_hard_estimation}
	\left(\frac{\gamma}{\nu} \right)^{\text{(h)}} = 1.765(4) \;\;  (L_{\text{min}}=200).
\end{equation}
\end{subequations}

%\begin{align}
 %   \label{eq:gamma_soft_estimation}
%	&\left(\frac{\gamma}{\nu} \right)^{\text{(s)}} = 1.814(5),&  %\chi^2/\text{d.o.f.} \approx 0.54,& L_{\text{min}}=200 \\
%	\label{eq:gamma_hard_estimation}
%	&\left(\frac{\gamma}{\nu} \right)^{\text{(h)}} = 1.765(4),&  %\chi^2/\text{d.o.f.} \approx 0.67,&  L_{\text{min}}=200
%\end{align}

We note that the estimated value of $\omega \approx 0.2$ for both soft- and hard-constraint clusters represents a rather slow decay of corrections, consistent with the slow convergence of $\gamma/\nu$ for most definitions of $S$ that is clearly visible in Fig.~\ref{fig:q2_gamma_soft_hard_Tc-no_corrections}. We are not aware of any theoretical estimates relating to the value of $\omega$.

\subsection{Fractal dimension and hyperscaling} \label{subsec:fractal_dimension_hyperscaling}

At the percolation point the incipient spanning cluster is a fractal object and its mass $M$ (i.e., the number of spins that belong to the spanning cluster) scales with the system size as $M \sim L^{D}$.  Here, $D$ denotes the fractal dimension of the incipient spanning cluster. Since from the discussion in Sec.~\ref{sec:observables} the percolation strength $P_\infty$ corresponds to the fraction of spins in the percolating cluster, it follows that $D = d - \beta/\nu$. From the estimates of $\beta/\nu$ for the soft- and hard-constraint clusters provided in Eqs.~\eqref{eq:beta_soft_estimation} and \eqref{eq:beta_hard_estimation} we hence obtain
\begin{equation}
    D^{\text{(s)}} =  1.9050(7); \;\; D^{\text{(h)}} =  1.8816(11),
    \label{eq:fractal_dimension_estimates}
\end{equation}
suggesting that the fractal dimension is different for the two cluster types. Also, the fact that $D^{\text{(s)}} > D^{\text{(h)}}$ highlights that the soft-constraint clusters are denser than the hard-constraint ones, a reasonable expectation as the hard-constraint clusters are a subset of the soft-constraint ones. Additionally, let us point out that the fractal dimensions for both cluster types are smaller than $D = 187/96 \approx 1.9479$, the fractal dimension of geometrical clusters in the Ising model~\cite{stella_scaling_1989}.

On the other hand, if hyperscaling is valid, the fractal dimension can also be estimated from $D = \beta / \nu + \gamma / \nu$~\cite{stauffer_introduction_1994}. Combining our estimates from Eqs.~\eqref{eq:beta_soft_estimation} and \eqref{eq:beta_hard_estimation} with \eqref{eq:gamma_soft_estimation} and \eqref{eq:gamma_hard_estimation}, we arrive at
\begin{equation}
\label{eq:other_fractal_dimension_estimates}
    D^{\text{(s)}} =  1.909(5); \;\; D^{\text{(h)}} =  1.883(4),
\end{equation}
which is consistent with the estimates of Eq.~\eqref{eq:fractal_dimension_estimates}, thus illustrating that hyperscaling is valid for both cluster types.

\section{Discussion} 
\label{sec:conclusions}

We have studied the percolation properties of clusters defined in the overlap space of two statistically independent systems of the square-lattice ferromagnetic Ising model. To this end, two distinct cluster types were introduced which we dubbed soft- and hard-constraint clusters. After a short exposition of the behavior of the main observables, i.e., the wrapping probabilities, average cluster sizes, and percolation strengths, the critical behavior
of the system was investigated. Our results indicate that both cluster types are described by the same correlation length exponent which is found to be in agreement with the value $\nu = 1$ of the Ising ferromagnet. Additionally, both cluster types percolate at a temperature that is indistinguishable from the transition point of the two-dimensional Ising model for the lattice sizes up to $L=2048$ that we considered in our study. In marked contrast with the exponent $\nu$, our analysis for the exponent ratios $\beta/\nu$ and $\gamma/\nu$ manifests the following: (i) the exponent values are clearly different from those that characterize the geometrical clusters of the Ising model, and (ii) there is a small but seemingly systematic difference in the exponents estimated for the soft-constraint and the hard-constraint clusters, cf.\ the overview of exponent estimates provided in Table~\ref{table:2-replica_critical_exponents}.

{
\setlength\arrayrulewidth{1pt}
\renewcommand{\arraystretch}{1.5}
\begin{table}[]\centering
	\caption{\label{table:2-replica_critical_exponents} Critical exponents of the soft- and hard-constraint clusters in comparison to the exact values of the geometrical clusters of the square-latice Ising model model.}
\begin{tabular}{|c|l|l||l|} 
	\cline{2-4} 
	\multicolumn{1}{c|}{}& \multicolumn{2}{c||}{overlap}  & \multicolumn{1}{c|}{Ising} \\
	\hline
	\diagbox{Exponent}{Constraint} & soft & hard & \multicolumn{1}{c|}{$ - $}  \\
	\hline\hline
  	$\nu$ & 1.005(5) & 1.00(3) & 1 \\
	$\beta / \nu$ & 0.0950(7) & 0.1184(11) & $ 5/96 \approx 0.052$  \\
	$\gamma / \nu$ & 1.814(5) & 1.765(4) &  $ 91/48 \approx 1.895 $   \\
	$ D=\gamma / \nu + \beta /\nu $ & 1.909(5) & 1.883(4) & $ 187/96 \approx 1.947 $  \\
	$ D=d- \beta /\nu $ & 1.9050(7) & 1.8816(11) & $ 187/96 $\\
	\hline
\end{tabular}
\end{table}	
}

Under the assumption that $T_\mathrm{p} = T_\mathrm{c}$ that is so well corroborated by our numerical results, the scaling exponents for the overlap clusters can be deduced from the following argument: according to the expected scaling of $P_\infty$, at $T_\mathrm{c}$ the probability of a randomly picked site to be in the percolating cluster of replica one scales as $\sim L^{-(\beta/\nu)_\mathrm{IG}}$, where $(\beta/\nu)_\mathrm{IG} = 5/96$ corresponds to the exponent of the geometrical clusters in the Ising model. As the same holds for replica two and the two replicas are uncorrelated, the probability of the site being in the percolating cluster of \emph{both} replicas decays as $\sim L^{-2(\beta/\nu)_\mathrm{IG}}$, such that $\beta/\nu = 2(\beta/\nu)_\mathrm{IG} = 10/96 \approx 0.1042$. As a consequence, one expects $\gamma/\nu = 2-4(\beta/\nu)_\mathrm{IG} = 43/24 \approx 1.792$ and $D = 2-2(\beta/\nu)_\mathrm{IG} = 91/48 \approx 1.8958$. As a glance at Table~\ref{table:2-replica_critical_exponents} shows, these values are quite compatible with our findings in particular for the hard-constraint clusters to which this type of argument applies. The small deviations observed might be a consequence of the slow decay of scaling corrections expressed in the small value $\omega \approx 0.2$ of the Wegner exponent. The clusters in the soft-constraint problem are, by construction, larger than those of the hard-constraint variant, but their scaling does not directly follow from the above argument, so it is possible that they show asymptotically different exponents.

At the same time, the density of the percolating cluster in the overlap is below that of the percolating spin cluster in a single Ising model, while the clusters of the former are also found to be more compact than those in the latter. This is a natural consequence of the superimposition of the two fractal structures such that the objects investigated here correspond to their intersection. It would be most intriguing to study how clusters in the mutual overlap of more than two copies of the system behave --- a task which is left for future work, however.

\begin{acknowledgments}
		
%N.G.~F. acknowledges support through the Visiting Scholar Program of Chemnitz University of Technology.
		
The authors thank L. M\"unster for useful discussions. We acknowledge the provision of computing time on the parallel computer cluster \emph{Zeus} of Coventry University. N.G. Fytas acknowledges support through the Visiting Scholar Program of Chemnitz University
of Technology and is grateful to the colleagues at the Institute of Physics for their warm hospitality, which he enjoyed while part of this work was completed. 
		
\end{acknowledgments}

%\nocite{*}
	
\bibliography{references}% Produces the bibliography via BibTeX.

\begin{thebibliography}{44}%
\makeatletter
\providecommand \@ifxundefined [1]{%
 \@ifx{#1\undefined}
}%
\providecommand \@ifnum [1]{%
 \ifnum #1\expandafter \@firstoftwo
 \else \expandafter \@secondoftwo
 \fi
}%
\providecommand \@ifx [1]{%
 \ifx #1\expandafter \@firstoftwo
 \else \expandafter \@secondoftwo
 \fi
}%
\providecommand \natexlab [1]{#1}%
\providecommand \enquote  [1]{``#1''}%
\providecommand \bibnamefont  [1]{#1}%
\providecommand \bibfnamefont [1]{#1}%
\providecommand \citenamefont [1]{#1}%
\providecommand \href@noop [0]{\@secondoftwo}%
\providecommand \href [0]{\begingroup \@sanitize@url \@href}%
\providecommand \@href[1]{\@@startlink{#1}\@@href}%
\providecommand \@@href[1]{\endgroup#1\@@endlink}%
\providecommand \@sanitize@url [0]{\catcode `\\12\catcode `\$12\catcode
  `\&12\catcode `\#12\catcode `\^12\catcode `\_12\catcode `\%12\relax}%
\providecommand \@@startlink[1]{}%
\providecommand \@@endlink[0]{}%
\providecommand \url  [0]{\begingroup\@sanitize@url \@url }%
\providecommand \@url [1]{\endgroup\@href {#1}{\urlprefix }}%
\providecommand \urlprefix  [0]{URL }%
\providecommand \Eprint [0]{\href }%
\providecommand \doibase [0]{https://doi.org/}%
\providecommand \selectlanguage [0]{\@gobble}%
\providecommand \bibinfo  [0]{\@secondoftwo}%
\providecommand \bibfield  [0]{\@secondoftwo}%
\providecommand \translation [1]{[#1]}%
\providecommand \BibitemOpen [0]{}%
\providecommand \bibitemStop [0]{}%
\providecommand \bibitemNoStop [0]{.\EOS\space}%
\providecommand \EOS [0]{\spacefactor3000\relax}%
\providecommand \BibitemShut  [1]{\csname bibitem#1\endcsname}%
\let\auto@bib@innerbib\@empty
%</preamble>
\bibitem [{\citenamefont {Janke}(1996)}]{wj:chem}%
  \BibitemOpen
  \bibfield  {author} {\bibinfo {author} {\bibfnamefont {W.}~\bibnamefont
  {Janke}},\ }\bibfield  {title} {\bibinfo {title} {{M}onte {C}arlo simulations
  of spin systems},\ }in\ \href@noop {} {\emph {\bibinfo {booktitle}
  {Computational Physics}}},\ \bibinfo {editor} {edited by\ \bibinfo {editor}
  {\bibfnamefont {K.~H.}\ \bibnamefont {Hoffmann}}\ and\ \bibinfo {editor}
  {\bibfnamefont {M.}~\bibnamefont {Schreiber}}}\ (\bibinfo  {publisher}
  {Springer},\ \bibinfo {address} {Berlin},\ \bibinfo {year} {1996})\ pp.\
  \bibinfo {pages} {10--43}\BibitemShut {NoStop}%
\bibitem [{\citenamefont {Wegner}(1971)}]{wegner:71}%
  \BibitemOpen
  \bibfield  {author} {\bibinfo {author} {\bibfnamefont {F.~J.}\ \bibnamefont
  {Wegner}},\ }\bibfield  {title} {\bibinfo {title} {Duality in generalized
  {I}sing models and phase transitions without local order parameters},\ }\href
  {https://pubs.aip.org/aip/jmp/article-abstract/12/10/2259/465334/Duality-in-Generalized-Ising-Models-and-Phase?redirectedFrom=fulltext}
  {\bibfield  {journal} {\bibinfo  {journal} {J. Math. Phys.}\ }\textbf
  {\bibinfo {volume} {12}},\ \bibinfo {pages} {2259} (\bibinfo {year}
  {1971})}\BibitemShut {NoStop}%
\bibitem [{\citenamefont {Kitaev}(2003)}]{kitaev:03}%
  \BibitemOpen
  \bibfield  {author} {\bibinfo {author} {\bibfnamefont {A.~Y.}\ \bibnamefont
  {Kitaev}},\ }\bibfield  {title} {\bibinfo {title} {Fault-tolerant quantum
  computation by anyons},\ }\href
  {https://www.sciencedirect.com/science/article/pii/S0003491602000180?casa_token=YVE8Lp60R_MAAAAA:ovZCW1Zee_HjLUFlE4AYfjqaxpw5jD9ocp2WCr3-RdZiT3igMvrdyW7yprcQ4j2QpVLM-AhZ}
  {\bibfield  {journal} {\bibinfo  {journal} {Ann. Phys.}\ }\textbf {\bibinfo
  {volume} {303}},\ \bibinfo {pages} {2} (\bibinfo {year} {2003})}\BibitemShut
  {NoStop}%
\bibitem [{\citenamefont {Castelnovo}\ and\ \citenamefont
  {Chamon}(2007)}]{castelnovo:07}%
  \BibitemOpen
  \bibfield  {author} {\bibinfo {author} {\bibfnamefont {C.}~\bibnamefont
  {Castelnovo}}\ and\ \bibinfo {author} {\bibfnamefont {C.}~\bibnamefont
  {Chamon}},\ }\bibfield  {title} {\bibinfo {title} {Topological order and
  topological entropy in classical systems},\ }\href
  {https://journals.aps.org/prb/abstract/10.1103/PhysRevB.76.174416} {\bibfield
   {journal} {\bibinfo  {journal} {Phys. Rev. B}\ }\textbf {\bibinfo {volume}
  {76}},\ \bibinfo {pages} {174416} (\bibinfo {year} {2007})}\BibitemShut
  {NoStop}%
\bibitem [{\citenamefont {Johnston}(2012)}]{johnston:12}%
  \BibitemOpen
  \bibfield  {author} {\bibinfo {author} {\bibfnamefont {D.~A.}\ \bibnamefont
  {Johnston}},\ }\bibfield  {title} {\bibinfo {title} {Gonihedric (and
  fuki-nuke) order},\ }\href
  {https://iopscience.iop.org/article/10.1088/1751-8113/45/40/405001/meta}
  {\bibfield  {journal} {\bibinfo  {journal} {J. Phys. A}\ }\textbf {\bibinfo
  {volume} {45}},\ \bibinfo {pages} {405001} (\bibinfo {year}
  {2012})}\BibitemShut {NoStop}%
\bibitem [{\citenamefont {Binder}\ and\ \citenamefont
  {Young}(1986)}]{binder:86a}%
  \BibitemOpen
  \bibfield  {author} {\bibinfo {author} {\bibfnamefont {K.}~\bibnamefont
  {Binder}}\ and\ \bibinfo {author} {\bibfnamefont {A.~P.}\ \bibnamefont
  {Young}},\ }\bibfield  {title} {\bibinfo {title} {Spin glasses: Experimental
  facts, theoretical concepts, and open questions},\ }\href
  {https://doi.org/10.1103/RevModPhys.58.801} {\bibfield  {journal} {\bibinfo
  {journal} {Rev. Mod. Phys.}\ }\textbf {\bibinfo {volume} {58}},\ \bibinfo
  {pages} {801} (\bibinfo {year} {1986})}\BibitemShut {NoStop}%
\bibitem [{\citenamefont {Parisi}(1983)}]{parisi:83}%
  \BibitemOpen
  \bibfield  {author} {\bibinfo {author} {\bibfnamefont {G.}~\bibnamefont
  {Parisi}},\ }\bibfield  {title} {\bibinfo {title} {Order parameter for
  spin-glasses},\ }\href {https://doi.org/10.1103/physrevlett.50.1946}
  {\bibfield  {journal} {\bibinfo  {journal} {Phys. Rev. Lett.}\ }\textbf
  {\bibinfo {volume} {50}},\ \bibinfo {pages} {1946} (\bibinfo {year}
  {1983})}\BibitemShut {NoStop}%
\bibitem [{\citenamefont {M{\'{e}}zard}\ \emph {et~al.}(1987)\citenamefont
  {M{\'{e}}zard}, \citenamefont {Parisi},\ and\ \citenamefont
  {Virasoro}}]{mezard:book}%
  \BibitemOpen
  \bibfield  {author} {\bibinfo {author} {\bibfnamefont {M.}~\bibnamefont
  {M{\'{e}}zard}}, \bibinfo {author} {\bibfnamefont {G.}~\bibnamefont
  {Parisi}},\ and\ \bibinfo {author} {\bibfnamefont {M.~A.}\ \bibnamefont
  {Virasoro}},\ }\href@noop {} {\emph {\bibinfo {title} {Spin Glass Theory and
  Beyond}}}\ (\bibinfo  {publisher} {World Scientific},\ \bibinfo {address}
  {Singapore},\ \bibinfo {year} {1987})\BibitemShut {NoStop}%
\bibitem [{\citenamefont {Boccaletti}\ \emph {et~al.}(2014)\citenamefont
  {Boccaletti}, \citenamefont {Bianconi}, \citenamefont {Criado}, \citenamefont
  {{del Genio}}, \citenamefont {Gómez-Gardeñes}, \citenamefont {Romance},
  \citenamefont {Sendiña-Nadal}, \citenamefont {Wang},\ and\ \citenamefont
  {Zanin}}]{boccaletti:14}%
  \BibitemOpen
  \bibfield  {author} {\bibinfo {author} {\bibfnamefont {S.}~\bibnamefont
  {Boccaletti}}, \bibinfo {author} {\bibfnamefont {G.}~\bibnamefont
  {Bianconi}}, \bibinfo {author} {\bibfnamefont {R.}~\bibnamefont {Criado}},
  \bibinfo {author} {\bibfnamefont {C.}~\bibnamefont {{del Genio}}}, \bibinfo
  {author} {\bibfnamefont {J.}~\bibnamefont {Gómez-Gardeñes}}, \bibinfo
  {author} {\bibfnamefont {M.}~\bibnamefont {Romance}}, \bibinfo {author}
  {\bibfnamefont {I.}~\bibnamefont {Sendiña-Nadal}}, \bibinfo {author}
  {\bibfnamefont {Z.}~\bibnamefont {Wang}},\ and\ \bibinfo {author}
  {\bibfnamefont {M.}~\bibnamefont {Zanin}},\ }\bibfield  {title} {\bibinfo
  {title} {The structure and dynamics of multilayer networks},\ }\href
  {https://doi.org/https://doi.org/10.1016/j.physrep.2014.07.001} {\bibfield
  {journal} {\bibinfo  {journal} {Phys. Rep.}\ }\textbf {\bibinfo {volume}
  {544}},\ \bibinfo {pages} {1} (\bibinfo {year} {2014})}\BibitemShut {NoStop}%
\bibitem [{\citenamefont {Stauffer}\ and\ \citenamefont
  {Aharony}(1994)}]{stauffer_introduction_1994}%
  \BibitemOpen
  \bibfield  {author} {\bibinfo {author} {\bibfnamefont {D.}~\bibnamefont
  {Stauffer}}\ and\ \bibinfo {author} {\bibfnamefont {A.}~\bibnamefont
  {Aharony}},\ }\href@noop {} {\emph {\bibinfo {title} {Introduction to
  percolation theory}}},\ \bibinfo {edition} {rev., 2nd}\ ed.\ (\bibinfo
  {publisher} {Taylor \& Francis},\ \bibinfo {address} {London ; Bristol, PA},\
  \bibinfo {year} {1994})\BibitemShut {NoStop}%
\bibitem [{\citenamefont {Fortuin}\ and\ \citenamefont
  {Kasteleyn}(1972)}]{fortuin_random-cluster_1972}%
  \BibitemOpen
  \bibfield  {author} {\bibinfo {author} {\bibfnamefont {C.~M.}\ \bibnamefont
  {Fortuin}}\ and\ \bibinfo {author} {\bibfnamefont {P.~W.}\ \bibnamefont
  {Kasteleyn}},\ }\bibfield  {title} {\bibinfo {title} {On the random-cluster
  model: {I}. {Introduction} and relation to other models},\ }\href
  {https://doi.org/10.1016/0031-8914(72)90045-6} {\bibfield  {journal}
  {\bibinfo  {journal} {Physica}\ }\textbf {\bibinfo {volume} {57}},\ \bibinfo
  {pages} {536} (\bibinfo {year} {1972})}\BibitemShut {NoStop}%
\bibitem [{\citenamefont {Coniglio}\ and\ \citenamefont
  {Klein}(1980)}]{coniglio:80a}%
  \BibitemOpen
  \bibfield  {author} {\bibinfo {author} {\bibfnamefont {A.}~\bibnamefont
  {Coniglio}}\ and\ \bibinfo {author} {\bibfnamefont {W.}~\bibnamefont
  {Klein}},\ }\bibfield  {title} {\bibinfo {title} {Clusters and {Ising}
  critical droplets: a renormalisation group approach},\ }\href
  {https://doi.org/10.1088/0305-4470/13/8/025} {\bibfield  {journal} {\bibinfo
  {journal} {J. Phys. A}\ }\textbf {\bibinfo {volume} {13}},\ \bibinfo {pages}
  {2775} (\bibinfo {year} {1980})}\BibitemShut {NoStop}%
\bibitem [{\citenamefont {Swendsen}\ and\ \citenamefont
  {Wang}(1987)}]{swendsen_nonuniversal_1987}%
  \BibitemOpen
  \bibfield  {author} {\bibinfo {author} {\bibfnamefont {R.~H.}\ \bibnamefont
  {Swendsen}}\ and\ \bibinfo {author} {\bibfnamefont {J.-S.}\ \bibnamefont
  {Wang}},\ }\bibfield  {title} {\bibinfo {title} {Nonuniversal critical
  dynamics in {Monte} {Carlo} simulations},\ }\href
  {https://doi.org/10.1103/PhysRevLett.58.86} {\bibfield  {journal} {\bibinfo
  {journal} {Phys. Rev. Lett.}\ }\textbf {\bibinfo {volume} {58}},\ \bibinfo
  {pages} {86} (\bibinfo {year} {1987})}\BibitemShut {NoStop}%
\bibitem [{\citenamefont {Wolff}(1989)}]{wolff_collective_1989}%
  \BibitemOpen
  \bibfield  {author} {\bibinfo {author} {\bibfnamefont {U.}~\bibnamefont
  {Wolff}},\ }\bibfield  {title} {\bibinfo {title} {Collective {M}onte {C}arlo
  updating for spin systems},\ }\href
  {https://doi.org/10.1103/PhysRevLett.62.361} {\bibfield  {journal} {\bibinfo
  {journal} {Phys. Rev. Lett.}\ }\textbf {\bibinfo {volume} {62}},\ \bibinfo
  {pages} {361} (\bibinfo {year} {1989})}\BibitemShut {NoStop}%
\bibitem [{\citenamefont {Metropolis}\ \emph {et~al.}(1953)\citenamefont
  {Metropolis}, \citenamefont {Rosenbluth}, \citenamefont {Rosenbluth},
  \citenamefont {Teller},\ and\ \citenamefont
  {Teller}}]{metropolis_equation_1953}%
  \BibitemOpen
  \bibfield  {author} {\bibinfo {author} {\bibfnamefont {N.}~\bibnamefont
  {Metropolis}}, \bibinfo {author} {\bibfnamefont {A.~W.}\ \bibnamefont
  {Rosenbluth}}, \bibinfo {author} {\bibfnamefont {M.~N.}\ \bibnamefont
  {Rosenbluth}}, \bibinfo {author} {\bibfnamefont {A.~H.}\ \bibnamefont
  {Teller}},\ and\ \bibinfo {author} {\bibfnamefont {E.}~\bibnamefont
  {Teller}},\ }\bibfield  {title} {\bibinfo {title} {Equation of state
  calculations by fast computing machines},\ }\href
  {https://doi.org/10.1063/1.1699114} {\bibfield  {journal} {\bibinfo
  {journal} {J. Chem. Phys.}\ }\textbf {\bibinfo {volume} {21}},\ \bibinfo
  {pages} {1087} (\bibinfo {year} {1953})}\BibitemShut {NoStop}%
\bibitem [{\citenamefont {Binder}(1976)}]{binder:76b}%
  \BibitemOpen
  \bibfield  {author} {\bibinfo {author} {\bibfnamefont {K.}~\bibnamefont
  {Binder}},\ }\bibfield  {title} {\bibinfo {title} {``{C}lusters'' in the
  {I}sing model, metastable states and essential singularity},\ }\href
  {https://www.sciencedirect.com/science/article/abs/pii/0003491676901597}
  {\bibfield  {journal} {\bibinfo  {journal} {Ann. Phys.}\ }\textbf {\bibinfo
  {volume} {98}},\ \bibinfo {pages} {390} (\bibinfo {year} {1976})}\BibitemShut
  {NoStop}%
\bibitem [{\citenamefont {Coniglio}\ and\ \citenamefont
  {Fierro}(2009)}]{coniglio_correlated_2009}%
  \BibitemOpen
  \bibfield  {author} {\bibinfo {author} {\bibfnamefont {A.}~\bibnamefont
  {Coniglio}}\ and\ \bibinfo {author} {\bibfnamefont {A.}~\bibnamefont
  {Fierro}},\ }\bibfield  {title} {\bibinfo {title} {Correlated
  {Percolation}},\ }in\ \href {https://doi.org/10.1007/978-0-387-30440-3_104}
  {\emph {\bibinfo {booktitle} {Encyclopedia of {Complexity} and {Systems}
  {Science}}}},\ \bibinfo {editor} {edited by\ \bibinfo {editor} {\bibfnamefont
  {R.~A.}\ \bibnamefont {Meyers}}}\ (\bibinfo  {publisher} {Springer},\
  \bibinfo {address} {New York},\ \bibinfo {year} {2009})\ pp.\ \bibinfo
  {pages} {1596--1615}\BibitemShut {NoStop}%
\bibitem [{\citenamefont {Saberi}(2015)}]{saberi_recent_2015}%
  \BibitemOpen
  \bibfield  {author} {\bibinfo {author} {\bibfnamefont {A.~A.}\ \bibnamefont
  {Saberi}},\ }\bibfield  {title} {\bibinfo {title} {Recent advances in
  percolation theory and its applications},\ }\href
  {https://doi.org/10.1016/j.physrep.2015.03.003} {\bibfield  {journal}
  {\bibinfo  {journal} {Phys. Rep.}\ }\textbf {\bibinfo {volume} {578}},\
  \bibinfo {pages} {1} (\bibinfo {year} {2015})}\BibitemShut {NoStop}%
\bibitem [{\citenamefont {Coniglio}\ \emph {et~al.}(1977)\citenamefont
  {Coniglio}, \citenamefont {Nappi}, \citenamefont {Peruggi},\ and\
  \citenamefont {Russo}}]{coniglio:77}%
  \BibitemOpen
  \bibfield  {author} {\bibinfo {author} {\bibfnamefont {A.}~\bibnamefont
  {Coniglio}}, \bibinfo {author} {\bibfnamefont {C.~R.}\ \bibnamefont {Nappi}},
  \bibinfo {author} {\bibfnamefont {F.}~\bibnamefont {Peruggi}},\ and\ \bibinfo
  {author} {\bibfnamefont {L.}~\bibnamefont {Russo}},\ }\bibfield  {title}
  {\bibinfo {title} {Percolation points and critical point in the {I}sing
  model},\ }\href
  {https://iopscience.iop.org/article/10.1088/0305-4470/10/2/010/meta?casa_token=ob92iCtNanYAAAAA:NpEELiJFMEsNz_AkKnNb4sy3TA7bx3uhtdPp1aaGtvqAVXzNxTnu728cxRLKfwC8r_1fYpuK}
  {\bibfield  {journal} {\bibinfo  {journal} {J. Phys. A}\ }\textbf {\bibinfo
  {volume} {10}},\ \bibinfo {pages} {205} (\bibinfo {year} {1977})}\BibitemShut
  {NoStop}%
\bibitem [{\citenamefont {Stella}\ and\ \citenamefont
  {Vanderzande}(1989)}]{stella_scaling_1989}%
  \BibitemOpen
  \bibfield  {author} {\bibinfo {author} {\bibfnamefont {A.~L.}\ \bibnamefont
  {Stella}}\ and\ \bibinfo {author} {\bibfnamefont {C.}~\bibnamefont
  {Vanderzande}},\ }\bibfield  {title} {\bibinfo {title} {Scaling and fractal
  dimension of {Ising} clusters at the \(d=2\) critical point},\ }\href
  {https://doi.org/10.1103/PhysRevLett.62.1067} {\bibfield  {journal} {\bibinfo
   {journal} {Phys. Rev. Lett.}\ }\textbf {\bibinfo {volume} {62}},\ \bibinfo
  {pages} {1067} (\bibinfo {year} {1989})}\BibitemShut {NoStop}%
\bibitem [{\citenamefont {Vanderzande}\ and\ \citenamefont
  {Stella}(1989)}]{vanderzande_bulk_1989}%
  \BibitemOpen
  \bibfield  {author} {\bibinfo {author} {\bibfnamefont {C.}~\bibnamefont
  {Vanderzande}}\ and\ \bibinfo {author} {\bibfnamefont {A.~L.}\ \bibnamefont
  {Stella}},\ }\bibfield  {title} {\bibinfo {title} {Bulk, surface and hull
  fractal dimension of critical {Ising} clusters in \(d=2\)},\ }\href
  {https://doi.org/10.1088/0305-4470/22/10/005} {\bibfield  {journal} {\bibinfo
   {journal} {J. Phys. A}\ }\textbf {\bibinfo {volume} {22}},\ \bibinfo {pages}
  {L445} (\bibinfo {year} {1989})}\BibitemShut {NoStop}%
\bibitem [{\citenamefont {Vanderzande}(1992)}]{vanderzande_fractal_1992}%
  \BibitemOpen
  \bibfield  {author} {\bibinfo {author} {\bibfnamefont {C.}~\bibnamefont
  {Vanderzande}},\ }\bibfield  {title} {\bibinfo {title} {Fractal dimensions of
  {Potts} clusters},\ }\href
  {https://iopscience.iop.org/article/10.1088/0305-4470/25/2/008/meta?casa_token=JcUqDP0uSnIAAAAA:ZkBeSqctg_bGk3GO8wd7xP5xSBStz8HsuQYKaraFVTqxyMYoiSQPffOs4XN-FZLpUOnFdRlb}
  {\bibfield  {journal} {\bibinfo  {journal} {J. Phys. A}\ }\textbf {\bibinfo
  {volume} {25}},\ \bibinfo {pages} {L75} (\bibinfo {year} {1992})}\BibitemShut
  {NoStop}%
\bibitem [{\citenamefont {Janke}\ and\ \citenamefont
  {Schakel}(2004)}]{janke_geometrical_2004}%
  \BibitemOpen
  \bibfield  {author} {\bibinfo {author} {\bibfnamefont {W.}~\bibnamefont
  {Janke}}\ and\ \bibinfo {author} {\bibfnamefont {A.~M.}\ \bibnamefont
  {Schakel}},\ }\bibfield  {title} {\bibinfo {title} {Geometrical vs.
  {Fortuin}–{Kasteleyn} clusters in the two-dimensional \(q\)-state {Potts}
  model},\ }\href {https://doi.org/10.1016/j.nuclphysb.2004.08.030} {\bibfield
  {journal} {\bibinfo  {journal} {Nucl. Phys. B}\ }\textbf {\bibinfo {volume}
  {700}},\ \bibinfo {pages} {385} (\bibinfo {year} {2004})}\BibitemShut
  {NoStop}%
\bibitem [{\citenamefont {Janke}\ and\ \citenamefont
  {Schakel}(2005)}]{janke_fractal_2005}%
  \BibitemOpen
  \bibfield  {author} {\bibinfo {author} {\bibfnamefont {W.}~\bibnamefont
  {Janke}}\ and\ \bibinfo {author} {\bibfnamefont {A.~M.~J.}\ \bibnamefont
  {Schakel}},\ }\bibfield  {title} {\bibinfo {title} {Fractal structure of spin
  clusters and domain walls in the two-dimensional {Ising} model},\ }\href
  {https://doi.org/10.1103/PhysRevE.71.036703} {\bibfield  {journal} {\bibinfo
  {journal} {Phys. Rev. E}\ }\textbf {\bibinfo {volume} {71}},\ \bibinfo
  {pages} {036703} (\bibinfo {year} {2005})}\BibitemShut {NoStop}%
\bibitem [{\citenamefont {De~Arcangelis}\ \emph {et~al.}(1991)\citenamefont
  {De~Arcangelis}, \citenamefont {Coniglio},\ and\ \citenamefont
  {Peruggi}}]{arcangelis:91}%
  \BibitemOpen
  \bibfield  {author} {\bibinfo {author} {\bibfnamefont {L.}~\bibnamefont
  {De~Arcangelis}}, \bibinfo {author} {\bibfnamefont {A.}~\bibnamefont
  {Coniglio}},\ and\ \bibinfo {author} {\bibfnamefont {F.}~\bibnamefont
  {Peruggi}},\ }\bibfield  {title} {\bibinfo {title} {Percolation transition in
  spin glasses},\ }\href
  {https://iopscience.iop.org/article/10.1209/0295-5075/14/6/003/meta}
  {\bibfield  {journal} {\bibinfo  {journal} {EPL}\ }\textbf {\bibinfo {volume}
  {14}},\ \bibinfo {pages} {515} (\bibinfo {year} {1991})}\BibitemShut
  {NoStop}%
\bibitem [{\citenamefont {Machta}\ \emph {et~al.}(2008)\citenamefont {Machta},
  \citenamefont {Newman},\ and\ \citenamefont
  {Stein}}]{machta_percolation_2008}%
  \BibitemOpen
  \bibfield  {author} {\bibinfo {author} {\bibfnamefont {J.}~\bibnamefont
  {Machta}}, \bibinfo {author} {\bibfnamefont {C.~M.}\ \bibnamefont {Newman}},\
  and\ \bibinfo {author} {\bibfnamefont {D.~L.}\ \bibnamefont {Stein}},\
  }\bibfield  {title} {\bibinfo {title} {The percolation signature of the spin
  glass transition},\ }\href {https://doi.org/10.1007/s10955-007-9446-2}
  {\bibfield  {journal} {\bibinfo  {journal} {J. Stat. Phys.}\ }\textbf
  {\bibinfo {volume} {130}},\ \bibinfo {pages} {113} (\bibinfo {year}
  {2008})}\BibitemShut {NoStop}%
\bibitem [{\citenamefont {Fajen}\ \emph {et~al.}(2020)\citenamefont {Fajen},
  \citenamefont {Hartmann},\ and\ \citenamefont
  {Young}}]{fajen_percolation_2020}%
  \BibitemOpen
  \bibfield  {author} {\bibinfo {author} {\bibfnamefont {H.}~\bibnamefont
  {Fajen}}, \bibinfo {author} {\bibfnamefont {A.~K.}\ \bibnamefont
  {Hartmann}},\ and\ \bibinfo {author} {\bibfnamefont {A.~P.}\ \bibnamefont
  {Young}},\ }\bibfield  {title} {\bibinfo {title} {Percolation of
  {F}ortuin-{K}asteleyn clusters for the random-bond {I}sing model},\ }\href
  {https://link.aps.org/doi/10.1103/PhysRevE.102.012131} {\bibfield  {journal}
  {\bibinfo  {journal} {Phys. Rev. E}\ }\textbf {\bibinfo {volume} {102}}
  (\bibinfo {year} {2020})}\BibitemShut {NoStop}%
\bibitem [{\citenamefont {Houdayer}(2001)}]{houdayer_cluster_2001}%
  \BibitemOpen
  \bibfield  {author} {\bibinfo {author} {\bibfnamefont {J.}~\bibnamefont
  {Houdayer}},\ }\bibfield  {title} {\bibinfo {title} {A cluster {M}onte
  {C}arlo algorithm for 2-dimensional spin glasses},\ }\href
  {https://doi.org/10.1007/PL00011151} {\bibfield  {journal} {\bibinfo
  {journal} {Eur. Phys. J. B}\ }\textbf {\bibinfo {volume} {22}},\ \bibinfo
  {pages} {479} (\bibinfo {year} {2001})}\BibitemShut {NoStop}%
\bibitem [{\citenamefont {J{\"o}rg}(2005)}]{joerg:05}%
  \BibitemOpen
  \bibfield  {author} {\bibinfo {author} {\bibfnamefont {T.}~\bibnamefont
  {J{\"o}rg}},\ }\bibfield  {title} {\bibinfo {title} {Cluster {M}onte {C}arlo
  algorithms for diluted spin glasses},\ }\href
  {https://academic.oup.com/ptps/article/doi/10.1143/PTPS.157.349/1860069}
  {\bibfield  {journal} {\bibinfo  {journal} {Prog. Theor. Phys. Supp.}\
  }\textbf {\bibinfo {volume} {157}},\ \bibinfo {pages} {349} (\bibinfo {year}
  {2005})}\BibitemShut {NoStop}%
\bibitem [{\citenamefont {M\"unster}\ and\ \citenamefont
  {Weigel}(2023)}]{munster_cluster_2023}%
  \BibitemOpen
  \bibfield  {author} {\bibinfo {author} {\bibfnamefont {L.}~\bibnamefont
  {M\"unster}}\ and\ \bibinfo {author} {\bibfnamefont {M.}~\bibnamefont
  {Weigel}},\ }\bibfield  {title} {\bibinfo {title} {Cluster percolation in the
  two-dimensional {I}sing spin glass},\ }\href
  {https://doi.org/10.1103/PhysRevE.107.054103} {\bibfield  {journal} {\bibinfo
   {journal} {Phys. Rev. E}\ }\textbf {\bibinfo {volume} {107}},\ \bibinfo
  {pages} {054103} (\bibinfo {year} {2023})}\BibitemShut {NoStop}%
\bibitem [{\citenamefont {Haake}\ \emph {et~al.}(1985)\citenamefont {Haake},
  \citenamefont {Lewenstein},\ and\ \citenamefont {Wilkens}}]{haake:85}%
  \BibitemOpen
  \bibfield  {author} {\bibinfo {author} {\bibfnamefont {F.}~\bibnamefont
  {Haake}}, \bibinfo {author} {\bibfnamefont {M.}~\bibnamefont {Lewenstein}},\
  and\ \bibinfo {author} {\bibfnamefont {M.}~\bibnamefont {Wilkens}},\
  }\bibfield  {title} {\bibinfo {title} {Relation of random and competing
  nonrandom couplings for spin-glasses},\ }\href
  {https://journals.aps.org/prl/abstract/10.1103/PhysRevLett.55.2606}
  {\bibfield  {journal} {\bibinfo  {journal} {Phys. Rev. Lett.}\ }\textbf
  {\bibinfo {volume} {55}},\ \bibinfo {pages} {2606} (\bibinfo {year}
  {1985})}\BibitemShut {NoStop}%
\bibitem [{\citenamefont {Chayes}\ \emph {et~al.}(1998)\citenamefont {Chayes},
  \citenamefont {Machta},\ and\ \citenamefont {Redner}}]{chayes:98}%
  \BibitemOpen
  \bibfield  {author} {\bibinfo {author} {\bibfnamefont {L.}~\bibnamefont
  {Chayes}}, \bibinfo {author} {\bibfnamefont {J.}~\bibnamefont {Machta}},\
  and\ \bibinfo {author} {\bibfnamefont {O.}~\bibnamefont {Redner}},\
  }\bibfield  {title} {\bibinfo {title} {Graphical representations for {I}sing
  systems in external fields},\ }\href
  {https://link.springer.com/article/10.1023/B:JOSS.0000026726.43558.80}
  {\bibfield  {journal} {\bibinfo  {journal} {J. Stat. Phys.}\ }\textbf
  {\bibinfo {volume} {93}},\ \bibinfo {pages} {17} (\bibinfo {year}
  {1998})}\BibitemShut {NoStop}%
\bibitem [{\citenamefont {Weigel}\ and\ \citenamefont
  {Janke}(2010)}]{weigel_error_2010}%
  \BibitemOpen
  \bibfield  {author} {\bibinfo {author} {\bibfnamefont {M.}~\bibnamefont
  {Weigel}}\ and\ \bibinfo {author} {\bibfnamefont {W.}~\bibnamefont {Janke}},\
  }\bibfield  {title} {\bibinfo {title} {Error estimation and reduction with
  cross correlations},\ }\href {https://doi.org/10.1103/PhysRevE.81.066701}
  {\bibfield  {journal} {\bibinfo  {journal} {Phys. Rev. E}\ }\textbf {\bibinfo
  {volume} {81}},\ \bibinfo {pages} {066701} (\bibinfo {year}
  {2010})}\BibitemShut {NoStop}%
\bibitem [{\citenamefont {Machta}\ \emph {et~al.}(1996)\citenamefont {Machta},
  \citenamefont {Choi}, \citenamefont {Lucke}, \citenamefont {Schweizer},\ and\
  \citenamefont {Chayes}}]{machta_invaded_1996}%
  \BibitemOpen
  \bibfield  {author} {\bibinfo {author} {\bibfnamefont {J.}~\bibnamefont
  {Machta}}, \bibinfo {author} {\bibfnamefont {Y.~S.}\ \bibnamefont {Choi}},
  \bibinfo {author} {\bibfnamefont {A.}~\bibnamefont {Lucke}}, \bibinfo
  {author} {\bibfnamefont {T.}~\bibnamefont {Schweizer}},\ and\ \bibinfo
  {author} {\bibfnamefont {L.~M.}\ \bibnamefont {Chayes}},\ }\bibfield  {title}
  {\bibinfo {title} {Invaded cluster algorithm for {Potts} models},\ }\href
  {https://doi.org/10.1103/PhysRevE.54.1332} {\bibfield  {journal} {\bibinfo
  {journal} {Phys. Rev. E}\ }\textbf {\bibinfo {volume} {54}},\ \bibinfo
  {pages} {1332} (\bibinfo {year} {1996})}\BibitemShut {NoStop}%
\bibitem [{\citenamefont {Press}\ \emph {et~al.}(2007)\citenamefont {Press},
  \citenamefont {Teukolsky}, \citenamefont {Vetterling},\ and\ \citenamefont
  {Flannery}}]{numrec}%
  \BibitemOpen
  \bibfield  {author} {\bibinfo {author} {\bibfnamefont {W.~H.}\ \bibnamefont
  {Press}}, \bibinfo {author} {\bibfnamefont {S.~A.}\ \bibnamefont
  {Teukolsky}}, \bibinfo {author} {\bibfnamefont {W.~T.}\ \bibnamefont
  {Vetterling}},\ and\ \bibinfo {author} {\bibfnamefont {B.~P.}\ \bibnamefont
  {Flannery}},\ }\href@noop {} {\emph {\bibinfo {title} {Numerical Recipes: The
  Art of Scientific Computing}}},\ \bibinfo {edition} {3rd}\ ed.\ (\bibinfo
  {publisher} {Cambridge University Press},\ \bibinfo {address} {Cambridge},\
  \bibinfo {year} {2007})\BibitemShut {NoStop}%
\bibitem [{\citenamefont {Newman}\ and\ \citenamefont
  {Ziff}(2001)}]{newman_fast_2001}%
  \BibitemOpen
  \bibfield  {author} {\bibinfo {author} {\bibfnamefont {M.~E.~J.}\
  \bibnamefont {Newman}}\ and\ \bibinfo {author} {\bibfnamefont {R.~M.}\
  \bibnamefont {Ziff}},\ }\bibfield  {title} {\bibinfo {title} {Fast {Monte}
  {Carlo} algorithm for site or bond percolation},\ }\href
  {https://doi.org/10.1103/PhysRevE.64.016706} {\bibfield  {journal} {\bibinfo
  {journal} {Phys. Rev. E}\ }\textbf {\bibinfo {volume} {64}},\ \bibinfo
  {pages} {016706} (\bibinfo {year} {2001})}\BibitemShut {NoStop}%
\bibitem [{\citenamefont {Privman}(1990)}]{privman:privman}%
  \BibitemOpen
  \bibfield  {author} {\bibinfo {author} {\bibfnamefont {V.}~\bibnamefont
  {Privman}},\ }\bibfield  {title} {\bibinfo {title} {{Finite-Size} scaling
  theory},\ }in\ \href@noop {} {\emph {\bibinfo {booktitle} {Finite Size
  Scaling and Numerical Simulation of Statistical Systems}}},\ \bibinfo
  {editor} {edited by\ \bibinfo {editor} {\bibfnamefont {V.}~\bibnamefont
  {Privman}}}\ (\bibinfo  {publisher} {World Scientific},\ \bibinfo {address}
  {Singapore},\ \bibinfo {year} {1990})\ pp.\ \bibinfo {pages}
  {1--98}\BibitemShut {NoStop}%
\bibitem [{\citenamefont {Martins}\ and\ \citenamefont
  {Plascak}(2003)}]{martins_percolation_2003}%
  \BibitemOpen
  \bibfield  {author} {\bibinfo {author} {\bibfnamefont {P.~H.~L.}\
  \bibnamefont {Martins}}\ and\ \bibinfo {author} {\bibfnamefont {J.~A.}\
  \bibnamefont {Plascak}},\ }\bibfield  {title} {\bibinfo {title} {Percolation
  on two- and three-dimensional lattices},\ }\href
  {https://doi.org/10.1103/PhysRevE.67.046119} {\bibfield  {journal} {\bibinfo
  {journal} {Phys. Rev. E}\ }\textbf {\bibinfo {volume} {67}},\ \bibinfo
  {pages} {046119} (\bibinfo {year} {2003})}\BibitemShut {NoStop}%
\bibitem [{\citenamefont {Binder}(1981)}]{binder_finite_1981}%
  \BibitemOpen
  \bibfield  {author} {\bibinfo {author} {\bibfnamefont {K.}~\bibnamefont
  {Binder}},\ }\bibfield  {title} {\bibinfo {title} {Finite size scaling
  analysis of {Ising} model block distribution functions},\ }\href
  {https://link.springer.com/article/10.1007/BF01293604} {\bibfield  {journal}
  {\bibinfo  {journal} {Z. Phys. B}\ }\textbf {\bibinfo {volume} {43}},\
  \bibinfo {pages} {119} (\bibinfo {year} {1981})}\BibitemShut {NoStop}%
\bibitem [{\citenamefont {Ferrenberg}\ and\ \citenamefont
  {Landau}(1991)}]{ferrenberg_critical_1991}%
  \BibitemOpen
  \bibfield  {author} {\bibinfo {author} {\bibfnamefont {A.~M.}\ \bibnamefont
  {Ferrenberg}}\ and\ \bibinfo {author} {\bibfnamefont {D.~P.}\ \bibnamefont
  {Landau}},\ }\bibfield  {title} {\bibinfo {title} {Critical behavior of the
  three-dimensional {Ising} model: {A} high-resolution {Monte} {Carlo} study},\
  }\href {https://doi.org/10.1103/PhysRevB.44.5081} {\bibfield  {journal}
  {\bibinfo  {journal} {Phys. Rev. B}\ }\textbf {\bibinfo {volume} {44}},\
  \bibinfo {pages} {5081} (\bibinfo {year} {1991})}\BibitemShut {NoStop}%
\bibitem [{\citenamefont {Ferrenberg}\ and\ \citenamefont
  {Swendsen}(1988)}]{ferrenberg_new_1988}%
  \BibitemOpen
  \bibfield  {author} {\bibinfo {author} {\bibfnamefont {A.~M.}\ \bibnamefont
  {Ferrenberg}}\ and\ \bibinfo {author} {\bibfnamefont {R.~H.}\ \bibnamefont
  {Swendsen}},\ }\bibfield  {title} {\bibinfo {title} {New {Monte} {Carlo}
  technique for studying phase transitions},\ }\href
  {https://doi.org/10.1103/PhysRevLett.61.2635} {\bibfield  {journal} {\bibinfo
   {journal} {Phys. Rev. Lett.}\ }\textbf {\bibinfo {volume} {61}},\ \bibinfo
  {pages} {2635} (\bibinfo {year} {1988})}\BibitemShut {NoStop}%
\bibitem [{\citenamefont {Ziff}\ and\ \citenamefont {Newman}(2002)}]{ziff:02}%
  \BibitemOpen
  \bibfield  {author} {\bibinfo {author} {\bibfnamefont {R.~M.}\ \bibnamefont
  {Ziff}}\ and\ \bibinfo {author} {\bibfnamefont {M.}~\bibnamefont {Newman}},\
  }\bibfield  {title} {\bibinfo {title} {Convergence of threshold estimates for
  two-dimensional percolation},\ }\href@noop {} {\bibfield  {journal} {\bibinfo
   {journal} {Phys. Rev. E}\ }\textbf {\bibinfo {volume} {66}},\ \bibinfo
  {pages} {016129} (\bibinfo {year} {2002})}\BibitemShut {NoStop}%
\bibitem [{\citenamefont {Akritidis}\ \emph {et~al.}(2022)\citenamefont
  {Akritidis}, \citenamefont {Fytas},\ and\ \citenamefont
  {Weigel}}]{akritidis_corrections_2022}%
  \BibitemOpen
  \bibfield  {author} {\bibinfo {author} {\bibfnamefont {M.}~\bibnamefont
  {Akritidis}}, \bibinfo {author} {\bibfnamefont {N.~G.}\ \bibnamefont
  {Fytas}},\ and\ \bibinfo {author} {\bibfnamefont {M.}~\bibnamefont
  {Weigel}},\ }\bibfield  {title} {\bibinfo {title} {Corrections to scaling in
  geometrical clusters of the {2D} {Ising} model},\ }\href
  {https://doi.org/10.1088/1742-6596/2207/1/012004} {\bibfield  {journal}
  {\bibinfo  {journal} {J. Phys.: Conf. Ser.}\ }\textbf {\bibinfo {volume}
  {2207}},\ \bibinfo {pages} {012004} (\bibinfo {year} {2022})}\BibitemShut
  {NoStop}%
\bibitem [{\citenamefont {Efron}(1979)}]{efron}%
  \BibitemOpen
  \bibfield  {author} {\bibinfo {author} {\bibfnamefont {B.}~\bibnamefont
  {Efron}},\ }\bibfield  {title} {\bibinfo {title} {Bootstrap methods: Another
  look at the jackknife},\ }\href {https://www.jstor.org/stable/2958830}
  {\bibfield  {journal} {\bibinfo  {journal} {The Annals of Statistics}\
  }\textbf {\bibinfo {volume} {7}},\ \bibinfo {pages} {1} (\bibinfo {year}
  {1979})}\BibitemShut {NoStop}%
\end{thebibliography}%
	
\clearpage

\end{document}